\documentclass[12pt]{article}
\pdfoutput=1
\usepackage{jheppub}
\usepackage{ucs}
\usepackage[utf8x]{inputenc}
\usepackage{amsmath}
\usepackage{amsfonts}
\usepackage{amssymb}
\usepackage{amsthm}
\usepackage{graphicx}
\usepackage{subfigure}
\usepackage[american]{babel}
\usepackage{fontenc}
\usepackage{graphicx}
\usepackage{mathtools}
\usepackage{float}
\usepackage{bm}
\usepackage{ulem}
\usepackage[sans]{dsfont}

\title{
Zero-Range Effective Field Theory \\
for Resonant Wino  Dark Matter\\  III. Annihilation Effects}
\author[a]{Eric Braaten,}
\author[a]{Evan Johnson,}
\author[a,b]{and Hong Zhang}

\affiliation[a]{Department of Physics,
         The Ohio State University, Columbus, OH\ 43210, USA}
\affiliation[b]{Physik Department T31, Technische Universit\"at M\"unchen,\\ James Franck Stra\ss e 1, D-85748 Garching, Germany}

\emailAdd{braaten.1@osu.edu}
\emailAdd{johnson.6036@osu.edu}
\emailAdd{hong.zhang@tum.de}

\abstract{
Near  a critical value of  the wino mass where there is a zero-energy S-wave resonance at the neutral-wino-pair threshold, low-energy winos can be described by a zero-range effective field theory (ZREFT) in which the winos interact nonperturbatively through a contact interaction and through Coulomb interactions. The effects of wino-pair annihilation  into electroweak gauge bosons are taken into account through the analytic continuation of the real parameters for the contact interaction to complex values. The parameters of ZREFT can be determined by matching wino-wino scattering amplitudes calculated  by solving the Schr\"odinger equation for  winos interacting through a real potential due to the exchange of electroweak gauge bosons and an imaginary potential due to wino-pair annihilation into electroweak gauge bosons. ZREFT at leading order gives an accurate analytic description of low-energy wino-wino scattering, inclusive wino-pair annihilation, and a wino-pair bound state. ZREFT can also be applied to partial annihilation rates, such as the Sommerfeld enhancement of the annihilation rate of wino pairs into monochromatic photons.}
\smallskip

\begin{document}
\maketitle
\flushbottom

\newpage

\section{Introduction}
\label{sec:Introduction}

A weakly interacting massive particle ({\it wimp}) is one of the best motivated candidates for a dark-matter particle that provides  most of the mass of the universe. A stable particle with weak interactions and whose mass is roughly at the electroweak scale is naturally produced in the early universe with a relic abundance comparable to the observed mass density of dark matter \cite{Kolb:1990vq,Steigman:2012nb}. If the wimp mass $M$ is in the TeV range, the  self-interactions of nonrelativistic wimps are complicated by a nonperturbative effect pointed out by Hisano et al.\ \cite{Hisano:2002fk}: ladder diagrams from the exchange of weak gauge bosons between wimps must be summed to all orders. There are critical values of the wimp mass where the resummation produces a resonance at the wimp-pair threshold (defined by the zero-energy limit for the scattering of a pair of wimps). If the wimp mass is near such a critical mass, the annihilation rate of a pair of wimps into electroweak gauge bosons can be enhanced by orders of magnitude  \cite{Hisano:2003ec,Hisano:2004ds}. This increase in the annihilation rate has been labeled a ``Sommerfeld enhancement'' \cite{ArkaniHamed:2008qn}. Wimp-wimp scattering cross sections at low relative velocity can also be increased by orders of magnitude near the critical mass, which can affect the relic abundance of dark matter \cite{Hisano:2006nn,Cirelli:2007xd}.

A resonance in an S-wave channel can generally produce a more dramatic enhancement over a broader range of $M$ than a resonance in a channel with higher orbital angular momentum. There is also a qualitative difference between a near-threshold resonance in an S-wave channel and in a channel for a higher partial wave. The S-wave resonance generates dynamically a length scale that is much larger than the range of the interactions. This length scale is the absolute value of the S-wave scattering length $a$, which can be orders of magnitude larger than the range. If there are no pair-annihilation channels, the scattering length can even be infinitely large. 

In a fundamental quantum field theory, wimps interact through the exchange of electroweak gauge bosons to which they couple through local gauge interactions. The enhancement of low-energy wimp-wimp cross sections and wimp-pair annihilation rates can be calculated by summing an infinite set of diagrams in that quantum field theory. The enhancement can be calculated more simply using a nonrelativistic effective field theory (NREFT) in which the wimps have instantaneous interactions at a distance through a potential generated by the exchange of the weak gauge bosons and in which charged wimps also have local couplings to the electromagnetic field. Few-body reaction rates of nonrelativistic wimps in NREFT can be calculated by the numerical solution of a Schr\"odinger equation \cite{Hisano:2002fk}. The effects of wimp-pair annihilation can be taken into account through a delta-function potential in the Schr\"odinger equation with an imaginary coefficient. A thorough development of NREFT for nearly degenerate neutralinos and charginos in the MSSM has been presented in ref.~\cite{Beneke:2012tg}.

The effects of wimp-pair annihilation on low-energy wimps are suppressed by a factor of $\alpha_2m_W/M$, where $\alpha_2$ is the $SU(2)$ gauge coupling constant, $m_W$ is the mass of the $W$ boson, and $M$ is the wimp mass. If $M$ is in the TeV region, this suppression factor is roughly $10^{-3}$. Most previous calculations of wimp-pair annihilation have been carried out at leading order in this suppression factor. At this order, a wimp-pair annihilation rate reduces to the product of the annihilation rate at leading order in the electroweak interactions and a velocity-dependent ``Sommerfeld enhancement factor'' $S(v)$ that can be calculated  numerically by solving a Schr\"odinger equation in NREFT with the real-valued potential only. This approximation breaks down sufficiently close to a resonance, because it does not take into account the unitarization of wimp-pair annihilation \cite{Blum:2016nrz}. The most glaring symptom of this breakdown is that at a critical mass for an S-wave resonance at the threshold, the Sommerfeld enhancement factor diverges as $1/v^2$ in the limit $v \to 0$.

In the case of an S-wave resonance near threshold, low-energy wimps can be described more simply using a zero-range effective field theory (ZREFT) in which the weak interactions between wimps from the exchange of weak gauge bosons are replaced by zero-range interactions between wimps. ZREFT exploits the large length scale that is generated dynamically  by an S-wave resonance. The real part of the inverse S-wave scattering length $1/a$ is zero at critical values of the wimp mass $M$. ZREFT is applicable if $M$ is close enough to a critical value that $1/|a|$  is  small compared to the inverse range $m_W$ of the weak interactions. The simplification provided by ZREFT is that the momentum scale $m_W$ is not treated explicitly, but is taken into account through the parameters of ZREFT.  ZREFT can therefore be used to calculate analytically observables that must be calculated numerically in NREFT, such as wimp-wimp cross sections for wimps with relative momentum less than $m_W$. There have been several previous applications of zero-range effective field theories to dark matter with resonant S-wave self-interactions. Braaten and Hammer pointed out that the elastic scattering cross section of the dark-matter particles, their total annihilation cross section, and the binding energy and width of a dark-matter bound state are all determined by the complex S-wave scattering length \cite{Braaten:2013tza}. Laha and Braaten studied the nuclear recoil energy spectrum in dark-matter direct detection experiments due to both elastic scattering and breakup scattering of an incident dark-matter bound state \cite{Laha:2013gva}. Laha extended that analysis to the angular recoil spectrum in directional detection experiments \cite{Laha:2015yoa}.

In ref.~\cite{Braaten:2017gpq}, we developed the ZREFT for wimps that consist of the neutral dark-matter particle $w^0$ and charged wimps $w^+$ and $w^-$ with a slightly larger mass. We refer to these wimps as {\it winos}, because the fundamental theory describing them could be the minimal supersymmetric standard model (MSSM) in a region of parameter space where the neutral wino is the lightest supersymmetric particle. The ZREFT for winos  can be organized into a systematically improvable effective field theory by expanding around a renormalization group fixed point. At the RG fixed point, the mass splitting between charged winos and neutral winos is zero, the electromagnetic interactions are turned off, the S-wave unitarity bound is saturated in a scattering channel that is a linear combination of $w^0 w^0$ and $w^+ w^-$, and there is no scattering in the orthogonal channel. In ref.~\cite{Braaten:2017gpq}, we calculated the wino-wino cross sections analytically in ZREFT without electromagnetism at leading order (LO) and at next-to-leading order (NLO) in the ZREFT power counting. The interaction parameters of ZREFT at LO and at NLO were determined by matching numerical results for scattering amplitudes obtained by solving the Schr\"odinger equation for NREFT. ZREFT at LO  gives fairly accurate predictions for most of the low-energy wino-wino cross sections. ZREFT at NLO  gives systematically improved predictions for all the wino-wino cross sections.

In ref.~\cite{Braaten:2017kci}, we extended the results in ref.~\cite{Braaten:2017gpq} to include electromagnetism by carrying out the Coulomb resummation of  diagrams in which photons are exchanged between pairs of charged winos. We showed that ZREFT at LO  gives good predictions for low-energy wino-wino cross sections. In particular, it reproduces the resonances in the neutral-wino elastic cross section just below the charged-wino-pair threshold. 

In this paper, we extend the results in refs.~\cite{Braaten:2017gpq} and \cite{Braaten:2017kci} by taking into account the effects of wino-pair annihilation in ZREFT through the analytic continuation of real interaction parameters to complex values. The two complex interaction parameters of ZREFT at LO are determined by matching scattering amplitudes with numerical results  obtained by solving the Schr\"odinger equation for NREFT with the imaginary delta-function potential included. We show that ZREFT at LO  gives good predictions for low-energy wino-wino scattering, inclusive wino-pair annihilation, and a wino-pair bound state. In particular, the analytic  expression for the Sommerfeld enhancement factor in ZREFT at LO correctly reproduces the velocity dependence from NREFT, even at very low velocities near the critical wino mass where the unitarization of wimp-pair annihilation is important.

This paper is organized as follows. We begin in section~\ref{sec:winoFTs} by summarizing various quantum field theories that can be used to describe nonrelativistic winos, including the fundamental theory, NREFT, and ZREFT. In section~\ref{sec:NREFT}, we  use the Schr\"odinger equation of NREFT with an imaginary delta-function potential to numerically calculate neutral-wino elastic cross sections and neutral-wino-pair annihilation rates. In section~\ref{sec:ZRMann}, we show how the effects of wino-pair annihilation can be taken into account analytically in a  field theory with zero-range interactions by analytically continuing real interaction parameters to complex values. In section~\ref{sec:ZREFTann}, we present analytic results for low-energy two-body observables in ZREFT at LO with wino-pair annihilation taken into account through the  analytic continuation of its parameters. We determine the two complex parameters of ZREFT at LO  by matching low-energy neutral-wino scattering amplitudes calculated  numerically in NREFT. We compare the resulting predictions of ZREFT at LO for  wino-wino cross sections, for inclusive wino-pair annihilation rates, and for the binding energy and width of a wino-pair bound state with numerical results from solving the Schr\"odinger equation for NREFT. Finally, we show how ZREFT can be used to calculate partial annihilation rates, such as the Sommerfeld enhancement of the annihilation rate of a neutral-wino pair into a monochromatic photon. Our results are summarized in section~\ref{sec:Conclusion}.

\section{Field Theories for Nonrelativistic Winos}
\label{sec:winoFTs}

In this Section, we summarize field theories that can be used to describe nonrelativistic winos, including the fundamental theory and the effective field theories NREFT and ZREFT.

\subsection{Fundamental theory}
\label{sec:QFT}

We assume the dark-matter particle is the neutral member of an $SU(2)$ triplet of Majorana fermions with zero hypercharge. The Lorentz-invariant quantum field theory that provides a fundamental  description of these fermions could simply be an extension of the Standard Model with this additional $SU(2)$ multiplet and with a symmetry that forbids the decay of the fermion into Standard Model particles. The fundamental theory could also be the Minimal Supersymmetric Standard Model (MSSM) in a region of parameter space where the lightest supersymmetric particle is a wino-like neutralino. In either case, we refer to  the particles in the $SU(2)$ multiplet as {\it winos}. We denote the neutral wino by $w^0$ and the charged winos by $w^+$ and $w^-$.

The relic density of the neutral wino is compatible with the observed mass density of dark matter if the neutral wino mass $M$  is roughly at the electroweak scale \cite{Kolb:1990vq}. We are particularly interested in a mass $M$ at the TeV scale so that effects from the exchange of electroweak gauge bosons between nonrelativistic winos must be summed to all orders. For the neutral wino to be stable, the charged wino must have a larger mass $M+\delta$. In the MSSM, the mass splitting $\delta$ arises from radiative corrections. The splitting is approximately  170~MeV, and it is insensitive to $M$ \cite{Cheng:1998hc,Feng:1999fu,Gherghetta:1999sw,Ibe:2012sx}.

The winos can be represented by a triplet of Majorana spinor fields. We take the neutral wino mass $M$ to be an adustable parameter, and we keep the wino mass splitting fixed at $\delta= 170$~MeV. The most important interactions of the winos are those with the electroweak gauge bosons: the photon, the $W^\pm$, and the $Z^0$. The relevant Standard Model parameters are the mass $m_W = 80.4$~GeV of the $W^\pm$,  the mass $m_Z = 91.2$~GeV of the $Z^0$, the $SU(2)$ coupling constant $\alpha_2= 1/29.5$, the electromagnetic coupling constant $\alpha = 1/137.04$, and the weak mixing angle, which is given by $\sin^2 \theta_w = 0.231$.

\begin{figure}[t]
\centering
\includegraphics[width=0.98\linewidth]{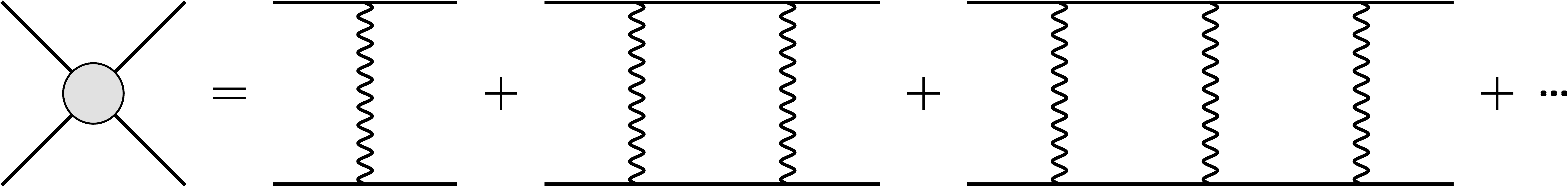}
\caption{Feynman diagrams in the fundamental theory for wino-wino scattering through the exchange of electroweak gauge bosons. The solid lines are neutral winos or charged winos, and the wavy lines are electroweak gauge bosons. If $M$ is sufficiently large and the winos are nonrelativistic, these ladder diagrams must be summed to all orders.}
\label{fig:FundamentalSum}
\end{figure}

Hisano, Matsumoto, and Nojiri pointed out that if the mass of the wino is large enough that $\alpha_2 M$ is of order $m_W$ or larger, loop diagrams in which electroweak gauge bosons are exchanged between nonrelativistic winos are not suppressed \cite{Hisano:2002fk}. The electroweak interactions between a pair of nonrelativistic winos must therefore be treated nonperturbatively by summing ladder diagrams from the exchange of electroweak bosons between the winos to all orders. For wino-wino scattering, the first few diagrams in the sum are shown in figure~\ref{fig:FundamentalSum}. Each of the wavy lines is summed over the gauge bosons $\gamma$, $W^+$, $W^-$, and $Z^0$ that can be exchanged between winos.

\begin{figure}[t]
\centering
\includegraphics[width=0.6\linewidth]{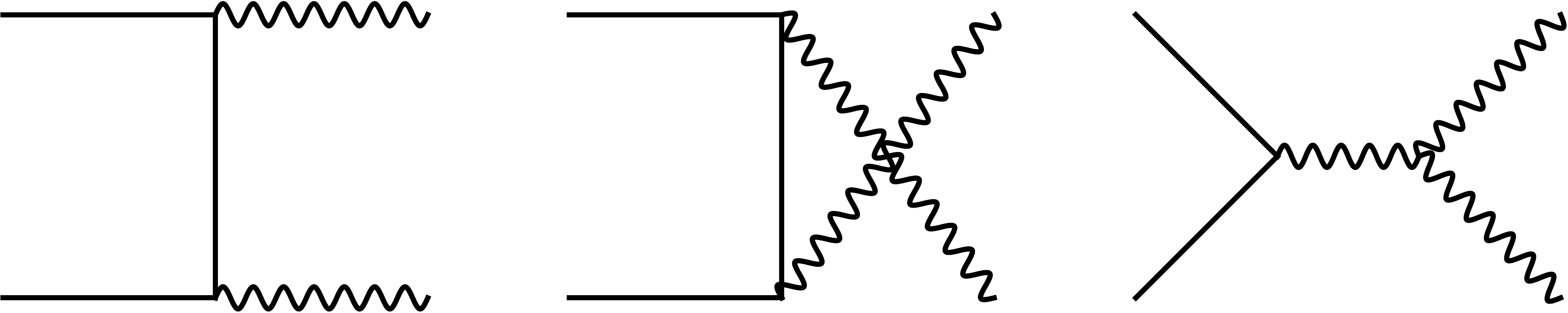}
\caption{Feynman diagrams in the fundamental theory for the annihilation of a wino pair into  electroweak gauge bosons at order $\alpha_2$. If $M$ is sufficiently large and the winos are nonrelativistic, the ladder diagrams from the exchange of electroweak gauge bosons between the incoming wino lines must  be summed to all orders.}
\label{fig:WinoPairAnnihilation}
\end{figure}

A pair of winos can annihilate into a pair of electroweak gauge bosons through the Feynman diagrams in figure~\ref{fig:WinoPairAnnihilation}. A neutral-wino pair $w^0 w^0$ can annihilate at order $\alpha_2^2$ into $W^+W^-$. A charged-wino pair $w^+ w^-$  can annihilate at order $\alpha_2^2$ into $\gamma\gamma$, $\gamma Z^0$, $Z^0Z^0$, and $W^+W^-$. The inclusive annihilation rates at threshold are \cite{Hisano:2003ec}
\begin{subequations}
\begin{eqnarray}
2v_0\sigma_{0,\mathrm{ann}} &=& \frac{2\pi \alpha_2^2}{M^2},
\label{eq:sig0ann-thresh}
\\
2v_1\sigma_{1,\mathrm{ann}} &=&   \frac{3\pi\alpha_2^2}{2M^2}.
\label{eq:sig1ann-thresh}
\end{eqnarray}
\label{eq:sig0,1ann-thresh}%
\end{subequations}
The prefactors $2v_i$ on the left sides are the relative velocities of the wino pairs. The annihilation rates of $w^0 w^0$ into $\gamma\gamma$, $\gamma Z^0$, and $Z^0Z^0$ at leading order come from one-loop diagrams in which a $W$ is exchanged between the incoming neutral winos. The annihilation rates into the final states that include a monochromatic photon are \cite{Bergstrom:1997fh,Bern:1997ng,Ullio:1997ke}
\begin{subequations}
\begin{eqnarray}
2v_0\sigma[w^0 w^0 \to \gamma \gamma] &=& \frac{4\pi s_w^4 \alpha_2^4}{m_W^2},
\label{eq:sig0-gammagamma}
\\
2v_0\sigma[w^0 w^0 \to \gamma Z^0]  &=&   \frac{8\pi s_w^2 c_w^2 \alpha_2^4}{m_W^2},
\label{eq:sig0-gammaZ}
\end{eqnarray}
\label{eq:sig0-monogamma}%
\end{subequations}
where $s_w = \sin \theta_w$ and $c_w = \cos \theta_w$. These annihilation rates differ from that for $w^0 w^0 \to W^+ W^-$ in eq.~\eqref{eq:sig0ann-thresh} by a factor of order $\alpha_2^2 M^2/m_W^2$.

If the winos are nonrelativistic, the corrections to the annihilation rates proportional to powers of $\alpha_2M/m_W$ are not suppressed if  $\alpha_2 M$ is order $m_W$. To include these corrections, the ladder diagrams from the exchange of electroweak gauge bosons between the incoming winos must  be summed to all orders. The effect on the wino-pair annihilation rates in eqs.~\eqref{eq:sig0,1ann-thresh} and \eqref{eq:sig0-monogamma} is to multiply them by Sommerfeld enhancement factors that depend on the relative velocity of the annihilating winos.

\subsection{Nonrelativistic  effective field theory}
\label{sec:NonrelativisticEFT}

Low-energy winos can be described by a nonrelativistic effective field theory in which they  interact through potentials from the exchange of weak gauge bosons and in which charged winos also have local couplings to the electromagnetic field. We call this effective field theory {\it NREFT}. In NREFT, the nonrelativistic wino fields are 2-component spinor fields: $\zeta$ which annihilates a neutral wino $w^0$, $\eta$ which annihilates a charged wino $w^-$, and $\xi$ which creates a charged wino $w^+$. The kinetic terms for winos in the Lagrangian density are
\begin{equation}
\mathcal{L}_{\rm kinetic} = \zeta^\dagger\left(i\partial_0+\frac{\bm{\nabla}^2}{2M}\right) \zeta 
+ \eta^\dagger\left(iD_0+\frac{\bm{D}^2}{2M} -\delta\right) \eta
+ \xi^\dagger\left(iD_0-\frac{\bm{D}^2}{2M} +\delta\right) \xi,
\label{eq:kineticLHMN}
\end{equation}
where $D_0$ and $\bm{D}$ are electromagnetic covariant derivatives acting on the charged wino fields. The neutral and charged winos have the same kinetic mass $M$, and the wino mass splitting $\delta$ is taken into account through the rest energy of the charged winos. The weak interaction terms in the Hamiltonian are instantaneous interactions at a distance through a potential produced by the exchange of the $W^\pm$ and $Z^0$ gauge bosons:
\begin{eqnarray}
H_{\rm weak} &=& - \frac12 \int\!\! d^3x \int\!\! d^3y
\bigg(   \frac{\alpha_2 \cos^2 \theta_w}{|\bm{x}-\bm{y}|} e^{-m_Z|\bm{x}-\bm{y}|} 
\eta^\dagger(\bm{x}) \xi(\bm{y})\,  \xi^\dagger(\bm{y}) \eta(\bm{x})~~
\nonumber\\
&&+ \frac{\alpha_2}{|\bm{x}-\bm{y}|}  e^{-m_W|\bm{x}-\bm{y}|} 
\left[ \zeta^\dagger(\bm{x}) \zeta^c(\bm{y})\,  \xi^\dagger(\bm{y}) \eta(\bm{x})
+ \zeta^{c\dagger}(\bm{x}) \zeta(\bm{y})\,  \eta^\dagger(\bm{y}) \xi(\bm{x})\right] \bigg),
\label{eq:LintHMN}
\end{eqnarray}
where $\zeta^c = -i \sigma_2 \xi^*$ and $\sigma_2$ is a Pauli matrix. The potentials from the exchange of $W^\pm$ and $Z^0$ have ranges of order $1/m_W$.

The amplitudes for wino-wino scattering in NREFT can be represented diagrammatically by the same sum of ladder diagrams as in figure~\ref{fig:FundamentalSum}, except that each wavy line is the sum of an exchanged photon line and instantaneous interactions at a distance through the potentials from $W^\pm$ and $Z^0$ exchange in eq.~\eqref{eq:LintHMN}.

\begin{figure}[t]
\centering
\includegraphics[width=1\linewidth]{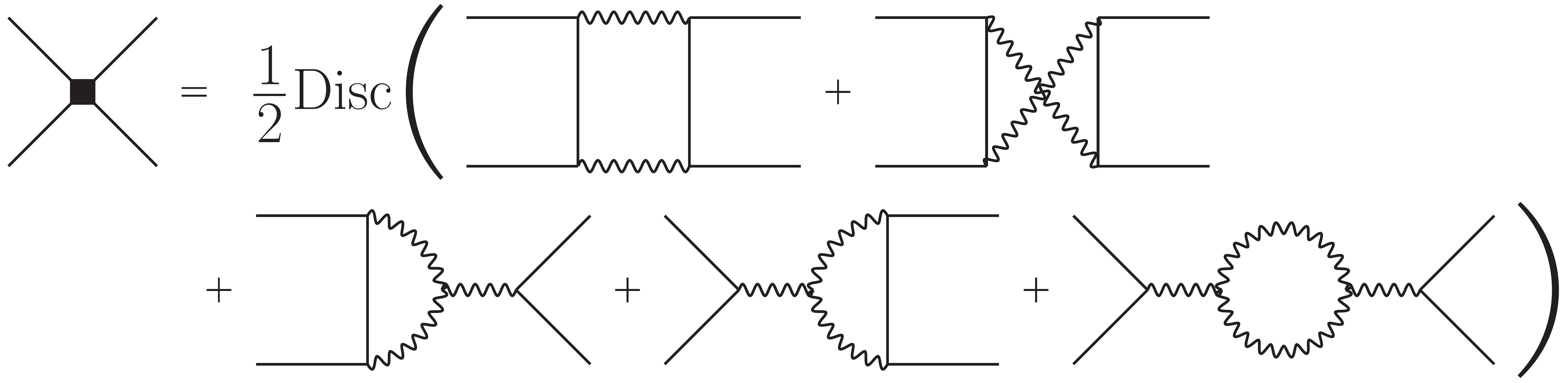}
\caption{Effective vertex of order $\alpha_2^2/M^2$ for a wino-pair contact interaction from wino-pair annihilation into two electroweak gauge bosons.}
\label{fig:AnnihilationVertex}
\end{figure}

Wino-pair annihilation into electroweak gauge bosons cannot be described explicitly in NREFT, because the  electroweak gauge bosons have momenta of order $M$, which is beyond the domain of validity of the effective theory. However the effects of wino-pair annihilation on low-energy winos can be taken into account through local anti-Hermitian terms in the Hamiltonian density. The leading term in the expansion in powers of the wino 3-momenta divided by $M$ is a wino-pair contact interaction term  \cite{Hisano:2004ds}: \footnote{The coefficients of $\zeta^\dagger \zeta^c\,  \xi^\dagger \eta$ and $\zeta^{c\dagger} \zeta \,  \eta^\dagger \xi$ in ref.~\cite{Hisano:2004ds} have been multiplied by  a factor of 2 to make them consistent with  a later paper by some of the same authors \cite{Hisano:2006nn}.}
\begin{eqnarray}
{\cal H}_{\rm annihilation} = -i \frac{\pi \alpha_2^2}{2 M^2} \Big[
4\, \zeta^\dagger\zeta^c \zeta^{c\dagger} \zeta
+ 2\left(\zeta^\dagger \zeta^c\,  \xi^\dagger \eta + \zeta^{c\dagger} \zeta \,  \eta^\dagger \xi \right)
+  3\,  \eta^\dagger \xi\,  \xi^\dagger \eta \Big],
\label{eq:Hintann1}
\end{eqnarray}
where $\alpha_2$ is the $SU(2)$ coupling constant. The coefficients in eq.~\eqref{eq:Hintann1} are obtained by calculating the cuts of the one-loop diagrams for wino-wino scattering in figure~\ref{fig:AnnihilationVertex}, with the cuts passing through two electroweak gauge bosons.

The effects of annihilation on wino-wino scattering can be taken into account in NREFT by replacing each wavy line in the ladder diagram in figure~\ref{fig:FundamentalSum} by the sum of an exchanged photon line, instantaneous weak interactions at a distance through the potentials from exchange of $W^\pm$ and $Z^0$, and contact interactions between winos from eq.~\eqref{eq:Hintann1}. The inclusive annihilation rate of a wino pair can be obtained at leading order in $\alpha_2^2/M^2$ by summing the ladder diagrams for the electroweak interactions of a wino pair before its annihilation into electroweak gauge bosons through the effective vertex in figure~\ref{fig:AnnihilationVertex}. This resummation of the annihilation rate for a pair of neutral winos was first carried out by Hisano, Matsumoto, and Nojiri \cite{Hisano:2003ec,Hisano:2004ds}. They discovered dramatic enhancements of the annihilation rate into a pair of electroweak gauge bosons at threshold at a sequence of critical values of $M$. Near these critical values, the annihilation rate at threshold is increased by orders of magnitude. For $\delta = 170$~MeV, the first such resonance is an S-wave resonance at $M_*=2.39$~TeV.

There are many important momentum scales for nonrelativistic winos. The momentum scale of the electroweak gauge bosons from wino-pair annihilation is $M$. The inverse range of the weak interactions is $m_W = 80.4$~GeV. The momentum scale below which weak interactions are nonperturbative is $\alpha_2 M$. The momentum scale below which electromagnetic interactions are nonperturbative is the  Bohr momentum $\alpha M$. Another important momentum scale is the scale $\sqrt{2M\delta}$ associated with transitions between a neutral-wino pair and a charged-wino pair. For $\delta = 170$~MeV and the first resonance mass $M_* = 2.39$~TeV, these momentum scales are $\alpha_2 M_* = 81.1$~GeV, $\alpha M_* = 17.5$~GeV, and $\sqrt{2M_*\delta}=28.5$~GeV. The dimensionless coupling constants are $\alpha_2$ and $\alpha$. Another important dimensionless factor is the suppression factor $\alpha_2 m_W/M$ for wino-pair annihilation effects, which is about $10^{-3}$ if $M = 2.39$~TeV.

\subsection{Zero-range model}
\label{sec:ZRM}

There can be a resonance near the neutral-wino-pair threshold (defined by the zero-energy limit for the scattering of a pair of neutral winos) in any partial wave. An S-wave resonance near the threshold is special, because there is a dynamically generated length scale that is much larger than the range $1/m_W$ of the weak interactions \cite{Braaten:2004rn}. This length scale is the absolute value of the neutral-wino scattering length $a_0$. The corresponding momentum scale $1/|a_0|$ can be much smaller than any of the other momentum scales provided by the interactions described above. For winos with relative momenta small compared to $m_W$, the effects of the exchange of weak bosons can be mimicked by zero-range interactions. Thus winos with sufficiently low energy can be described by a nonrelativistic field theory with local interactions between winos and with local couplings of the charged winos to the electromagnetic field. This remains true even if there is an S-wave resonance near the neutral-wino-pair threshold. In this case, the zero-range interactions must be nonperturbative, because otherwise they cannot generate the large length scale $|a_0|$.

A simple nonrelativistic field theory for low-energy winos with local interactions is the {\it Zero-Range Model} introduced in ref.~\cite{Braaten:2017gpq}. The winos are described by nonrelativistic two-component spinor fields $w_0$, $w_+$, and $w_-$ that annihilate $w^0$, $w^+$, and $w^-$, respectively. They can be identified with the fields $\zeta$, $\xi^\dagger$, and $\eta$ in NREFT, respectively. The kinetic terms for winos in the Lagrangian for the Zero-Range Model are
\begin{equation}
\mathcal{L}_{\rm kinetic} = w_0^\dagger\left(i\partial_0+\frac{\bm{\nabla}^2}{2M}\right) w_0  
+ \sum_\pm w_\pm^\dagger\left(iD_0+\frac{\bm{D}^2}{2M}-\delta\right) w_\pm ,
\label{eq:kineticL}
\end{equation}
where the electromagnetic covariant derivatives are
\begin{eqnarray}
D_0 w_\pm = (\partial_0 \pm ieA_0)w_\pm,
\qquad
\bm{D} w_\pm = (\bm{\nabla} \mp ie\bm{A})w_\pm.
\end{eqnarray}
The neutral and charged winos have the same kinetic mass $M$, and the mass splitting $\delta$ is taken into account through the rest energy of the charged wino. Since the neutral wino is a Majorana fermion, a pair of neutral winos can have an S-wave resonance at threshold only in the spin-singlet channel. That channel is coupled to the spin-singlet channel for charged winos. The Lagrangian for zero-range interactions in the spin-singlet channels can be expressed as
\begin{eqnarray}
\mathcal{L}_{\rm zero-range} &=& 
-\tfrac{1}{4} \lambda_{00} ( w_0^{c\dagger} w_0^{d\dagger} )
\tfrac12 ( \delta^{ac}\delta^{bd}- \delta^{ad}\delta^{bc}) ( w_0^a w_0^b )
\nonumber\\
&& 
-\tfrac{1}{2} \lambda_{01} (  w_+^{c\dagger} w_-^{d\dagger} )
\tfrac12 ( \delta^{ac}\delta^{bd}- \delta^{ad}\delta^{bc}) ( w_0^a w_0^b )
\nonumber\\
&& 
-\tfrac{1}{2} \lambda_{01} (  w_0^{c\dagger} w_0^{d\dagger} )
\tfrac12 ( \delta^{ac}\delta^{bd}- \delta^{ad}\delta^{bc}) ( w_+^a w_-^b )
\nonumber\\
&& 
- \lambda_{11} ( w_+^{c\dagger} w_-^{d\dagger} )
\tfrac12 ( \delta^{ac}\delta^{bd}- \delta^{ad}\delta^{bc})
( w_+^a w_-^b ),
\label{eq:ZRint}
\end{eqnarray}
where $\lambda_{00}$, $\lambda_{01}$, and $\lambda_{11}$ are bare coupling constants. The factor $\frac12( \delta^{ac}\delta^{bd}- \delta^{ad}\delta^{bc})$ is the projector onto the spin-singlet channel. In the absence of wino-pair annihilation, the bare coupling constants $\lambda_{ij}$ are constrained by unitarity to be real valued. If wino-pair annihilation is taken into account, the bare coupling constants are complex valued, with the imaginary parts suppressed by $\alpha_2 m_W/M$.

\begin{figure}[t]
\centering
\includegraphics[width=0.98\linewidth]{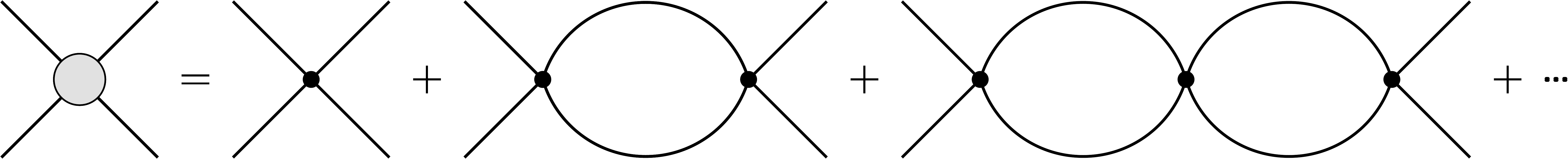}
\caption{Diagrams for wino-wino scattering in the Zero-Range Model without electromagnetism. The solid lines are either neutral winos $w^0w^0$ or charged winos $w^+ w^-$. The solid lines in each bubble are summed over $w^0w^0$ and $w^+ w^-$. The bubble diagrams  must be summed to all orders.}
\label{fig:ZREFTSum0}
\end{figure}

In the Zero-Range Model, the zero-range interactions must be treated nonperturbatively by summing bubble diagrams involving the vertices from the interaction terms in eq.~\eqref{eq:ZRint} to all orders. In the absence of electromagnetic interactions, the first few diagrams for wino-wino scattering are shown in figure~\ref{fig:ZREFTSum0}. In the Zero-Range Model with electromagnetism, the electromagnetic interactions must also be treated nonperturbatively by summing to all orders ladder diagrams in which Coulomb photons are exchanged between charged winos. The ladder diagrams inside each $w^+ w^-$ bubble in the diagrams in figure~\ref{fig:ZREFTSum0} must be summed to all orders. The ladder diagrams on external $w^+ w^-$ lines must also be summed to all orders. Finally, if the incoming and outgoing external lines  are both $w^+ w^-$, the ladder diagrams from Coulomb exchange with no zero-range interactions  must be summed to all orders.

\subsection{Zero-range effective field theory}
\label{sec:ZREFT}

An {\it effective field theory} can be  defined as a sequence of models with an increasing finite number of parameters that take into account corrections with systematically improving accuracy. A nonrelativistic effective field theory called {\it ZREFT} for winos that have an S-wave resonance near the neutral-wino-pair threshold was introduced in ref.~\cite{Braaten:2017gpq}. The winos interact through zero-range self-interactions and through local couplings to the electromagnetic field. Range corrections for S-wave interactions can be incorporated to all orders through terms in the Lagrangian with increasing numbers of  time derivatives acting on the fields \cite{Birse:1998dk}. If all possible range corrections are included, the theory has infinitely many parameters. 

An effective field theory can be defined most rigorously through deformations of a renormalization-group (RG) fixed point. Systematically improving accuracy is  ensured by adding to the Lagrangian operators with increasingly higher scaling dimensions. In ref.~\cite{Lensky:2011he}, Lensky and Birse carried out a careful RG analysis of the two-particle sector for a nonrelativistic field theory for distinguishable particles with two coupled scattering channels and with zero-range S-wave interactions. They identified three distinct RG fixed points. The first RG fixed point is the {\it noninteracting fixed point} at which the $2\times 2$ T-matrix is zero at all energies $E$: $\bm{\mathcal{T}}_*(E) = 0$. The second RG fixed point is the {\it two-channel-unitarity fixed point}, in which the cross sections saturate the S-wave unitarity bounds in both scattering channels. At this fixed point, the two scattering channels have the same threshold at $E=0$ and the T-matrix with the standard normalization of states in a nonrelativistic field theory is
\begin{equation}
\label{eq:Tfp2}
\bm{\mathcal{T}}_*(E) = \frac{4\pi i}{M \sqrt{ME}} \,
\begin{pmatrix} 1~  & ~0\\ 0~ & ~1 \end{pmatrix} ,
\end{equation}
where $M$ is the mass of the particle. The cross sections have the scaling behavior $1/E$. The power-law dependence on $E$ reflects the scale invariance of the interactions. In ref.~\cite{Lensky:2011he}, Lensky and Birse pointed out that there is a third RG fixed point: the {\it single-channel-unitarity fixed point}. At this fixed point, the two scattering channels have the same threshold at $E=0$ and the T-matrix is
\begin{equation}
\label{eq:Tfp3}
\bm{\mathcal{T}}_*(E) = \frac{4\pi i}{M \sqrt{ME}}
\begin{pmatrix} \cos^2\phi  & \cos\phi  \sin\phi \\ 
 \cos\phi  \sin\phi & \sin^2\phi \end{pmatrix}.
\end{equation}
There is nontrivial scattering in a single channel that is a linear combination of the two scattering channels  with mixing angle $\phi$. In that channel, the cross section saturates the S-wave unitarity bound. There is no scattering in the orthogonal channel. The single-channel-unitarity fixed point is the most natural one for describing a system with a single fine tuning, such as the tuning of the wino mass $M$ to a unitarity value where there is an S-wave resonance at the threshold.

In ref.~\cite{Lensky:2011he}, Lensky and Birse diagonalized the RG flow near the single-channel-unitarity fixed point whose T-matrix $\bm{\mathcal{T}}_*(E)$ is given  in eq.~\eqref{eq:Tfp3}, identifying all the scaling perturbations and their scaling dimensions. The coefficients of the scaling perturbations provide a complete parametrization of the T-matrix. There is one relevant scaling perturbation that corresponds to changing $\sqrt{ME}$ in the denominator in eq.~\eqref{eq:Tfp3} to  $\sqrt{ME} - i \gamma$, where $\gamma$ is a real parameter that can be interpreted as an inverse scattering length. There are two marginal scaling perturbations. One of them corresponds to turning on the splitting $2\delta$ between the thresholds in the two channels, and the other corresponds to changing the mixing angle $\phi$. All the other scaling perturbations are irrelevant. The inclusion of scaling perturbations with increasingly higher scaling dimensions defines the successive improvements of ZREFT. The parameters in ZREFT at leading order (LO) are $M$, $\delta$, the mixing angle $\phi$, and the inverse scattering length $\gamma$. There are two additional parameters in ZREFT at next-to-leading order (NLO), and there is one additional parameter at next-to-next-to-leading order  (NNLO).

ZREFT can  be extended to a nonrelativistic effective field theory for neutral winos, charged winos, and photons. In ZREFT at LO, the only electromagnetic coupling is that of the charged winos through the covariant derivatives acting on the charged wino fields in eq.~\eqref{eq:kineticL}. Thus including electromagnetism does not introduce any additional adjustable parameters at LO. In ZREFT beyond LO, there are additional parameters from interaction terms involving the gauge-invariant electromagnetic field strengths $\bm{E}$ and $\bm{B}$.
 
\section{NREFT}
\label{sec:NREFT}

In this section, we use NREFT to calculate the neutral-wino elastic cross section and the inclusive neutral-wino-pair annihilation rate. 

\subsection{Schr\"odinger equation}
\label{sec:SchrEq}

In NREFT, ladder diagrams from the exchange of electroweak gauge bosons between a pair of wimps can be summed to all orders by solving a Schr\"odinger equation. In the absence of wino-pair annihilation, the coupled-channel radial Schr\"odinger equation for S-wave scattering in the spin-singlet channel is
\begin{equation}
\left[ -\frac{1}{M} \begin{pmatrix} 1~  & ~0 \\ 0~ & ~1 \end{pmatrix}\left( \frac{d\ }{dr} \right)^2
+ 2\delta \begin{pmatrix} 0~  & ~0 \\ 0~ & ~1 \end{pmatrix}
+\bm{V}(r) \right] r \binom{R_0(r)}{R_1(r)} = E\,  r \binom{R_0(r)}{R_1(r)},
\label{eq:radialSchrEq}
\end{equation}
where $R_0(r)$ and $R_1(r)$ are the radial wavefunctions for a  pair of neutral winos and a pair of charged winos, respectively. The $2 \times 2$ matrix of potentials is
\begin{equation}
\bm{V}(r) = -  
\alpha_2 \begin{pmatrix}                0               & \sqrt{2}\, e^{-m_Wr}/r \\ 
                         \sqrt{2}\, e^{-m_Wr}/r  &  c_w^2\, e^{-m_Zr}/r   \end{pmatrix}
-  \alpha \begin{pmatrix} 0~  & ~0 \\ 0~ & ~1/r \end{pmatrix},
\label{eq:V-matrix}
\end{equation}
where $\alpha_2$ is the $SU(2)$ coupling constant and $c_w = \cos \theta_w$. The Schr\"odinger equation in eq.~\eqref{eq:radialSchrEq} has a continuum of positive energy eigenvalues $E$ that correspond to S-wave scattering states. There may also be discrete negative eigenvalues that correspond to S-wave bound states.

If wino-pair annihilation is taken into account, the coupled-channel radial Schr\"odinger equation for S-wave scattering in the spin-singlet channel is \cite{Hisano:2003ec}
\begin{equation}
\left[ -\frac{1}{M} \begin{pmatrix} 1~  & ~0 \\ 0~ & ~1 \end{pmatrix}\left( \frac{d\ }{dr} \right)^2
+ 2\delta \begin{pmatrix} 0~  & ~0 \\ 0~ & ~1 \end{pmatrix}
+\bm{V}(r) - i\frac{\delta(r)}{2 \pi r^2} \bm{\Gamma} \right] r \binom{R_0(r)}{R_1(r)} = E\,  r \binom{R_0(r)}{R_1(r)}.
\label{eq:radialSchrEqann}
\end{equation}
Wino-pair annihilation is taken into account through the imaginary delta-function potential multiplied by the $2\times2$ matrix\footnote{In Refs.~\cite{Hisano:2003ec} and \cite{Hisano:2004ds}, the off-diagonal elements of $\bm{\Gamma}$ differ by a factor of $\frac12$ from those in the later paper \cite{Hisano:2006nn} by some of the same authors.}
\cite{Hisano:2006nn}
\begin{equation}
\bm{\Gamma} = \frac{\pi \alpha_2^2}{2 M^2} \begin{pmatrix} 2  & \sqrt{2} \\ \sqrt2 & 3 \end{pmatrix}.
\label{eq:Gamma}
\end{equation}
The upper and lower diagonal entries of this matrix are half the leading-order annihilation rate of $w^0 w^0$ in eq.~\eqref{eq:sig0ann-thresh} and the leading-order annihilation rate of $w^+ w^-$  in eq.~\eqref{eq:sig1ann-thresh}, respectively. The Schr\"odinger equation in eq.~\eqref{eq:radialSchrEqann} has a continuum of positive energy eigenvalues $E$ that correspond to S-wave scattering states. There may also be discrete complex eigenvalues that correspond to  unstable S-wave bound states.

The coupled-channel radial Schr\"odinger equation in eq.~\eqref{eq:radialSchrEq} or eq.~\eqref{eq:radialSchrEqann} can be solved for the radial wavefunctions $R_0(r)$ and $R_1(r)$. For energy $E$ above the charged-wino-pair threshold $2 \delta$, the asymptotic solutions for $R_0(r)$ and $R_1(r)$ as $r \to \infty$ determine a dimensionless and symmetric $2 \times 2$ S-matrix $\bm{S}(E)$. The dimensionless $2\times2$ T-matrix $\bm{T}(E)$ is defined by
\begin{equation}
\bm{S}(E) =
\mathds{1} + i \, \bm{T}(E),
\label{eq:S-high}
\end{equation}
where $\mathds{1}$ is the $2\times 2$ unit matrix. If $\bm{\Gamma} = 0$  as in eq.~\eqref{eq:radialSchrEq}, the S-matrix is unitary and the T-matrix satisfies the unitarity equation
\begin{equation}
2 \, \textrm{Im}\, \bm{T}(E) = \bm{T} ^\dagger(E)\, \bm{T}(E) .
\label{eq:T-unitarity}
\end{equation}
If $\bm{\Gamma} \neq 0$  as in eq.~\eqref{eq:radialSchrEqann}, the difference between the left side and the right side of eq.~\eqref{eq:T-unitarity} can be interpreted as the annihilation contribution from intermediate states without any winos. These states include electroweak gauge bosons with momenta of order $M$, which are not described explicitly in NREFT. For energy in the range $0<E<2\delta$, the asymptotic solution for $R_0(r)$ determines a $1 \times 1$ S-matrix that can be expressed as $S_{00}(E) = \exp\big(2 i \delta_0(E)\big)$, where $\delta_0(E)$ is the S-wave phase shift. If $\bm{\Gamma} = 0$, the phase shift is  real valued.

We denote by $\sigma_{i \to j}(E)$ the contribution from the S-wave spin-singlet channel to the cross section for scattering from channel $i$ to channel $j$ at energy $E$, averaged over initial spins and summed over final spins. The expressions for these cross sections in terms of the T-matrix elements $T_{ji}$ are
\begin{subequations}
\begin{eqnarray}
\sigma_{0 \to j}(E) &=& \frac{2\pi}{M^2 v_0(E)^2} \big| T_{j0}(E) \big|^2,
\label{eq:sig0j-T}
\\
\sigma_{1\to j}(E) &=&   \frac{\pi}{M^2 v_1(E)^2}\big| T_{j1}(E) \big|^2,
\label{eq:sig1j-T}
\end{eqnarray}
\label{eq:sigij-T}
\end{subequations}
where $v_0(E)$ and $v_1(E)$ are the wino velocities in the center-of-mass frame for a neutral-wino pair and a charged-wino pair with total energy $E$:
\begin{subequations}
\label{eq:v0,1-E}
\begin{eqnarray}
v_0(E) &=&  \sqrt{E/M},
\label{eq:v0-E}
 \\
v_1(E) &=&  \sqrt{(E-2 \delta)/M}.
\label{eq:v1-E}
\end{eqnarray}
\end{subequations}
For the neutral-wino elastic cross section $\sigma_{0 \to 0}$, the energy threshold is $E=0$. For the other three cross sections $\sigma_{1 \to 0}$, $\sigma_{0 \to 1}$,  and $\sigma_{1 \to 1}$, the energy threshold is $E=2\delta$. The {\it S-wave unitarity bounds} for the scattering of $w^0w^0$, which are identical spin-$\tfrac12$ particles, and for the scattering of $w^+w^-$, which are distinguishable spin-$\tfrac12$ particles, are
\begin{subequations}
\begin{eqnarray}
\sigma_{0\to 0}(E)  &\le& \frac{8 \pi}{ME},
\label{sigma-unitarity0}
\\
\sigma_{1\to 1}(E)  &\le& \frac{4 \pi}{M(E-2\delta)}.
\label{sigma-unitarity1}
\end{eqnarray}
\label{sigma-unitarity}
\end{subequations}

By the optical theorem, the total cross sections for a neutral-wino pair and for a charged-wino pair are proportional to the imaginary parts of the appropriate T-matrix elements:
\begin{subequations}
\begin{eqnarray}
\sigma_{0,\mathrm{tot}}(E) &=& \frac{4\pi}{M^2 v_0(E)^2}  \mathrm{Im}T_{00}(E) ,
\label{eq:sig0tot-T}
\\
\sigma_{1,\mathrm{tot}}(E) &=&   \frac{2\pi}{M^2 v_1(E)^2} \mathrm{Im}T_{11}(E).
\label{eq:sig1tot-T}
\end{eqnarray}
\label{eq:sig01tot-T}
\end{subequations}
If there is wino-pair annihilation, the inclusive annihilation cross sections for a neutral-wino pair and for a charged-wino pair can be obtained by subtracting the cross sections into wino-pair final states, which are given in eqs.~\eqref{eq:sigij-T}:
\begin{subequations}
\begin{eqnarray}
\sigma_{0,\mathrm{ann}}(E) &=& \frac{2\pi}{M^2 v_0(E)^2} 
\left( 2\,  \mathrm{Im}T_{00}(E) -  \big| T_{00}(E) \big|^2 -   \big| T_{10}(E) \big|^2 \right),
\label{eq:sig0ann-T}
\\
\sigma_{1,\mathrm{ann}}(E) &=&   \frac{\pi}{M^2 v_1(E)^2}
\left( 2\,  \mathrm{Im}T_{11}(E) -  \big| T_{01}(E) \big|^2 -   \big| T_{11}(E) \big|^2 \right).
\label{eq:sig1ann-T}
\end{eqnarray}
\label{eq:sig0,1ann-T}%
\end{subequations}
The thresholds for $\sigma_{0,\mathrm{ann}}$ and $\sigma_{1,\mathrm{ann}}$ are $E>0$ and $E>2 \delta$, respectively. In eq.~\eqref{eq:sig0ann-T}, $T_{10}(E)$ should be interpreted as 0 if $0 < E < 2 \delta$.

For most observables that are nonzero in the absence of wino-pair annihilation, the effects of wino-pair annihilation are very small. To determine the suppression factor associated with wino-pair annihilation, we compare the annihilation potential with the weak-interaction potential. The prefactor of the delta function in the annihilation potential is proportional to $\alpha_2^2/M^2$. The weak-interaction potential can be modeled by a delta function together with a short-distance cutoff of order $1/m_W$. The prefactor of the delta function is proportional to $\alpha_2/m_WM$. The factor of $1/m_W$ takes into account the linear ultraviolet divergence of the delta-function interaction between nonrelativistic particles. Comparing the prefactors, we find that the suppression factor for wino-pair annihilation is $\alpha_2 m_W/M$, which is about $10^{-3}$. Observables can be expanded in powers of $\bm{\Gamma}$ except near a resonance where there are denominators that vanish in the limit $\bm{\Gamma} \to 0$. Most previous  calculations of Sommerfeld enhancement factors have been at leading order in $\bm{\Gamma}$ and were calculated by solving a Schr\"odinger equation with the real-valued potential only. The calculations break down at low relative velocity, where the unitarization of wino-pair annihilation becomes important.

\subsection{Neutral-wino scattering}
\label{sec:w0Scattering}

\begin{figure}[t]
\centering
\includegraphics[width=0.6\linewidth]{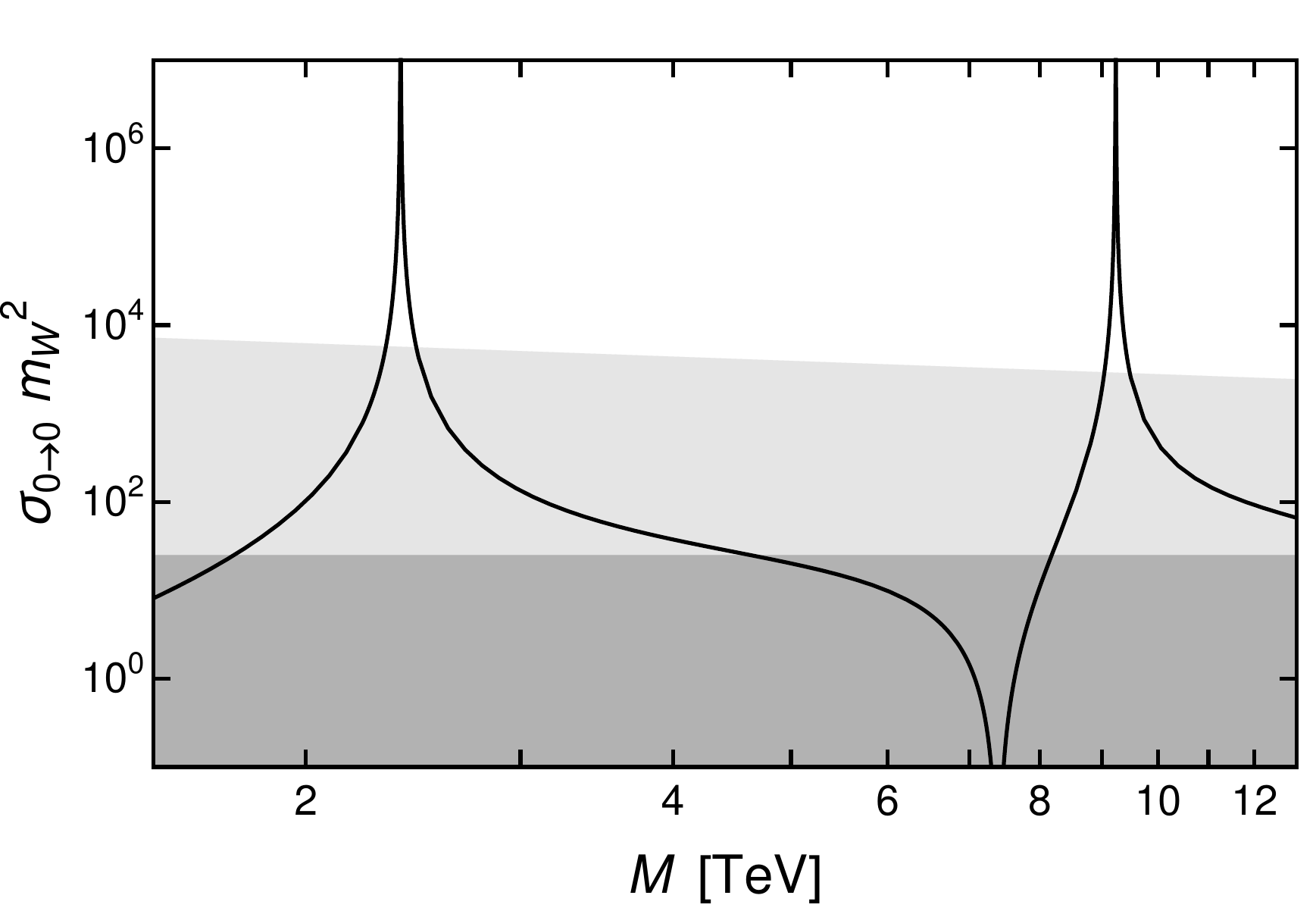}
\caption{Neutral-wino elastic cross section $\sigma_{0 \to 0}$ at zero energy as a function of the wino mass $M$. The peaks have finite maxima that cannot be seen in the plot. In a range of $M$ where $\sigma_{0 \to 0}$ is above the  darker shaded region ($\sigma_{0 \to 0} < 8 \pi/m_W^2$), the ZREFT for neutral and charged winos is applicable. In a range of $M$ where $\sigma_{0 \to 0}$ is above the lighter shaded region ($\sigma_{0 \to 0} < 8 \pi/\Delta^2$), a simpler ZREFT for neutral winos only is applicable.}
\label{fig:sigma00vsM}
\end{figure}

\begin{figure}[th]
\centering
\includegraphics[width=0.6\linewidth]{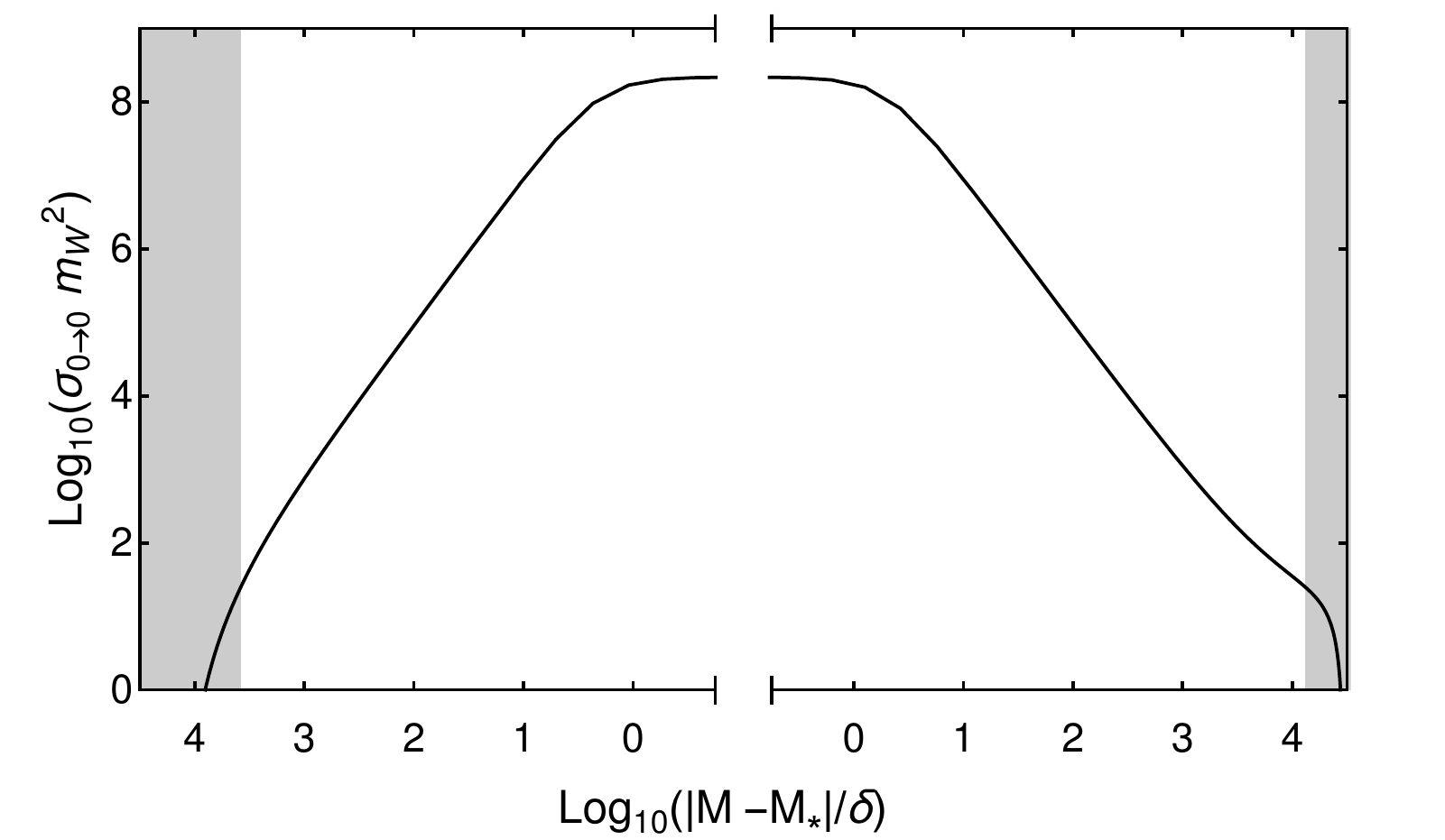}
\caption{Log-log plot of the neutral-wino elastic cross section $\sigma_{0 \to 0}$ at zero energy as a function of the wino mass $M$ near the unitarity mass $M_*=2.39$~TeV. The horizontal axis is the logarithm of $|M-M_*|/\delta$, with $M$ increasing to the right towards $M_*$ on the left side of the plot and increasing to the right away from $M_*$ on the right side of the plot. The grey regions are the ranges of $M$ in which $\sigma_{0 \to 0} < 8 \pi/m_W^2$, so a ZREFT for neutral and charged winos is not applicable.}
\label{fig:sigma00vsMlog}
\end{figure}

The neutral-wino elastic cross section $\sigma_{0 \to 0}(E=0)$ at zero energy for $\delta=170$~MeV is shown as a function of the wino mass $M$ in figure~\ref{fig:sigma00vsM}. The cross section has dramatic peaks at critical values of $M$. The first critical mass is $M_*=2.39$~TeV and the second is 9.24~TeV. The dramatic peaks indicate that there is a resonance near the neutral-wino-pair threshold. In the absence of wino-pair annihilation, the zero-energy cross section diverges as $|M-M_*|^2$ as $M$ approaches the critical mass $M_*$. If wino-pair annihilation is taken into account, the zero-energy cross section is finite at $M_*$, but the local maximum is orders of magnitude too large to be visible in figure~\ref{fig:sigma00vsM}. For $\delta = 170$~MeV, the zero-energy cross section  at the critical mass $M_*=2.39$~TeV is
\begin{equation}
\sigma_{0 \to 0}(E=0)\Big|_{M = M_*} = (2.17 \times 10^{8})/m_W^2.
\label{eq:sigma00-max}
\end{equation}
A log-log plot of the zero-energy cross section as a function of $M$ is shown in figure~\ref{fig:sigma00vsMlog}. There are scaling regions of $M$ just above and below $M_*$ where the cross section scales as $|M-M_*|^{-2}$. As $M \to M_*$, the cross section increases to its maximum value in eq.~\eqref{eq:sigma00-max}.

\begin{figure}[t]
\centering
\includegraphics[width=0.6\linewidth]{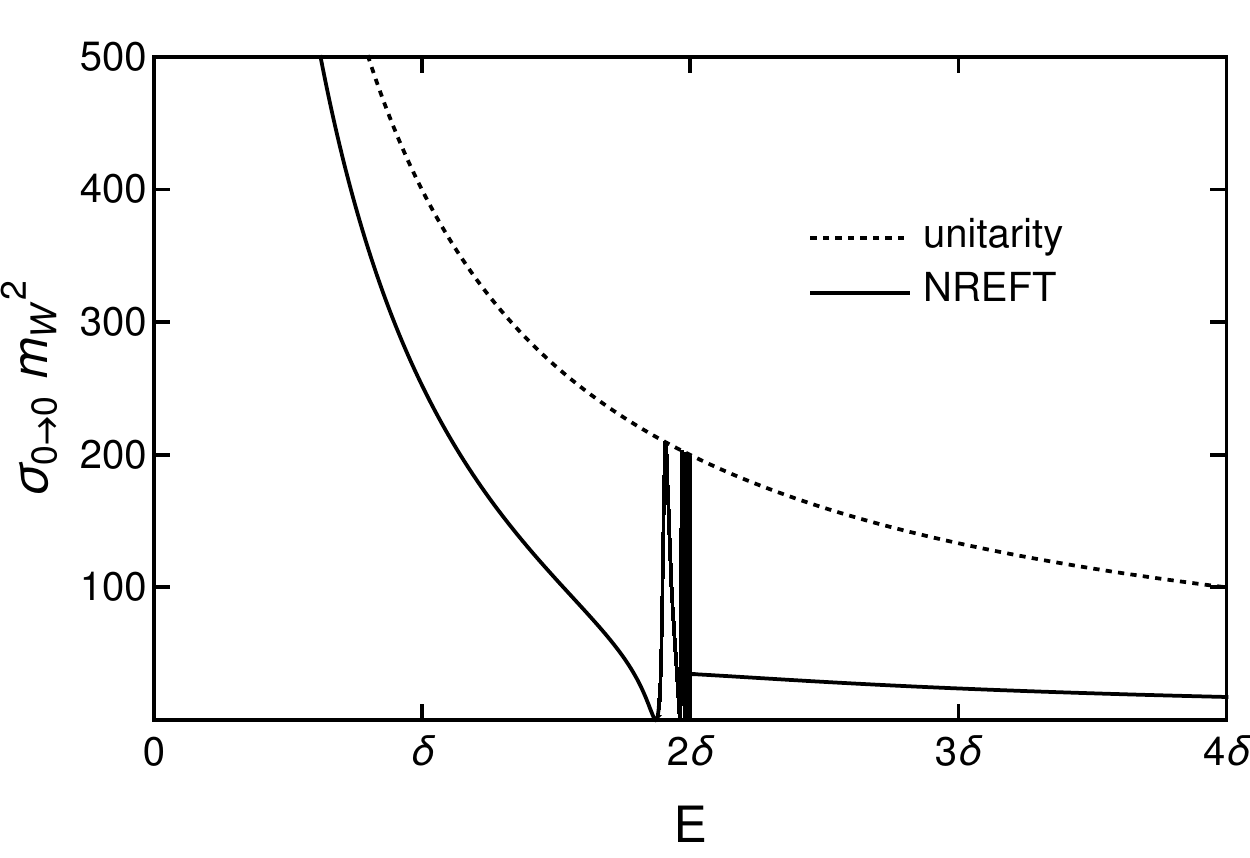}
\caption{Neutral-wino elastic cross section $\sigma_{0 \to 0}(E)$ at the unitarity mass $M_*=2.39$~TeV as a function of the energy $E$. The S-wave unitarity bound is shown as a dotted curve.}
\label{fig:sigma00-NREFT}
\end{figure}

\begin{figure}[th]
\centering
\includegraphics[width=0.6\linewidth]{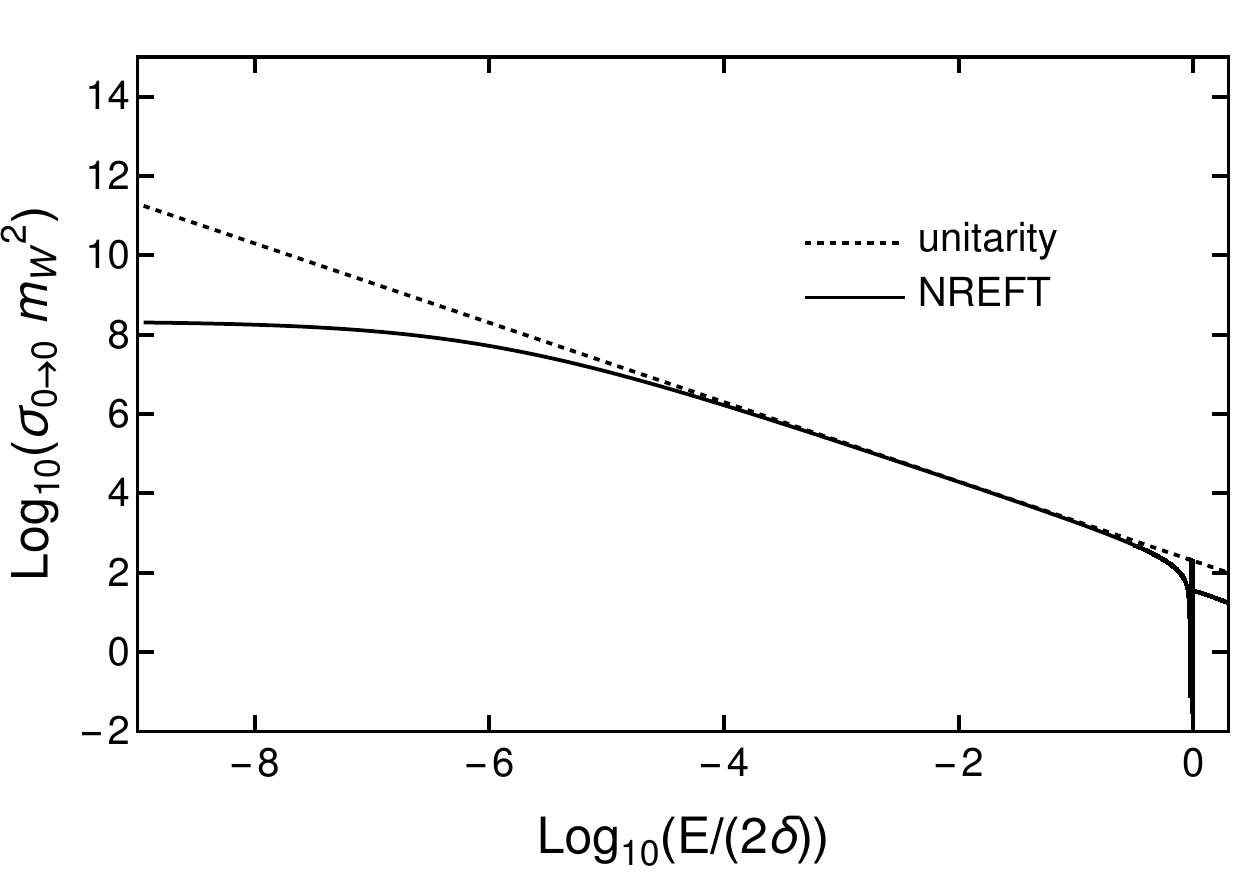}
\caption{Log-log plot of the neutral-wino elastic cross section $\sigma_{0 \to 0}(E)$ at the unitarity mass $M_*=2.39$~TeV as a function of the energy $E$. The S-wave unitarity bound is shown as a dotted line.}
\label{fig:sigma00-NREFT-lowE}
\end{figure}

The energy dependence of the neutral-wino elastic cross section $\sigma_{0 \to 0}(E)$ is most dramatic at a critical mass where the zero-energy cross section has a sharp peak. The cross section at $M_*=2.39$~TeV is shown as a function of energy in figure~\ref{fig:sigma00-NREFT}. Just above the charged-wino-pair threshold at $2\delta$, the cross section is $34.6/m_W^2$, and it decreases slowly as $E$ increases. Just below $2\delta$, the cross section has a sequence of narrow resonances whose peaks come very close to saturating the unitarity bound. The resonances can be interpreted as bound states in the Coulomb potential between the charged winos $w^+$ and $w^-$. As $E$ decreases further, the cross section approaches the unitarity bound in eq.~\eqref{sigma-unitarity0} from below. In the absence of wino-pair annihilation, the elastic cross section at the critical mass $M_*$ saturates the unitarity bound in eq.~\eqref{sigma-unitarity0} in the limit $E\to 0$. We therefore refer to such a critical mass as a {\it unitarity mass}, and we refer to a system with such a mass as being {\it at unitarity}. If wino-pair annihilation is taken into account, the elastic cross section at the unitarity mass $M_*$ has the maximum value in eq.~\eqref{eq:sigma00-max}, which is orders of magnitude too large to be visible in figure~\ref{fig:sigma00-NREFT}. A log-log plot of the elastic cross section near the neutral-wino-pair threshold is shown in figure~\ref{fig:sigma00-NREFT-lowE}. As $E$ increases from 0, the cross section decreases from the value in eq.~\eqref{eq:sigma00-max}, while approaching the unitarity bound. There is a range of about 3 decades in energy where the cross section nearly saturates the unitarity bound, and therefore scales as $E^{-1}$. The cross section begins to deviate significantly from the unitarity bound at an energy $E$ about an order of magnitude below the charged-wino-pair threshold $2\delta$.

Neutral winos with energies well below the charged-wino-pair threshold $2\delta$ have short-range interactions, because the Coulomb interaction enters only through virtual charged winos. The short-range interactions guarantee that $v_0(E)/T_{00}(E)$ can be expanded in powers of the relative momentum $p = \sqrt{ME}$:
\begin{eqnarray}
\frac{2 M v_0(E)}{T_{00}(E)} =
-\gamma_0  - i p + \frac12 r_0\,p^2 + \frac18 s_0\,p^4 + {\cal O}(p^6).
\label{eq:T00NRinv}
\end{eqnarray}
The only odd power of $p$ in the expansion is the imaginary term $-i p$. If $\bm{\Gamma} = 0$, the coefficients of the  even powers of $p$ are real; otherwise they have small imaginary parts. The leading term in the expansion in eq.~\eqref{eq:T00NRinv} defines the complex inverse scattering length $\gamma_0$. We refer to the vanishing of Re$[\gamma_0]$ as {\it unitarity}. The complex coefficients of the $p^2$ and $p^4$ terms define the {\it effective range}  $r_0$ and the {\it shape parameter} $s_0$.

\begin{figure}[t]
\centering
\includegraphics[width=0.48\linewidth]{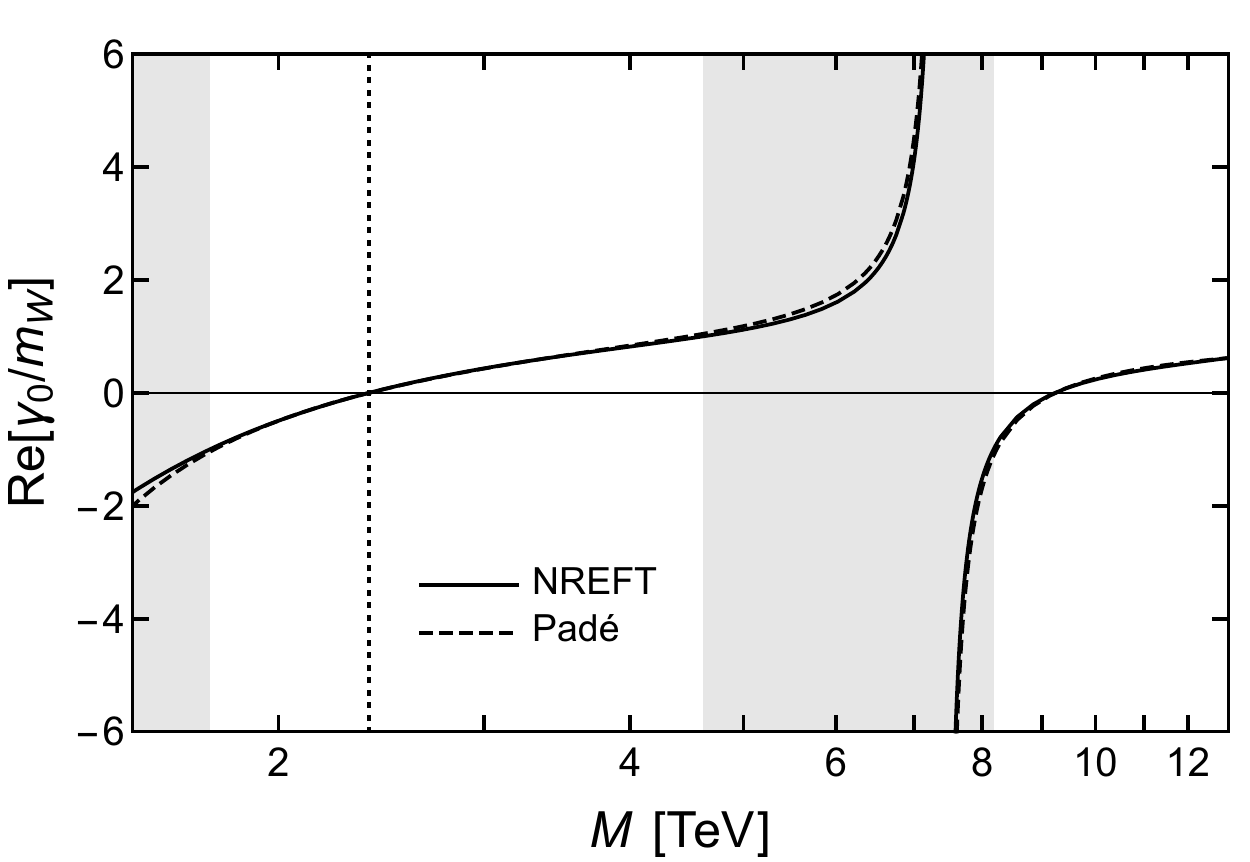}
~
\includegraphics[width=0.48\linewidth]{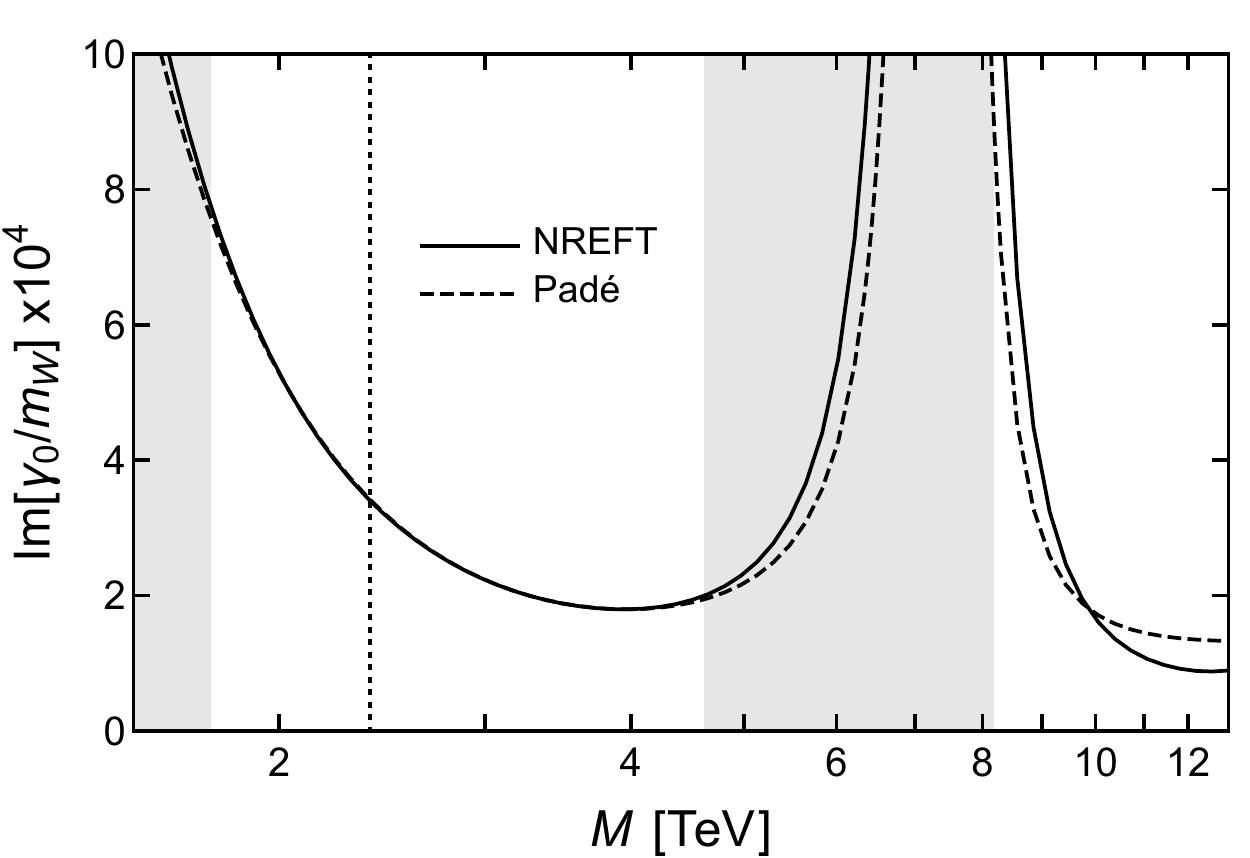}
\caption{Real and imaginary parts (left and right panels) of the inverse neutral-wino scattering length $\gamma_0= 1/a_0$ as functions of the wino mass $M$ (solid curves). In the left and right panels, the dashed curves are the Pad\'e approximants  in eqs.~\eqref{eq:Regamma0Pade} and \eqref{eq:Imgamma0Pade}, respectively. The vertical dotted line indicates the unitarity mass at $M_* = 2.39$~TeV. The grey regions are the ranges of $M$ in which $|\gamma_0| > m_W$, so a ZREFT for neutral and charged winos is not applicable.}
\label{fig:Real_a_vsM}
\end{figure}

The coefficients in the range expansion in eq.~\eqref{eq:T00NRinv} can be determined numerically by solving the Schr\"odinger equation. The real and imaginary parts of the inverse scattering length $\gamma_0$ for $\delta = 170$~MeV are shown as functions of the wino mass $M$ in figure~\ref{fig:Real_a_vsM}. The real part of $\gamma_0(M)$  can be fit surprisingly well by a Pad\'e approximant in $M$ of order [2,2], with zeros at  the first and second resonances at $M_*=2.39$~TeV and  $M_*' = 9.24$~TeV and with poles at the first and second poles of the real part of $\gamma_0(M)$ at $M_0 = 0.0027$~TeV and $M_0' = 7.39$~TeV. The only adjustable parameter is an overall prefactor. We choose to improve the fit near $M_*$ by fitting $M_0$ as well as the prefactor. The imaginary part of $\gamma_0(M)$ can be fit by a Pad\'e approximant in $M$ of order [4,4]. If a constant offset equal to the local minimum at $M' = 3.97$~TeV is subtracted, the remainder can be fit by a [3,4] Pad\'e approximant with double poles at the poles at $M_0$ and $M_0'$ of the real part of $\gamma_0(M)$, a double zero at $M'$, and a single zero at $M'' = 9.88$~TeV. We choose to improve the fit near $M_*$ by fitting $M_0$ as well as the prefactor. The resulting Pad\'e approximants are
\begin{subequations}
\begin{eqnarray}
\label{eq:Regamma0Pade}
 \mathrm{Re}\big[\gamma_0(M)\big] &=& (1.05\, m_W) \, \frac{(M-M_*)(M-M_*')}{(M-M_0)(M-M_0')},
 \\
\label{eq:Imgamma0Pade}
 \mathrm{Im}\big[\gamma_0(M)\big] &=& -(3.89 \times 10^{-4} \,m_W)\, 
 \left(\frac{M_*(M-M')^2 (M-M'')}{(M-M_0)^2(M-M_0')^2} -0.449 \right),
\end{eqnarray}
\label{eq:gamma0Pade}%
\end{subequations}
where $M_0 = 0.886$~TeV in eq.~\eqref{eq:Regamma0Pade} and $M_0=0.322$~TeV in eq.~\eqref{eq:Imgamma0Pade}. The ratio of the imaginary and real parts of $\gamma_0$ is consistent with a suppression factor of $\alpha_2 m_W/M$. At the unitarity mass $M_*= 2.39$~TeV, $\gamma_0$ is pure imaginary:
\begin{equation}
\gamma_0(M_*) = \left( 0 + 9.59 \times 10^{-4}\, i \right) \Delta_*,
\label{eq:gamma0*}
\end{equation}
where  $\Delta_*  = \sqrt{2 M_* \delta} = 28.5$~GeV. 

The winos can be described by a zero-range effective field theory (ZREFT) for neutral and charged winos if the neutral-wino inverse scattering length is smaller than the inverse range of the weak interactions: $|\gamma_0| < m_W$. Figure~\ref{fig:sigma00vsM} shows that the region of $M$ near $M_* = 2.39$~TeV in which the two-channel ZREFT is applicable is roughly from 1.8~TeV to 4.6~TeV. The energy region in which it is applicable is total wino-pair energy $E$ below $m_W^2/M$, which at $M_*$ is about 2700~MeV. There is a narrower range of $M$ in which neutral winos can be described by a single-channel ZREFT for neutral winos only. The neutral-wino inverse scattering length must be not only smaller than $m_W$ but also smaller than the range associated with the transition between a neutral-wino pair and a virtual charged-wino pair: $|\gamma_0|< \Delta$, where $\Delta = (2 M \delta)^{1/2}$. Figure~\ref{fig:sigma00vsM} shows that the region of $M$ in which the ZREFT for neutral winos only is applicable is roughly from 2.1~TeV to 2.9~TeV. The energy region in which it is applicable is $E$ below about $\delta=170$~MeV.

The coefficients of terms with higher powers of $p^2$ in the range expansion in eq.~\eqref{eq:T00NRinv} can also be determined numerically by solving the Schr\"odinger equation in eq.~\eqref{eq:radialSchrEqann}. For $\delta = 170$~MeV, the effective range and the shape parameter at the unitarity mass $M_* = 2.39$~TeV are
\begin{subequations}
\begin{eqnarray}
r_0(M_*) &=& \left(-1.653 + 5.33 \times 10^{-4} i \right) /\Delta_*,
\label{eq:r0*EM}
\\
s_0(M_*) &=& \left(- 2.653 + 9.31 \times 10^{-4} i \right)/\Delta_*^3,
\label{eq:s0*EM}
\end{eqnarray}
\label{eq:r0,s0*EM}%
\end{subequations}
The absolute values of the real parts of the coefficients are order 1, indicating that $\Delta_*  = \sqrt{2 M_* \delta}$ is an appropriate momentum scale for the real parts. The imaginary parts are smaller by about 4 orders of magnitude, consistent with a suppression factor of $\alpha_2 m_W/M$. The ratio of the real parts of $r_0$ and $s_0$ is nearly the same as the ratio of the imaginary parts. This apparent coincidence is explained in section~\ref{sec:ZREFTann}.

\begin{figure}[t]
\centering
\includegraphics[width=0.48\linewidth]{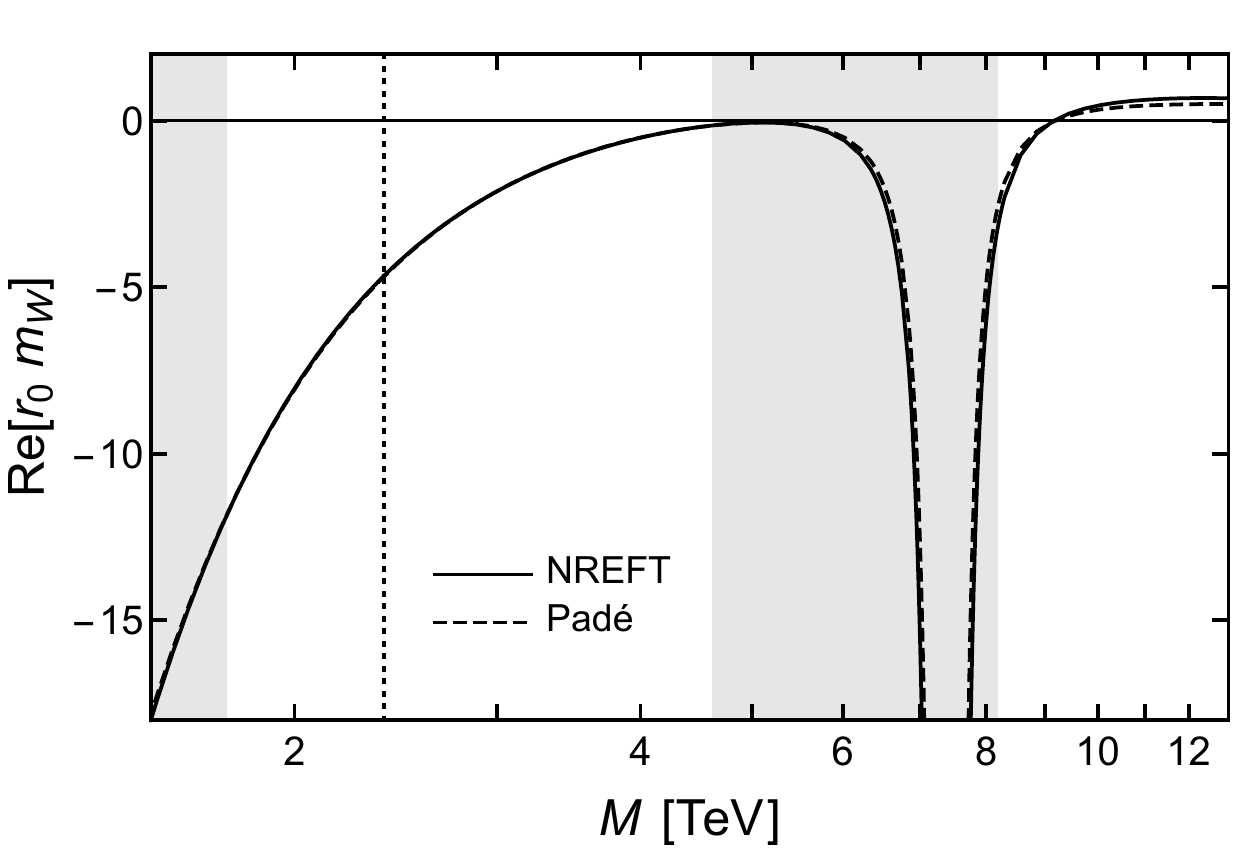}
~
\includegraphics[width=0.48\linewidth]{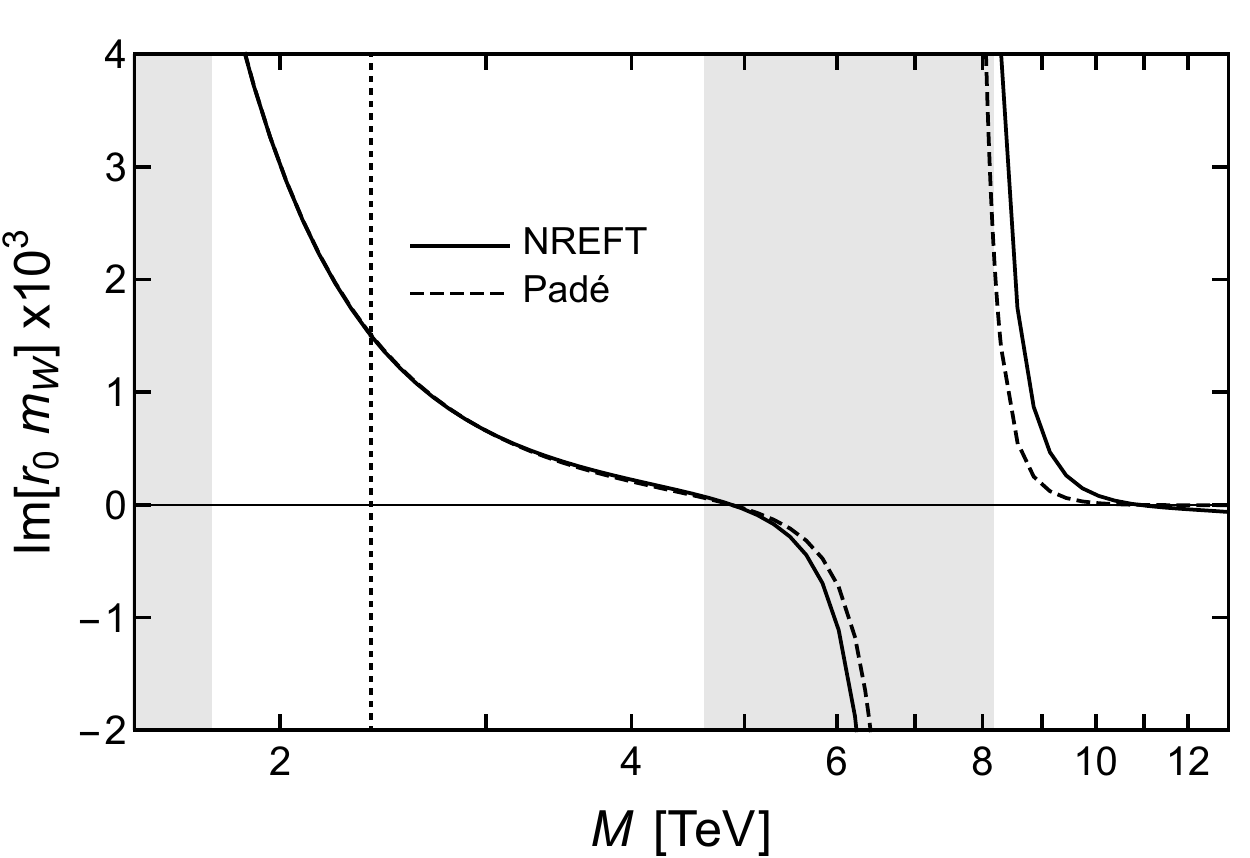}
\caption{Real  and imaginary parts (left and right panels) of the neutral-wino effective range $r_0$ as functions of the wino mass $M$ (solid curve). In the left and right panels, the dashed curves are the Pad\'e approximants in eqs.~\eqref{eq:Rer0Pade} and \eqref{eq:Imr0Pade}, respectively. The vertical dotted line indicates the unitarity mass $M_* = 2.39$~TeV. The grey regions are the ranges of $M$ in which  $|\gamma_0| > m_W$, so a ZREFT for neutral and charged winos is not applicable.}
\label{fig:r0vsM}
\end{figure}

The real and imaginary parts of the effective range $r_0$ for $\delta = 170$~MeV are shown as functions of the mass $M$ in figure~\ref{fig:r0vsM}. The real part of $r_0(M)$ can be fit surprisingly well by a Pad\'e approximant in $M$ of order [4,4].  If an offset equal to the local maximum near $M' = 5.13$~TeV is subtracted, the remainder can be fit by a [3,4] Pad\'e approximant with double poles at the poles $M_0$ and $M_0'$ of the real part of $\gamma_0(M)$, a double zero at $M' = 5.13$~TeV, and a single zero at $M'' = 9.11$~TeV. The only adjustable parameter in the Pad\'e approximant is an overall prefactor. We choose to improve the fit near $M_*$ by fitting $M_0$ as well as the prefactor. The imaginary part of $r_0(M)$ can be fit by a [2,6] Pad\'e approximant with triple poles at $M_0 $ and $M_0'$, and single zeros at $M'=4.88$~TeV and $M''=10.91$~TeV. The only adjustable parameter is an overall prefactor. We choose to improve the fit near $M_*$ by fitting $M_0$ as well as the prefactor. The resulting Pad\'e approximants are
\begin{subequations}
\begin{eqnarray}
\label{eq:Rer0Pade}
\mathrm{Re}\big[r_0(M)\big] &=& (5.06/m_W) \left( \frac{M_*(M-M')^2 (M-M'')}{(M-M_0)^2(M-M_0')^2} - 0.0108 \right),
 \\
\mathrm{Im}\big[r_0(M)\big] &=& -(1.65 \times 10^{-3}/m_W)  \frac{M_*^4(M-M') (M-M'')}{(M-M_0)^3(M-M_0')^3} ,
\label{eq:Imr0Pade}
\end{eqnarray}
\label{eq:r0Pade}%
\end{subequations}
where $M_0 = 0.096$~TeV in eq.~\eqref{eq:Rer0Pade} and $M_0=0.569$~TeV in eq.~\eqref{eq:Imr0Pade}.

\subsection{Neutral-wino-pair annihilation}
\label{sec:w+/-Scattering}

In NREFT, wino-pair annihilation is taken into account through the imaginary delta-function potential in the Schr\"odinger equation in eq.~\eqref{eq:radialSchrEqann}. The annihilation cross section $\sigma_{0,\mathrm{ann}}$ for a neutral-wino pair is expressed in terms of elements of the $2\times 2$ T-matrix in eq.~\eqref{eq:sig0ann-T}. The annihilation rate $2v_0\, \sigma_{0,\mathrm{ann}}$ is obtained by multiplying by the relative velocity, which is twice the velocity $v_0(E)$ of a neutral wino in eq.~\eqref{eq:v0-E}.

\begin{figure}[t]
\centering
\includegraphics[width=0.6\linewidth]{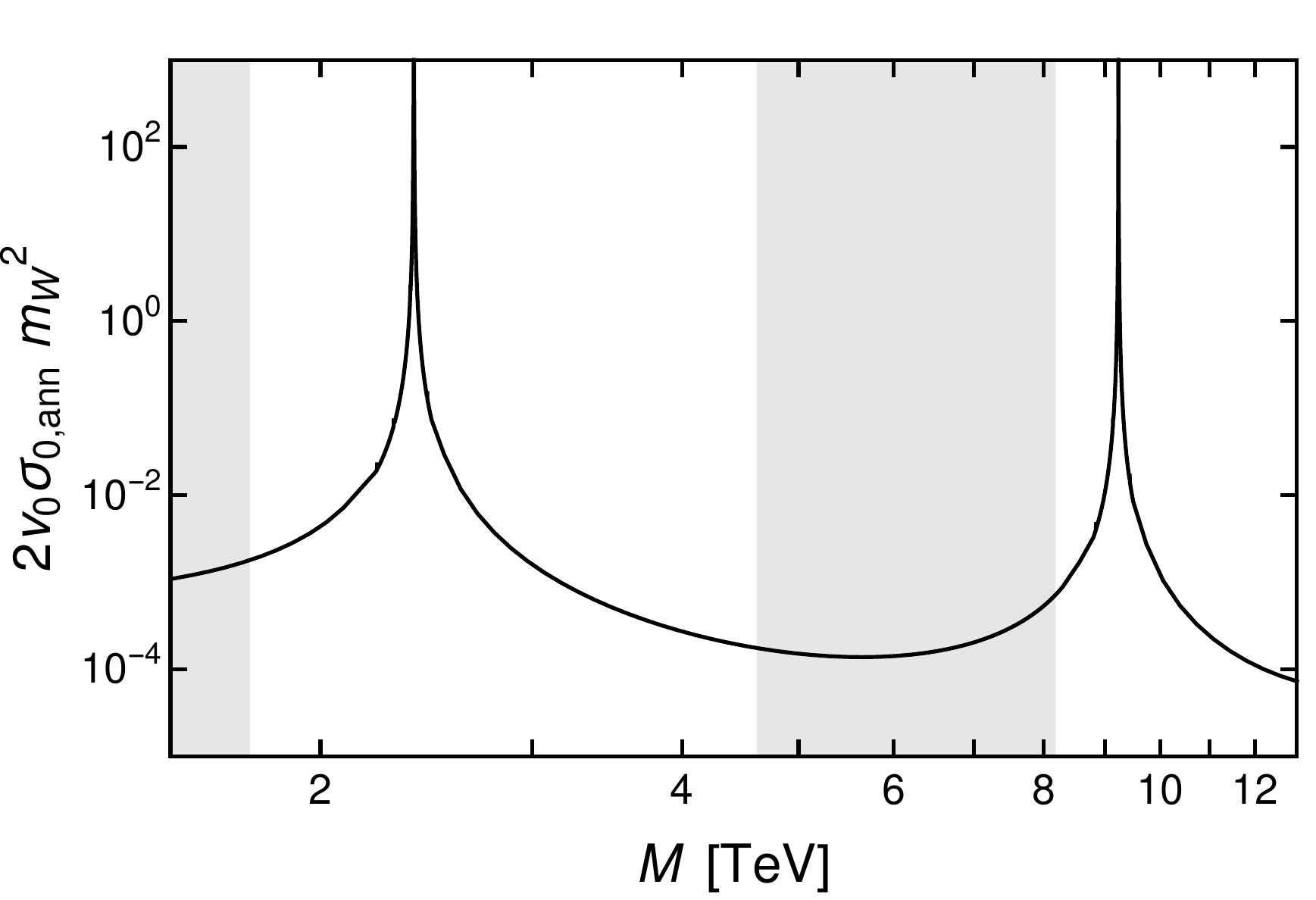}
\caption{Neutral-wino-pair annihilation rate $2 v_0 \,\sigma_{0,\mathrm{ann}}$ at zero energy as a function of the wino mass $M$. The peaks have finite maxima that are too large to be visible. The grey regions are the ranges of $M$ in which $|\gamma_0| > m_W$, so a ZREFT for neutral and charged winos is not applicable.} 
\label{fig:2v0sigma0annvsM}
\end{figure}

\begin{figure}[th]
\centering
\includegraphics[width=0.6\linewidth]{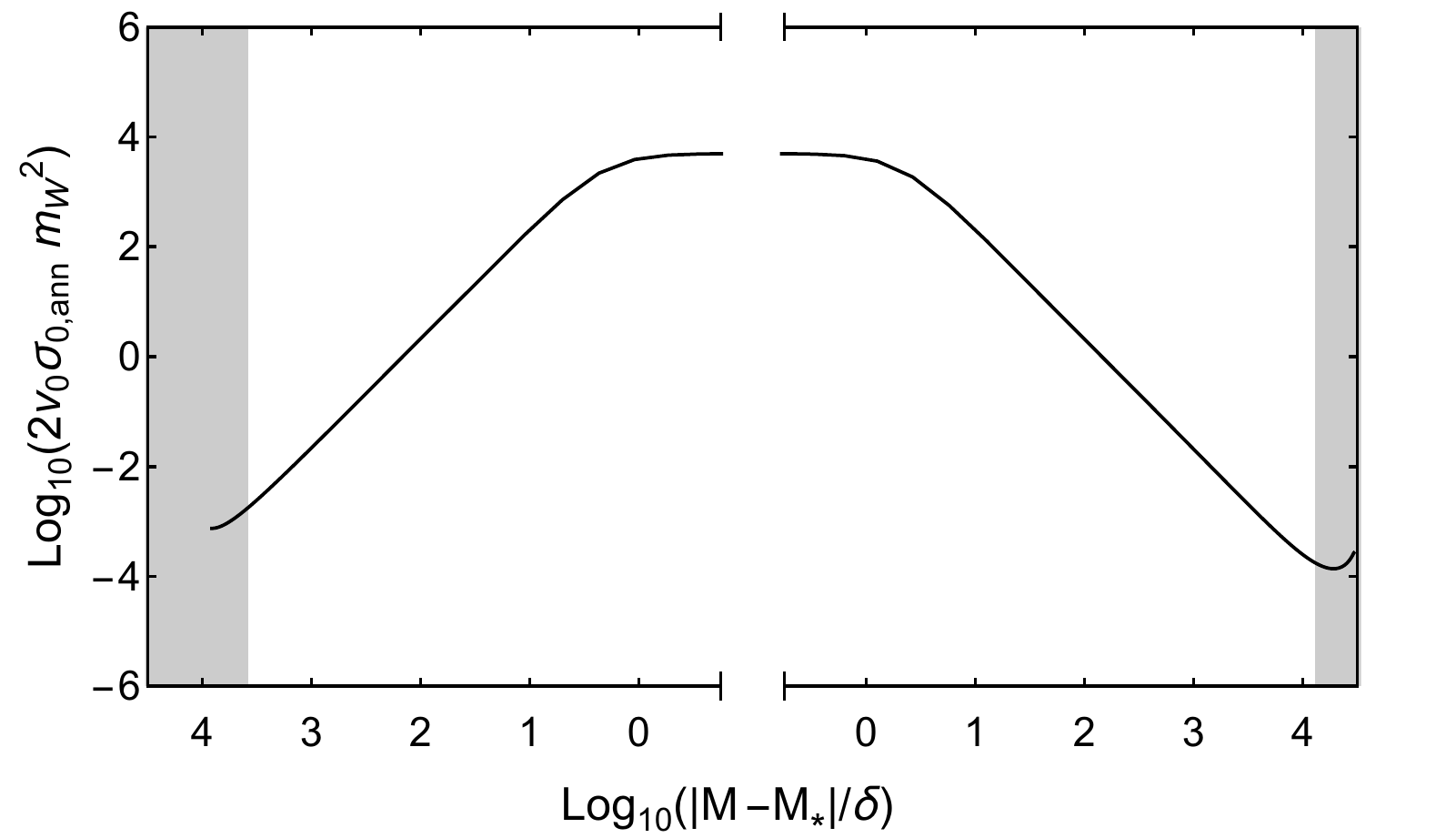}
\caption{Log-log plot of the neutral-wino-pair annihilation rate $2 v_0 \,\sigma_{0,\mathrm{ann}}$ at zero energy as a function of the wino mass $M$. The horizontal axis is the logarithm of $|M-M_*|/\delta$, with $M$ increasing towards $M_*$ on the left side of the plot and with $M$ increasing away from $M_*$ on the right side of the plot. The grey regions are the ranges of $M$ in which  $|\gamma_0| > m_W$, so a ZREFT for neutral and charged winos is not applicable.}
\label{fig:2v0sigma0annvsMlog}
\end{figure}

The neutral-wino-pair annihilation rate $2 v_0 \, \sigma_{0,\mathrm{ann}}$ at zero energy for $\delta=170$~MeV is shown as a function of the wino mass $M$ in figure~\ref{fig:2v0sigma0annvsM}. The annihilation rate has dramatic peaks at the same critical values of $M$ as the neutral-wino elastic cross section in figure~\ref{fig:sigma00vsM}. The local maxima are too large to be visible in figure~\ref{fig:2v0sigma0annvsM}. For $\delta = 170$~MeV, the annihilation rate at the first unitarity mass $M_*=2.39$~TeV is
\begin{equation}
2v_0\, \sigma_{0,\mathrm{ann}}(E=0)\Big|_{M = M_*} = (4.97 \times 10^3)/m_W^2.
\label{eq:2v0sigma0ann-max}
\end{equation}
A log-log plot of the annihilation rate  as a function of $M$ is shown in figure~\ref{fig:2v0sigma0annvsMlog}. There are scaling regions of $M$ just above and below $M_*$ where the annihilation rate scales as $|M-M_*|^{-2}$.  As $M \to M_*$, the annihilation rate increases to its  maximum value in eq.~\eqref{eq:2v0sigma0ann-max}.

\begin{figure}[t]
\centering
\includegraphics[width=0.6\linewidth]{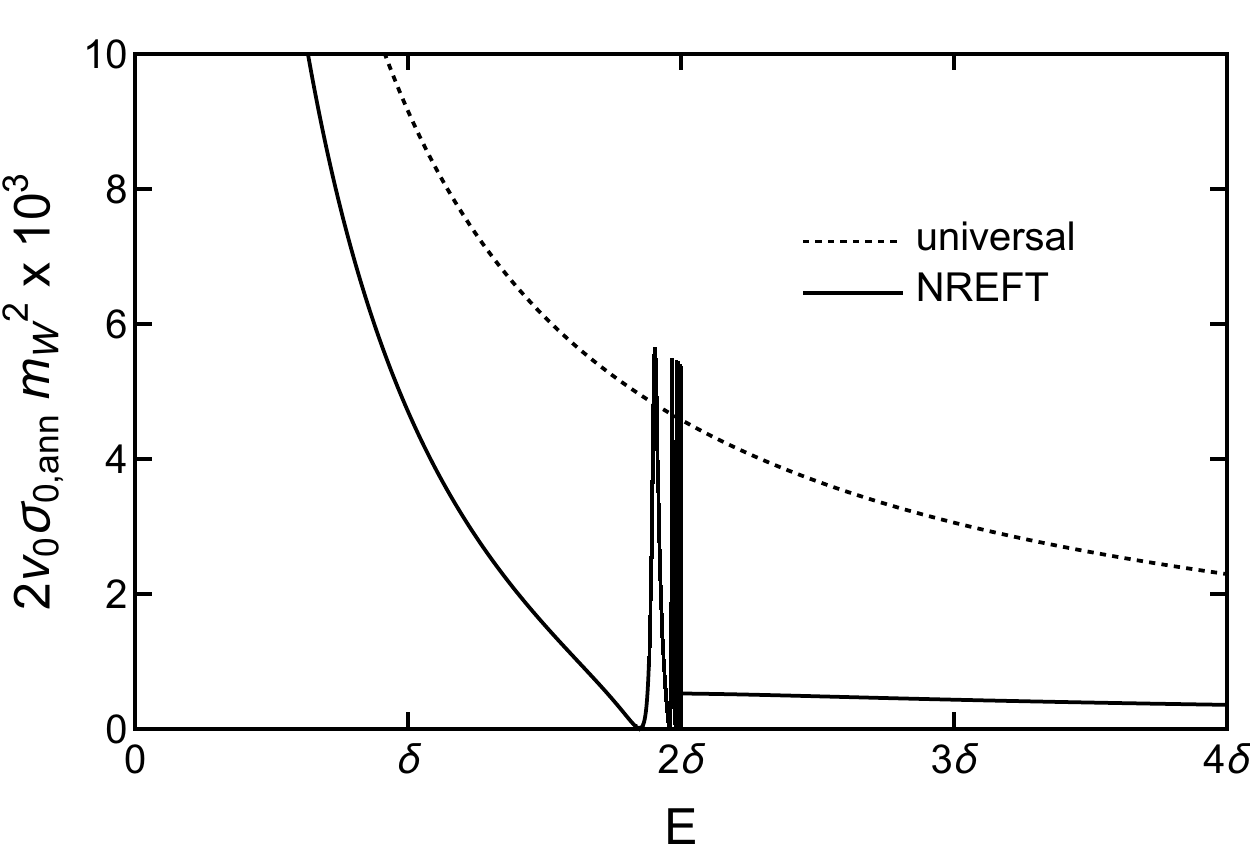}
\caption{Neutral-wino-pair annihilation rate at the unitarity mass $M_*=2.39$~TeV as a function of the energy $E$: NREFT (solid curve) and the universal approximation in eq.~\eqref{eq:sigma0ann-uni} (dashed curve).}
\label{fig:2v0sigma0ann-E}
\end{figure}

\begin{figure}[th]
\centering
\includegraphics[width=0.6\linewidth]{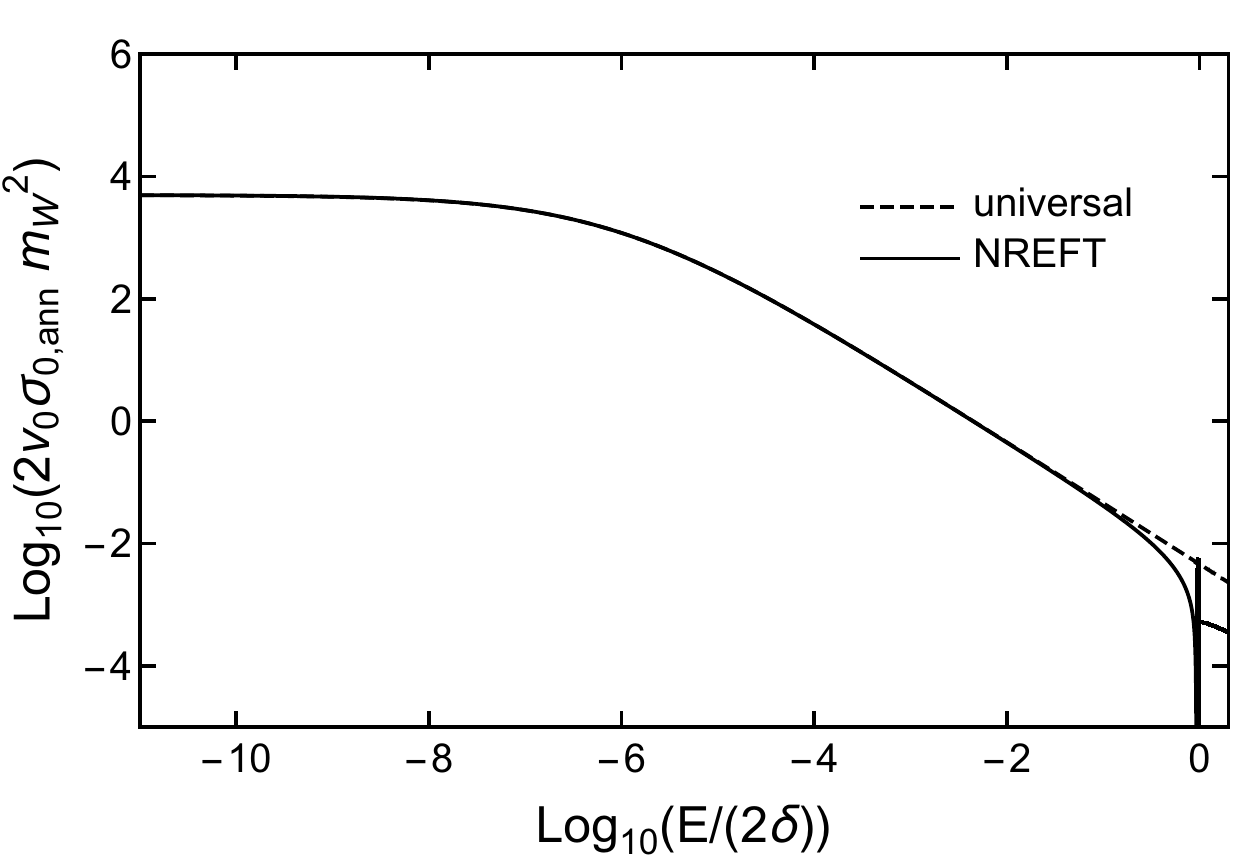}
\caption{Log-log plot of the neutral-wino-pair annihilation rate $2 v_0 \,\sigma_{0,\mathrm{ann}}$ at the unitarity mass $M_*=2.39$~TeV as a function of the energy $E$: NREFT (solid curve) and the universal approximation in eq.~\eqref{eq:sigma0ann-uni} (dashed curve).}
\label{fig:2v0sigma0ann-Elog}
\end{figure}

The energy dependence of the neutral-wino-pair annihilation rate $2 v_0 \, \sigma_{0,\mathrm{ann}}$ is most dramatic at a unitarity mass. The annihilation rate $2 v_0 \, \sigma_{0,\mathrm{ann}}(E)$ at $M_*=2.39$~TeV is shown as a function of the energy $E$ in  figure~\ref{fig:2v0sigma0ann-E}. Just above the charged-wino-pair threshold at $2\delta$, the annihilation rate is $(5.2 \times 10^{-4})/m_W^2$, and it decreases slowly as $E$ increases. Just below $2\delta$, the annihilation rate has a sequence of narrow peaks that can be associated with bound states in the Coulomb potential between the charged winos $w^+$ and $w^-$. As $E$ decreases further, the annihilation rate approaches the scaling behavior $E^{-1}$. In the limit $E \to 0$, it approaches the maximum value in eq.~\eqref{eq:2v0sigma0ann-max}, which is orders of magnitude too large to be visible in figure~\ref{fig:2v0sigma0ann-E}. A log-log plot of the annihilation rate near the neutral-wino-pair threshold is shown in figure~\ref{fig:2v0sigma0ann-Elog}. As $E$ increases from 0, the annihilation rate decreases from the value in eq.~\eqref{eq:2v0sigma0ann-max}, and it approaches the scaling behavior $E^{-1}$. There is a range of about 4 decades in the energy  in which the annihilation rate has that scaling behavior. The annihilation rate begins to deviate significantly from the scaling behavior at an energy $E$ about half an order of magnitude below the threshold $2\delta$.

\subsection{Universal approximations}
\label{sec:Universal}

Particles with short-range interactions that produce an S-wave resonance sufficiently close to their scattering threshold have universal low-energy behavior that is completely determined by their S-wave scattering length $a_0$ \cite{Braaten:2004rn}. The universal predictions are the LO predictions of the single-channel zero-range effective field theory. The universality holds even if there are highly inelastic scattering channels, such as annihilation channels \cite{Braaten:2013tza}. In this case, the universal low-energy behavior is completely determined by the complex inverse scattering length $\gamma_0$. The universal dimensionless T-matrix is
\begin{equation}
T_{0 0}(E) = \frac{2p}{- \gamma_0 - i p},
\label{eq:T-largea}
\end{equation}
where $p =\sqrt{ME}$. 
The universal approximations for the neutral-wino elastic cross section and the neutral-wino-pair annihilation rate are
\begin{subequations}
\begin{eqnarray}
\sigma_{0 \to 0}(E) &=& \frac{8\pi}{[\mathrm{Re}(\gamma_0)]^2+[p + \mathrm{Im}(\gamma_0)]^2},
\label{eq:sigma00-uni}
\\
2 v_0 \sigma_{0,\mathrm{ann}}(E) &=& 
\frac{(16 \pi/M)\, \mathrm{Im}(\gamma_0)}{[\mathrm{Re}(\gamma_0)]^2+[p + \mathrm{Im}(\gamma_0)]^2}.
\label{eq:sigma0ann-uni}
\end{eqnarray}
\label{eq:sigma-uni}%
\end{subequations}
These universal expressions were presented in ref.~\cite{Braaten:2004rn}. The universal region of parameters is where $|\gamma_0|$ is small compared to the inverse of the range set by the interactions. The universal region of the energy is where $|E|$ is small compared to the energy scale set by that range and the mass $M$.

For neutral winos, the range set by the interactions is the maximum of the range $1/m_W$ of the weak interactions and the length scale $1/\Delta$ associated with the transition to a charged-wino pair. The universal region of $M$ near the unitarity mass $M_*=2.39$~TeV is inside the region from 2.1~TeV to 2.9~TeV. The universal region of the energy is $|E| \ll 2\delta$. The universal approximations become increasingly accurate as $M \to M_*$ and as $E \to 0$. The neutral-wino elastic cross section at zero energy and the neutral-wino-pair annihilation rate at zero energy are shown as functions of $M$ in figures~\ref{fig:sigma00vsMlog} and \ref{fig:2v0sigma0annvsMlog}. Their dependence on $M$ is reproduced exactly by the universal approximations in eqs.~\eqref{eq:sigma-uni} by definition of the complex inverse scattering length $\gamma_0$ in eq.~\eqref{eq:T00NRinv}. The neutral-wino elastic cross section and the neutral-wino-pair annihilation rate at the unitarity mass $M_*=2.39$~TeV are shown as functions of $E$ in figures~\ref{fig:sigma00-NREFT-lowE} and \ref{fig:2v0sigma0ann-Elog}. As the energy $E$ increases from 0 to the scaling region, the universal approximations track the results from NREFT. The universal approximations begin to break down about an order of magnitude below the charged-wino-pair threshold. Above the scaling region of $E$, the universal approximation to $\sigma_{0 \to 0}$ continues to  decrease as $1/E$, tracking the unitarity bound, and the universal approximation to $2 v_0 \sigma_{0,\mathrm{ann}}$ also continues to decrease as $1/E$. Very near the resonance, the real and imaginary parts of the inverse scattering length in eqs.~\eqref{eq:gamma0Pade} can be simplified by setting $M = M_*$ except in the factor $M - M_*$ in Re$[\gamma_0]$:
\begin{subequations}
\begin{eqnarray}
 \mathrm{Re}\big[\gamma_0(M)\big] &\approx& (2.29 \,m_W)(M - M_*)/M_*,
\label{eq:Regamm0-M:uni}
\\
 \mathrm{Im}\big[\gamma_0(M)\big] &\approx& 3.42\times 10^{-4}\, m_W.
\label{eq:Imgamm0-M:uni}
\end{eqnarray}
\label{eq:gamm0-M:uni}
\end{subequations}

In ref.~\cite{Blum:2016nrz}, Blum, Sato, and Slatyer gave expressions for the elastic cross section and the pair annihilation rate similar to the universal expressions in eqs.~\eqref{eq:sigma00-uni} and \eqref{eq:sigma0ann-uni}. The expressions in eqs.~(2.30) and (2.31) of ref.~\cite{Blum:2016nrz} can be obtained from those in eqs.~\eqref{eq:sigma00-uni} and \eqref{eq:sigma0ann-uni} by replacing the complex constant $\gamma_0$ in the denominator $|\gamma_0+ i p|^2$  by a function $k_{p_0} + \mathcal{V}_{-1}\log(p/p_0)$, where  $p_0$ is a reference momentum, $\mathcal{V}_{-1}$ is a real constant, and $k_{p_0}$ is a complex constant that depends on $p_0$. Since a pair of neutral winos with energy well below the charged-wino-pair threshold have only short-range interactions, the corrections to $\gamma_0+ i p$ at small $p$  must be a power series in $p^2$, as in eq.~\eqref{eq:T00NRinv}. The logarithmic dependence on $p$ in ref.~\cite{Blum:2016nrz} is therefore unphysical.

In ref.~\cite{MarchRussell:2008tu}, March-Russell and West introduced an approximation to  the annihilation rate based on a treatment of threshold resonances by Bethe and Placek. Their approximation in eq.~(5) of ref.~\cite{MarchRussell:2008tu} can be obtained by replacing the complex constant $\gamma_0$ in the denominator $|\gamma_0 + i p|^2$ in eq.~\eqref{eq:sigma0ann-uni} by the function $\gamma_0 - \tfrac12 r_0 p^2$, where  $r_0$ is a real effective range. There is a very narrow range of energy where the effective-range term $\tfrac12 r_0 p^2$ in eq.~\eqref{eq:T00NRinv} is significant but the shape term $\tfrac18 s_0 p^4$ remains negligible. The approximation in ref.~\cite{MarchRussell:2008tu} is therefore not a significant improvement over the universal approximation in eq.~\eqref{eq:sigma0ann-uni}.

\section{Zero-Range Model}
\label{sec:ZRMann}

In this Section, we present the analytic results for the unitary transition amplitudes for $w^0 w^0$ and $w^+ w^-$ in the Zero-Range Model with Coulomb resummation that were calculated in ref.~\cite{Braaten:2017kci}. We then take into account the effects of wino-pair annihilation through the analytic continuation of real interaction parameters.

\subsection{Unitary transition amplitudes}
\label{sec:ZRMuta}

Observables in the sector consisting of two neutral winos $w^0w^0$ or two charged winos $w^+ w^-$ are conveniently encoded in the amplitudes for transitions among the two coupled channels. We denote the neutral channel $w^0 w^0$ by the index 0 and the charged channel $w^+ w^-$ by the index 1. In the Zero-Range Model, the S-wave spin-singlet transition amplitudes $\mathcal{A}_{ij}(E)$ are functions of the total energy $E$ of the wino pair only. The T-matrix elements $\mathcal{T}_{ij}(E)$ for  S-wave wino-wino scattering are obtained by evaluating the transition amplitudes $\mathcal{A}_{ij}(E)$ on the energy shell. The constraints on the T-matrix elements from S-wave unitarity can be derived from the unitarity condition for the amplitude matrix $\bm{\mathcal{A}}(E)$ at real $E$, which can be expressed as
\begin{equation}
\bm{\mathcal{A}}(E)- \bm{\mathcal{A}}(E)^* = 
- \frac{1}{8 \pi} \bm{\mathcal{A}}(E) \bm{M}^{1/2} \Big[ \bm{\kappa}(E)  - \bm{\kappa}(E)^* \Big]  
\bm{M}^{1/2} \bm{\mathcal{A}}(E)^*,
\label{eq:A-unitarity}
\end{equation}
where $\bm{M}$ is the $2\times2$ diagonal matrix
\begin{equation}
\label{eq:Mmatrix}
\bm{M}=
\begin{pmatrix}  M &   0   \\ 
                          0  & 2M   
\end{pmatrix}
\end{equation}
and $\bm{\kappa}$ is a diagonal matrix whose entries are functions of $E$:
\begin{equation}
\bm{\kappa}(E) =
\begin{pmatrix} \kappa_0(E)  &          0       \\ 
                                 0            & \kappa_1(E)
\end{pmatrix} .
\label{eq:kappamatrix}
\end{equation}
The functions $\kappa_0$ and $\kappa_1$ of the complex energy $E$ have branch points at 0 and $2\delta$, respectively:
\begin{subequations}
\begin{eqnarray}
\kappa_0(E) &=& \sqrt{-ME-i\varepsilon},
\label{eq:kappa0}
\\
\kappa_1(E) &=& \sqrt{-M(E-2\delta)-i\varepsilon}.
\label{eq:kappa1}
\end{eqnarray}
\label{eq:kappa01}%
\end{subequations}
The different diagonal entries of the matrix $\bm{M}$ in eq.~\eqref{eq:Mmatrix} are a convenient way to take into account that the neutral channel $w^0w^0$ consists of a pair of identical fermions while the charged channel $w^+w^-$ consists of two distinguishable fermions.

\begin{figure}[t]
\centering
\includegraphics[width=0.90\linewidth]{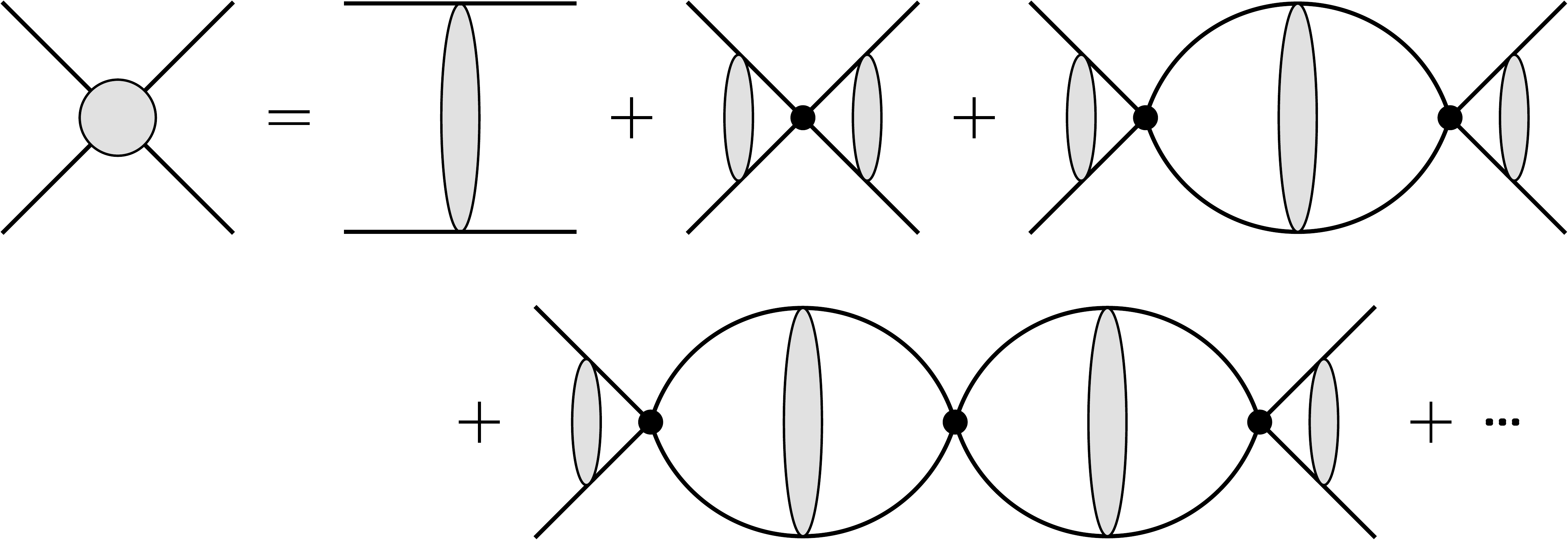}
\caption{Diagrams for the transition amplitudes $\mathcal{A}_{ij}(E)$ expressed as a sum over the number of zero-range interactions. A solid line represents either a neutral wino or a charged wino. The bubble diagrams must be summed to all orders. Each bubble is summed over a neutral-wino pair $w^0 w^0$ and a charged-wino pair $w^+ w^-$. On the right side of the equation, a shaded blob represents Coulomb resummation.}
\label{fig:ZREFTSumC}
\end{figure}

\begin{figure}[t]
\centering
\includegraphics[width=\linewidth]{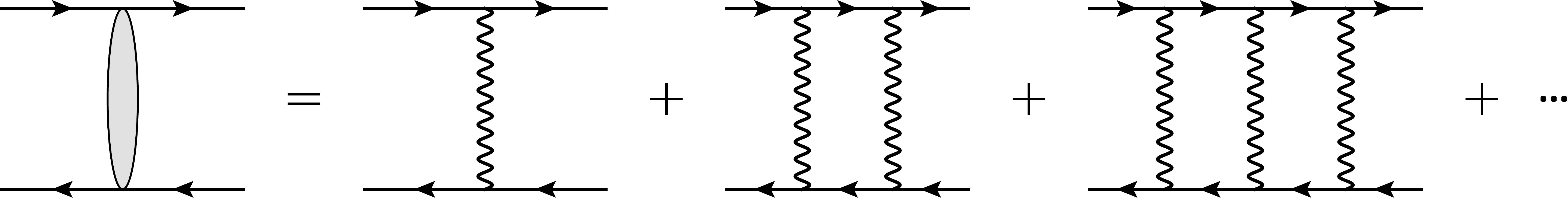}
\caption{Diagrams for $w^+ w^- \to w^+ w^-$ with no zero-range interactions. A solid line with a forward arrow or a backward arrow represents a $w^+$ or a $w^-$, respectively. The ladder diagrams from the exchange of Coulomb photons must be summed to all orders.}
\label{fig:Coulomb}
\end{figure}

\begin{figure}[th]
\centering
\includegraphics[width=0.75\linewidth]{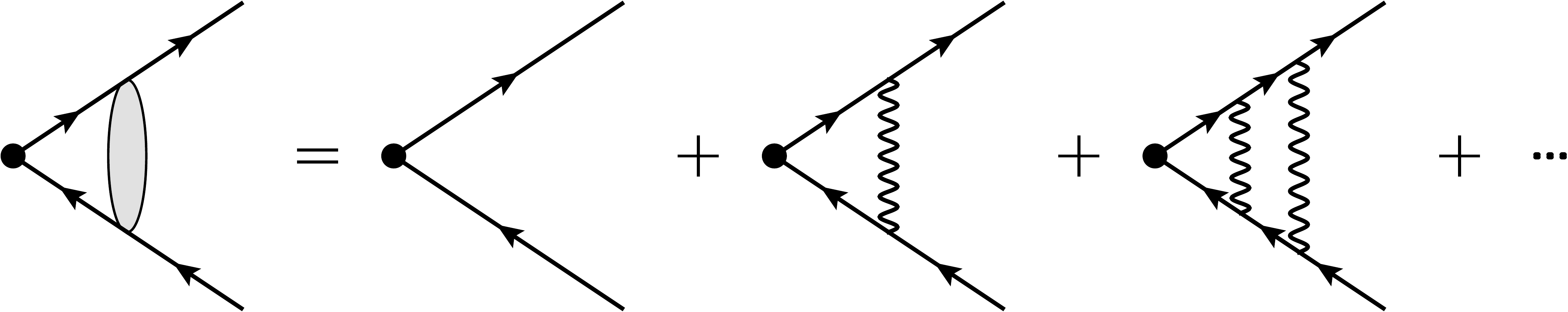}
\caption{Diagrams for the creation of $w^+ w^-$ at a point. The ladder diagrams from the exchange of Coulomb photons must be summed to all orders.}
\label{fig:pointC}
\end{figure}

\begin{figure}[th]
\centering
\includegraphics[width=0.9\linewidth]{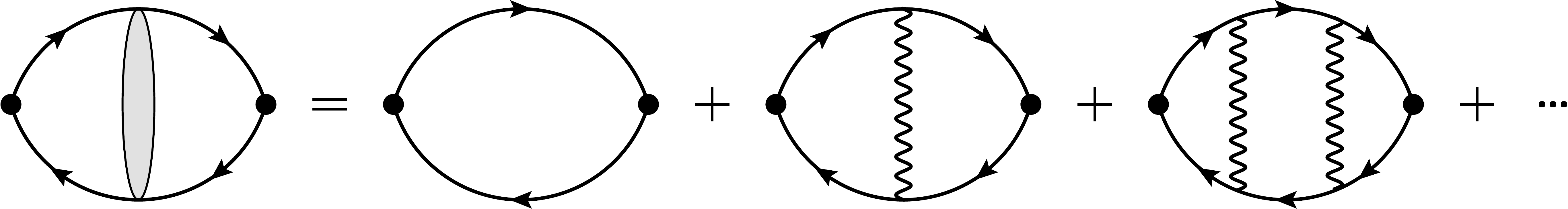}
\caption{Bubble diagrams for $w^+ w^-$. The ladder diagrams from the exchange of Coulomb photons must be summed to all orders.}
\label{fig:bubbleC}
\end{figure}

The amplitude $\mathcal{A}_{ij}(E)$ with Coulomb resummation is given by the sum of all diagrams with the appropriate incoming and outgoing pairs of wino lines specified by $i$ and $j$ and with the intermediate upper and lower wino lines connected by zero-range interactions and/or exchanges of Coulomb photons. In figure~\ref{fig:ZREFTSumC}, the amplitude is expressed as a sum over the number of zero-range interactions. The pair of incoming wino lines or outgoing wino lines is either $w^0w^0$ or $w^+ w^-$. Adjacent zero-range vertices are connected by a bubble whose upper and lower wino lines are summed over $w^0 w^0$ and $w^+ w^-$. The bubble diagrams must be summed to all orders. Each pair of upper and lower lines in figure~\ref{fig:ZREFTSumC} is connected by a blob that represents the sum of all ladder diagrams with the exchange of Coulomb photons. If the pair of lines is $w^0w^0$, the Coulomb exchange diagrams are 0. If the pair of lines  in the first diagram on the right side of figure~\ref{fig:ZREFTSumC} is $w^+w^-$, the blob represents the sum of the Coulomb-exchange diagrams in figure~\ref{fig:Coulomb}. If the pair of outgoing lines from the last zero-range vertex is $w^+w^-$, the blob connecting those lines represents the sum of the Coulomb-exchange diagrams in figure~\ref{fig:pointC}. If the pair of lines  in any bubble is $w^+w^-$, the blob represents the sum of the Coulomb-exchange diagrams in figure~\ref{fig:bubbleC}.

In ref.~\cite{Braaten:2017kci}, the transition amplitudes in the Zero-Range Model with Coulomb resummation were determined analytically by solving Lippmann-Schwinger equations. We used results for the Coulomb-resummed bubble diagrams by Kong and Ravndal \cite{Kong:1999sf}. We proceed to present the results of ref.~\cite{Braaten:2017kci}. The matrix of S-wave transition amplitudes for wino pairs with total energy $E$ can be expressed in the form
\begin{equation}
\label{eq:Amatrixann}
\bm{\mathcal{A}}(E) =
\bm{\mathcal{A}}_C(E) + \bm{W}(E) \,\bm{\mathcal{A}}_s(E)\, \bm{W}(E) .
\end{equation}
The first term on the right side is the matrix amplitude for Coulomb scattering, whose only nonzero entry is the second diagonal entry for $w^+ w^-$ scattering:
\begin{equation}
\label{eq:ACmatrix}
\bm{\mathcal{A}}_C(E) =
\begin{pmatrix} ~0~  & 0  \\  0  & \mathcal{A}_C(E) \end{pmatrix} .
\end{equation}
The S-wave Coulomb transition amplitude $\mathcal{A}_C(E)$ is given by the sum of diagrams in figure~\ref{fig:Coulomb}:
\begin{equation}
\label{eq:ACoulomb}
\mathcal{A}_C(E) = 
\left( 1 - \frac{\Gamma(1+i \eta)}{\Gamma(1-i \eta)}  \right) \frac{2\pi}{M\, \kappa_1(E)} ,
\end{equation}
where $\kappa_1$ is defined in eq.~\eqref{eq:kappa1} and $\eta$ is another energy variable defined by
\begin{equation}
\label{eq:eta-def}
\eta(E) \equiv 
 \frac{i\, \alpha \, M}{2\, \kappa_1(E)} 
 = i \frac{\alpha M}{2} \big[ - M(E-2 \delta) - i \epsilon \big]^{-1/2}.
\end{equation}
For a real energy $E = 2 \delta + p^2/M$ above the charged-wino-pair threshold, $\eta$ is real and negative: $\eta = - \alpha M/2p$. For a real energy $E$ below the charged-wino-pair threshold, $\eta$ is pure imaginary. The matrix $\bm{W}(E)$ in eq.~\eqref{eq:Amatrixann} is diagonal:
\begin{equation}
\label{eq:Wmatrix}
\bm{W}(E)= 
\begin{pmatrix} ~~1~~   & 0 \\  0 & W_1(E) \end{pmatrix}.
\end{equation}
Its second diagonal entry $W_1(E)$ is the dimensionless amplitude for $w^+w^-$ created at a point to produce $w^+w^-$ with total energy $E$ in the presence of Coulomb interactions. It can be obtained diagrammatically by expressing the sum of diagrams in figure~\ref{fig:pointC} as the vertex multiplied by $W_1(E)$:
\begin{equation}
W_1(E) = C(E) \, \left( \frac{\Gamma(1 + i \eta)}{\Gamma(1 - i \eta)} \right)^{1/2} ,
\label{eq:W1-eta}
\end{equation}
where $\eta$ is the function of $E$ in eq.~\eqref{eq:eta-def} and  $C$ is the square root of the Sommerfeld factor:
\begin{equation}
C^2(E)  = \frac{2 \pi \eta}{\exp(2 \pi \eta) - 1}.
\label{eq:Sommerfeld}
\end{equation}
The matrix  $\bm{\mathcal{A}}_s(E)$ in eq.~\eqref{eq:Amatrixann} is a matrix of short-distance transition amplitudes. It is the contribution to $\bm{\mathcal{A}}(E)$ from diagrams in which the first interaction and the last interaction are both zero-range interactions. It can be expressed most simply by giving its inverse:
\begin{equation}
\label{eq:Ainverse0C}
\bm{\mathcal{A}}_s^{-1}(E) = \frac{1}{8\pi} \bm{M}^{1/2}
\Big[ - \bm{\gamma} + \bm{K}(E) \Big] \bm{M}^{1/2},
\end{equation}
where  $\bm{\gamma}$ is a symmetric matrix of renormalized parameters,
\begin{equation}
\label{eq:gammamatrix}
\bm{\gamma}= 
\begin{pmatrix} \gamma_{00}   & \gamma_{01} \\ 
 \gamma_{01} & \gamma_{11} 
\end{pmatrix},
\end{equation}
and $\bm{K}$ is a diagonal matrix that depends on $E$:
\begin{equation}
\bm{K}(E) =
\begin{pmatrix} \kappa_0(E)  &          0       \\ 
                                 0            & K_1(E)
\end{pmatrix} .
\label{eq:Kmatrix}
\end{equation}
Its first diagonal entry is the function $\kappa_0$ in eq.~\eqref{eq:kappa0}, and its second diagonal entry is ~\cite{Kong:1999sf}
\begin{equation}
K_1(E) =\alpha M \left[ \psi(i \eta) + \frac{1}{2 i \eta} - \log(-i\eta)  \right],
\label{eq:K1-E}
\end{equation}
where $\psi(z)=(d/dz)\log \Gamma(z)$ and $\eta(E)$ is  defined in eq.~\eqref{eq:eta-def}. The function $K_1(E)$ has a branch point at $E=2\delta$. At real energies $E$, $K_1(E)$ is real for $E<2\delta$ and  complex for $E>2\delta$.

The matrix $\bm{\mathcal{A}}(E)$ of transition amplitudes in eq.~\eqref{eq:Amatrixann} satisfies the unitarity condition in eq.~\eqref{eq:A-unitarity} provided the parameters $\gamma_{ij}$ in the matrix $\bm{\gamma}$ in ~\eqref{eq:gammamatrix} are real valued. The unitarity condition is derived in appendix~\ref{app:unitarity}.

\subsection{Transition amplitudes with annihilation effects}
\label{sec:ZRMtaa}

The Zero-Range Model as described above does not take into account reactions in which a pair of winos annihilate into electroweak gauge bosons. These are highly inelastic reactions in which the final-state particles have momenta of order $M$, which is much larger than the momenta of the nonrelativistic winos. As far as low-energy winos are concerned, the annihilation reactions are localized to a region with size comparable to the de Broglie wavelengths of the annihilation products, which is order $1/M$ \cite{Braaten:2016sja}. Thus they can be taken into account in a nonrelativistic effective field theory through local interactions. In the  Zero-Range Model, the leading interaction terms that take into account annihilation reactions have the same form as the zero-range interactions in eq.~\eqref{eq:ZRint}. They can therefore be obtained by adding imaginary parts to the bare coupling constants $\lambda_{00}$, $\lambda_{01}$, and $\lambda_{11}$.  Equivalently, the annihilation reactions can be taken into account by analytically continuing the real-valued physical scattering parameters $\gamma_{00}$, $\gamma_{01}$, and $\gamma_{11}$ in eq.~\eqref{eq:gammamatrix} to complex values. We denote the analytically continued value of $\gamma_{ij}$ by $\gamma_{ij} + i \beta_{ij}$, where $\beta_{ij}$ is also real. The resulting matrix of transition amplitudes has the form in eq.~\eqref{eq:Amatrixann}, except that the matrix $\mathcal{A}_{s}(E)$ of short-distance transition amplitudes in eq.~\eqref{eq:Ainverse0C} is replaced by
\begin{equation}
\label{eq:Asinverse}
\bm{\mathcal{A}}_s^{-1}(E) = \frac{1}{8\pi} \bm{M}^{1/2}
\Big[ - \bm{\gamma} - i\bm{\beta} + \bm{K}(E) \Big] \bm{M}^{1/2},
\end{equation}
where  $\bm{\beta}$ is a symmetric matrix of real parameters:
\begin{equation}
\label{eq:betamatrix}
\bm{\beta}= 
\begin{pmatrix} \beta_{00}   & \beta_{01} \\ 
 \beta_{01} & \beta_{11} 
\end{pmatrix}.
\end{equation}
Thus the leading effects of wino-pair annihilation can be taken into account through three real parameters.  

Since wino-pair annihilation products are not described explicitly, the matrix of transition amplitudes obtained by inserting the matrix $\mathcal{A}_{s}(E)$ in eq.~\eqref{eq:Asinverse} into  eq.~\eqref{eq:Amatrixann} does not satisfy the unitarity condition in eq.~\eqref{eq:A-unitarity}. The correct unitarity equation is derived in appendix~\ref{app:unitarity}:
\begin{eqnarray}
\bm{\mathcal{A}}(E)- \bm{\mathcal{A}}(E)^* &=& 
- \frac{1}{8 \pi} \bm{\mathcal{A}}(E) \bm{M}^{1/2} \big[ \bm{\kappa}(E)  - \bm{\kappa}(E)^* \big] 
\bm{M}^{1/2} \bm{\mathcal{A}}(E)^*
\nonumber\\
&&\hspace{0cm}
+ \frac{i}{4 \pi} \bm{W}(E)\, \bm{\mathcal{A}}_s(E)\, \bm{M}^{1/2} \bm{\beta} \bm{M}^{1/2} \,
\bm{\mathcal{A}}_s(E)^*\, \bm{W}^*(E).
\label{eq:A-nonunitarity}
\end{eqnarray}
The term on the right side with the factor $\bm{\kappa}(E)  - \bm{\kappa}(E)^*$ is the contribution from intermediate wino-pair states. The term with the factor  $\bm{\beta}$ is the contribution from intermediate states that are wino-pair annihilation products, which are not described explicitly in the Zero-Range Model.

At a real energy $E$, the imaginary parts of the amplitudes $\mathcal{A}_{ij}(E)$ have physical interpretations in terms of reaction rates. The total reaction rate of a neutral-wino pair $w^0w^0$  with $E>0$ is twice the imaginary part of $\mathcal{A}_{00}(E)$. The total reaction rate of a charged-wino pair $w^+w^-$  with $E>2 \delta$ is twice the imaginary part of $\mathcal{A}_{11}(E)$. A superposition of $w^0w^0$  and $w^+w^-$ can be represented by a density matrix $\rho$ with unit trace. The total reaction rate for such a superposition is $2\,\mathrm{Tr}[\mathrm{Im}(\bm{\mathcal{A}}(E))\,\rho]$. The annihilation contribution to the reaction rate can be obtained by inserting the contribution to Im$\bm{\mathcal{A}}(E)$ from the last term in eq.~\eqref{eq:A-nonunitarity}. The annihilation contribution must be positive for any density matrix $\rho$. This implies that $\bm{\beta}$ must be a positive matrix.

\section{ZREFT}
\label{sec:ZREFTann}

In this section, we present the transition amplitudes for $w^0 w^0$ and $w^+ w^-$ in ZREFT at LO with Coulomb resummation that were calculated in ref.~\cite{Braaten:2017kci}. We take into account the effects of wino-pair annihilation through the analytic continuation of real interaction parameters. We determine the adjustable parameters of ZREFT at LO by matching low-energy $w^0 w^0$ scattering amplitudes from NREFT. We compare predictions of ZREFT at LO for wino-wino scattering, inclusive wino-pair annihilation, and  a wino-pair bound state with results from NREFT.

\subsection{Transition amplitudes}
\label{sec:ZREFTuta}

In order to give an explicit parametrization of the transition amplitudes $\mathcal{A}_{ij}(E)$ for ZREFT, we introduce two 2-component unit vectors that depend on the mixing angle $\phi$ in eq.~\eqref{eq:Tfp3}:
\begin{equation}
\label{eq:u,v-def}
\bm{u}(\phi) = \binom{\cos\phi}{\sin\phi}, \qquad \bm{v}(\phi) = \binom{-\sin\phi}{~~\cos\phi}.
\end{equation}
We use these vectors to define two projection matrices and another symmetric matrix:
\begin{subequations}
\begin{eqnarray}
\bm{\mathcal{P}}_u(\phi) &=& \bm{u}(\phi)\, \bm{u}(\phi)^T
= \begin{pmatrix} \cos^2\phi  & \cos\phi  \sin\phi \\  \cos\phi  \sin\phi & \sin^2\phi \end{pmatrix}, 
\label{eq:Pu}
\\ 
\bm{\mathcal{P}}_v(\phi) &=& \bm{v}(\phi)\, \bm{v}(\phi)^T
= \begin{pmatrix} \sin^2\phi  & - \cos\phi  \sin\phi \\  - \cos\phi  \sin\phi & \cos^2\phi \end{pmatrix},
\label{eq:Pv}
\\ 
\bm{\mathcal{P}}_m(\phi) &=& \bm{u}(\phi)\, \bm{v}(\phi)^T + \bm{v}(\phi)\, \bm{u}(\phi)^T
= \begin{pmatrix} -\sin(2\phi)  & ~\cos(2\phi) \\  \cos(2\phi) & ~\sin(2\phi) \end{pmatrix}.
\label{eq:Pm}
\end{eqnarray}
\label{eq:Pu,Pv,Pm}%
\end{subequations}
The superscript $T$ on $\bm{u}$ or $\bm{v}$ indicates the transpose of the column vector. The three matrices defined in eqs.~\eqref{eq:Pu,Pv,Pm} form a basis for $2 \times2$ symmetric matrices. This set of matrices is closed under differentiation:
\begin{subequations}
\begin{eqnarray}
\bm{\mathcal{P}}_u'(\phi) &=& \bm{\mathcal{P}}_m(\phi), 
\label{eq:dPu}
\\ 
\bm{\mathcal{P}}_v'(\phi) &=& -\bm{\mathcal{P}}_m(\phi), 
\label{eq:dPv}
\\ 
\bm{\mathcal{P}}_m'(\phi) &=& -2\, \bm{\mathcal{P}}_u(\phi) +2\, \bm{\mathcal{P}}_v(\phi).
\label{eq:dPm}
\end{eqnarray}
\label{eq:dPu,dPv,dPm}%
\end{subequations}
The three matrices are orthogonal with respect to the trace: Tr$(\bm{\mathcal{P}}_i \bm{\mathcal{P}}_j) = 0$ if $i \ne j$. The traces of the squares of these matrices are
\begin{equation}
\label{eq:TrP^2}
\mathrm{Tr}(\bm{\mathcal{P}}_u^2) = 1, \qquad \mathrm{Tr}(\bm{\mathcal{P}}_v^2) = 1, 
\qquad \mathrm{Tr}(\bm{\mathcal{P}}_m^2) = 2.
\end{equation}
The T-matrix at the RG fixed point for ZREFT is
\begin{equation}
\label{eq:Tfp3wino}
\bm{\mathcal{T}}_*(E) = \frac{8\pi i}{\sqrt{ME}} \, 
\bm{M}^{-1/2} \, \bm{\mathcal{P}}_u(\phi) \, \bm{M}^{-1/2},
\end{equation}
where $\bm{M}$ is the diagonal matrix in eq.~\eqref{eq:Mmatrix}.

In ZREFT with Coulomb resummation, the matrix of transition amplitudes $\bm{\mathcal{A}}(E)$ has the form in eq.~\eqref{eq:Amatrixann}, where $\bm{\mathcal{A}}_C(E)$ is the Coulomb amplitude matrix in eq.~\eqref{eq:ACmatrix}, $\bm{W}(E)$ is the diagonal matrix in eq.~\eqref{eq:Wmatrix}, and $\bm{\mathcal{A}}_s(E)$ is the matrix of short-distance transition amplitudes. A convenient choice for the interaction parameters of ZREFT are the coefficients in the expansion of the inverse of $\bm{\mathcal{A}}_s(E)$ in powers of $E$:
\begin{eqnarray}
\label{eq:Asinverseann}
\bm{\mathcal{A}}_s^{-1}(E) = \frac{1}{8\pi} \bm{M}^{1/2} 
\Big[ \big(- \gamma_u + \tfrac12 r_u p^2 + \ldots \big) \bm{\mathcal{P}}_u(\phi)
 + \big(-1/a_v + \ldots \big) \bm{\mathcal{P}}_v(\phi)
 \nonumber
 \\
 + \big(\tfrac12 r_m p^2 + \ldots \big) \bm{\mathcal{P}}_m(\phi) + \bm{K}(E) \Big] \bm{M}^{1/2},
\end{eqnarray}
where $p^2 = ME$ and $\bm{K}(E)$ is the diagonal matrix in eq.~\eqref{eq:Kmatrix}. The effects of Coulomb resummation are in the Coulomb transition amplitude $\mathcal{A}_C(E)$ in eq.~\eqref{eq:ACoulomb}, the function $W_1(E)$  in eq.~\eqref{eq:W1-eta}, and the function $K_1(E)$  in eq.~\eqref{eq:K1-E}. In eq.~\eqref{eq:Asinverseann}, the coefficients of $\bm{\mathcal{P}}_u$, $\bm{\mathcal{P}}_v$, and $\bm{\mathcal{P}}_m$ have been expanded in powers of $p^2$. The mixing angle $\phi$ has been chosen so that the $p^0$ term in the expansion of the coefficient of $\bm{\mathcal{P}}_m$ is 0. The successive improvements of ZREFT can be obtained by successive truncations of the expansions in $p^2$. At leading order (LO), the only nonzero term in the three expansions is the coefficient $-\gamma_u$ of $\bm{\mathcal{P}}_u(\phi)$. At next-to-leading order (NLO), there are two additional interaction parameters: $a_v$ and $r_u$. At NNLO, there is one additional interaction parameter: $r_m$.

In ZREFT, the leading effects of the annihilation of a wino pair into electroweak gauge bosons can be taken into account through the analytic continuation of the real parameters in the expansion of the short-distance amplitude in eq.~\eqref{eq:Asinverseann} in powers of $p^2$. If we keep only the imaginary parts of the coefficients of $(p^2)^0$, the analytic continuation corresponds to adding a constant matrix $-i \bm{\beta}$ inside the square brackets in eq.~\eqref{eq:Asinverseann}. That matrix can be expanded in the basis of symmetric matrices $\bm{\mathcal{P}}_u(\phi)$, $\bm{\mathcal{P}}_v(\phi)$, and $\bm{\mathcal{P}}_m(\phi)$ defined in Eqs.~\eqref{eq:Pu,Pv,Pm}:
\begin{equation}
\label{eq:betauvm}
\bm{\beta} = \beta_u \bm{\mathcal{P}}_u(\phi) + \beta_v \bm{\mathcal{P}}_v(\phi) 
+ \beta_m \bm{\mathcal{P}}_m(\phi) .
\end{equation}
The parameter $\beta_u$ can be regarded as the imaginary part of a complex parameter $\gamma_u$: $\beta_u = \mathrm{Im}(\gamma_u)$. The parameter $\beta_v$ can be absorbed into the imaginary part of a complex parameter $a_v$: $\beta_v \approx -\mathrm{Im}(a_v)/a_v^2$. Because of the expression for $\bm{\mathcal{P}}_u'(\phi)$ in eq.~\eqref{eq:dPu}, the parameter $\beta_m$ can be absorbed into  the imaginary part of a complex mixing angle $\phi$: $\beta_m \approx \gamma_u\mathrm{Im}(\phi)$. We therefore choose to replace the real parameters $\gamma_u$, $a_v$, and $\phi$ in the expression for $\bm{\mathcal{A}}_s^{-1}$ in eq.~\eqref{eq:Asinverseann} by complex parameters with small imaginary parts. The matrix $\bm{\beta}$ can then be expressed as
\begin{equation}
\label{eq:betaphi}
\bm{\beta} = \mathrm{Im} \left[ \gamma_u \bm{\mathcal{P}}_u(\phi) + (1/a_v) \bm{\mathcal{P}}_v(\phi)  \right] .
\end{equation}

The matrix of short-distance transition amplitudes for ZREFT at LO is obtained by setting the coefficients of positive powers of $p^2$ to zero in $\bm{\mathcal{A}}_s^{-1}(E)$ in eq.~\eqref{eq:Asinverseann}, inverting the matrix, and then taking the limit $a_v \to 0$:
\begin{equation}
\label{eq:AsmatrixLOC}
\bm{\mathcal{A}}_s(E) =  \lim_{a_v \to 0} 8\pi  \, \bm{M}^{-1/2}
\big[ - \gamma_u  \bm{\mathcal{P}}_u(\phi) -(1/a_v) \bm{\mathcal{P}}_v(\phi) + \bm{K}(E) \big]^{-1}
\bm{M}^{-1/2}.
\end{equation}
The inverse of the matrix is given to order $a_v$ in eq.~(6.2) of ref.~\cite{Braaten:2017gpq}. Inserting that expression into eq.~\eqref{eq:AsmatrixLOC}, it reduces to
\begin{equation}
\label{eq:AsmatrixLOC2}
\bm{\mathcal{A}}_s(E) =  \lim_{a_v \to 0} 8\pi  \, \bm{M}^{-1/2}
\left[ \frac{1}{L_u(E)} \bm{\mathcal{P}}_u(\phi) - a_v \bm{V}(\phi,E) \bm{V}(\phi,E)^T \right]
\bm{M}^{-1/2}.
\end{equation}
The 2-component column vector $\bm{V}(\phi,E)$ is
\begin{equation}
\bm{V}(\phi,E)=\frac{1}{L_u(E)} \big[-\gamma_u \mathds{1}+\bm{\tilde K}(E) \big] \bm{v}(\phi),
\label{eq:Vvector}
\end{equation}
where  $\bm{\tilde K}(E)$ is the matrix obtained from $\bm{K}(E)$ in eq.~\eqref{eq:Kmatrix} by interchanging the diagonal entries:
\begin{equation}
\bm{\tilde K}(E) =
\begin{pmatrix} K_1(E)  &          0       \\ 
                                 0            & \kappa_0(E)
\end{pmatrix} .
\label{eq:Ktilde}
\end{equation}
The denominator in eqs.~\eqref{eq:AsmatrixLOC2} and \eqref{eq:Vvector} is
\begin{equation}
L_u(E)=-\gamma_u +\cos^2\phi\,  \kappa_0(E)  +\sin^2\phi \, K_1(E),
\label{eq:Ku}
\end{equation}
where $\kappa_0(E)$ is given in eq.~\eqref{eq:kappa0} and $K_1(E)$ is given in eq.~\eqref{eq:K1-E}. The NLO term proportional to $a_v$ in eq.~\eqref{eq:AsmatrixLOC2} does not contribute to the cross sections for wino-wino scattering at LO, which are given in section~\ref{sec:CrossSectionLO}, but it does contribute to the wino-pair annihilation rates at LO, which are given in section~\ref{sec:Annihilation}.

\subsection{Wino-wino scattering}
\label{sec:CrossSectionLO}

The matrix of transition amplitudes $\bm{\mathcal{A}}(E)$ at LO is obtained by setting $a_v=0$ in the short-distance amplitude matrix in eq.~\eqref{eq:AsmatrixLOC2} and inserting it into the expression in eq.~\eqref{eq:Amatrixann}:
\begin{equation}
\label{eq:AmatrixLOC}
\bm{\mathcal{A}}(E) = 
\bm{\mathcal{A}}_C(E)
+\frac{8\pi}{L_u(E)}  \, \bm{W}(E) \,
\bm{M}^{-1/2} \, \bm{\mathcal{P}}_u(\phi)\,  \bm{M}^{-1/2}\, \bm{W}(E) ,
\end{equation}
where $\bm{\mathcal{A}}_C(E)$ is the Coulomb amplitude matrix in eq.~\eqref{eq:ACmatrix}, $\bm{W}(E)$ is the diagonal matrix in eq.~\eqref{eq:Wmatrix}, $\bm{\mathcal{P}}_u(\phi)$ is the matrix in eq.~\eqref{eq:Pu}, and $L_u(E)$ is the function in eq.~\eqref{eq:Ku}. The effects of wino-pair annihilation are taken into account through the imaginary parts of the parameters $\gamma_u$ and $\phi$.

The neutral-wino inverse scattering length $\gamma_0$ can be obtained by evaluating the transition amplitude ${\cal A}_{00}(E)$ at the neutral-wino-pair threshold:
\begin{equation}
{\cal A}_{00}(E=0) = - 8\pi  /(M \gamma_0).
\label{eq:T00-a0}
\end{equation}
The inverse neutral-wino scattering length is
\begin{equation}
\gamma_0 = (1 +  t_\phi^2)\gamma_u - t_\phi^2\, K_1(0) ,
\label{eq:a0-gammauLO}
\end{equation}
where $t_\phi \equiv \tan \phi$. This equation can be solved for $\gamma_u$ as a function of $\gamma_0$:
\begin{equation}
\gamma_u = \frac{\gamma_0 + t_\phi^2\, K_1(0)}{1 +  t_\phi^2}.
\label{eq:gammau-a0LO}
\end{equation}
If $|\gamma_0| \ll \sqrt{2M\delta}$, there are large cancellations in the denominator $L_u(E)$ in eq.~\eqref{eq:Ku}. These cancellations can be avoided by eliminating $\gamma_u$ in favor of $\gamma_0$. The resulting expression for the matrix of transition amplitudes is
\begin{equation}
\label{eq:TLO-hiE}
\bm{\mathcal{A}}(E) = 
 \begin{pmatrix} ~0~ & 0 \\ 0 & {\cal A}_C(E)
\end{pmatrix} 
 + \frac{8\pi}{L_0(E)} \begin{pmatrix} 1  & 0 \\  0 & W_1(E) \end{pmatrix} \bm{M}^{-1/2}   
\begin{pmatrix}    ~1~     & t_\phi\\  t_\phi & t_\phi^2 \end{pmatrix} 
\bm{M}^{-1/2} \begin{pmatrix} ~1~   & 0 \\  0 & W_1(E) \end{pmatrix},
\end{equation}
where ${\cal A}_C(E)$ is the Coulomb transition amplitude in eq.~\eqref{eq:ACoulomb} and $W_1(E)$ is the amplitude in eq.~\eqref{eq:W1-eta} for $w^+ w^-$ created at a point to have total energy $E$. The denominator in the second term is
\begin{equation}
L_0(E) =-\gamma_0  + t_\phi^2\, \big[ K_1(E) - K_1(0) \big]\, + \kappa_0(E),
\label{eq:L0-E}
\end{equation}
where $\kappa_0(E)$ is given in eq.~\eqref{eq:kappa0} and $K_1(E)$ is given in eq.~\eqref{eq:K1-E}. The effects of wino-pair annihilation are taken into account through the imaginary parts of the parameters $\gamma_0$ and $t_\phi$ in eqs.~\eqref{eq:TLO-hiE} and \eqref{eq:L0-E}. The diagonal entries of $\bm{\mathcal{A}}(E)$ are functions of $t_\phi^2$, and its off-diagonal entries are functions of $t_\phi^2$ multiplied by $t_\phi$. Thus the absolute squares $|\mathcal{A}_{ij}(E)|^2$ do not depend on the sign of $t_\phi$.

We denote by  $\sigma_{i \to j}(E)$ the cross section for scattering from channel $i$ to channel $j$ at energy $E$, averaged over initial spins and summed over final spins. The expressions for these cross sections in terms of the T-matrix elements ${\cal T}_{ij}(E)$ for states with the standard normalizations of a nonrelativistic field theory are
\begin{subequations}
\begin{eqnarray}
\sigma_{i \to 0}(E) &=&\frac{M^2}{8\pi}
\big| {\cal T}_{i0}(E) \big|^2 \frac{v_0(E)}{v_i(E)},
\label{eq:sig0E-calT}
\\
\sigma_{i\to 1}(E) &=&\frac{M^2}{4\pi}
\big| {\cal T}_{i1}(E) \big|^2 \frac{v_1(E)}{v_i(E)},
\label{eq:sig1E-calT}
\end{eqnarray}
\label{eq:sigE-calT}%
\end{subequations}
where the velocities $v_i(E)$  of the incoming winos are given in eqs.~\eqref{eq:v0,1-E}. The extra factor of $1/2$ in the cross sections $\sigma_{i \to 0}$ in eq.~\eqref{eq:sig0E-calT} for producing a neutral-wino pair compensates for overcounting by integrating over the entire phase space of the two identical particles. The T-matrix elements ${\cal T}_{ij}(E)$ are obtained by evaluating the transition amplitudes $\mathcal{A}_{ij}(E)$ in eq.~\eqref{eq:TLO-hiE} at a real energy $E$ above the appropriate threshold, which is $E=0$ for a neutral-wino pair $w^0 w^0$ and $E=2 \delta$ for a charged-wino pair $w^+ w^-$. 

A pair of neutral winos has short-range interactions at energies well below the charged-wino-pair threshold, because  the effects of electromagnetic interactions only enter through a virtual $w^+ w^-$ pair. The reciprocal of the T-matrix element ${\cal T}_{00}(E)$ for neutral-wino elastic scattering can therefore be expanded in powers of the relative momentum $p = \sqrt{ME}$:
\begin{eqnarray}
\frac{8\pi}{M {\cal T}_{00}(E)} =
-\gamma_0  - i p + \tfrac12 r_0 \, p^2+ \tfrac18 s_0 \, p^4  + {\cal O}(p^6).
\label{eq:T00NLOinv}
\end{eqnarray}
The only odd power of $p$ in the expansion is the imaginary term $-i p$. The coefficients of the even powers of $p$ have small imaginary parts. The leading term in the expansion in eq.~\eqref{eq:T00NLOinv} defines the complex inverse scattering length $\gamma_0$. We refer to a wino mass $M$ where the real part of $\gamma_0$ vanishes as a {\it unitarity mass}. The complex coefficients of the $p^2$ and $p^4$ terms define the effective range  $r_0$ and the shape parameter $s_0$.

For center-of-mass energy in the range $0 \leq E < 2\delta$, only the neutral-wino-pair channel is open. The T-matrix element for  $w^0w^0 \to w^0w^0$ in ZREFT at LO is given by the 00 entry of the matrix in eq.~\eqref{eq:TLO-hiE}:
\begin{equation}
{\cal T}_{00}(E) = \frac{8\pi/M}{L_0(E)},
\label{eq:T00LO}
\end{equation}
where $L_0(E)$ is given in eq.~\eqref{eq:L0-E}. The predictions for the effective range $r_0$ and the shape parameter $s_0$ can be determined by expanding $L_0(E)$ in powers of $p$ and comparing to the expansion in eq.~\eqref{eq:T00NLOinv}:
\begin{subequations}
\begin{eqnarray}
r_0 &=& 2 t_\phi^2\, K_1'(0)/M,
\label{eq:r0LO}
\\
s_0 &=& 4 t_\phi^2\, K_1''(0)/M^2,
\label{eq:s0LO}
\end{eqnarray}
\label{eq:r0s0LO}%
\end{subequations}
where $K_1(E)$ is defined in eq.~\eqref{eq:K1-E}. These predictions for $r_0$ and $s_0$ do not depend on $\gamma_0$, so their imaginary parts come only from the factor of $t_\phi^2$. This implies that the ratio of Re$[s_0]$ to Re$[r_0]$ and the ratio of Im$[s_0]$ to  Im$[r_0]$ are equal and independent of the complex interaction parameters $\gamma_0$ and $t_\phi$. The prediction for these ratios are
\begin{equation}
\frac{\mathrm{Re}[s_0]}{\mathrm{Re}[r_0]}
=\frac{\mathrm{Im}[s_0]}{\mathrm{Im}[r_0]}
=\frac{2 K_1''(0)}{K_1'(0) \, M}.
\label{eq:s0r0ratio}
\end{equation}
The last term depends on $M$ through the dimensionless variable $\alpha M/\Delta$. Thus ZREFT at LO gives a parameter-free prediction for these ratios as functions of $M$.

For center-of-mass energy in the range $E > 2\delta$, the neutral-wino-pair channel and the charged-wino-pair channel are both open. The T-matrix element for  $w^0w^0 \to w^0w^0$  in ZREFT at LO is given in eq.~\eqref{eq:T00LO}. The T-matrix elements for  $w^0w^0 \to w^+w^-$ and $w^+w^- \to w^+w^-$ in ZREFT at LO are given by the 01 and 11 entries of the matrix in eq.~\eqref{eq:TLO-hiE}:
\begin{subequations}
\begin{eqnarray}
{\cal T}_{01}(E) &=& \frac{(4\sqrt2\, \pi/M) t_\phi W_1(E)}{L_0(E)},
\label{eq:T01LO}
\\
{\cal T}_{11}(E) &=& \mathcal{A}_C(E) + \frac{(4 \pi/M) t_\phi^2 W_1^2(E)}{L_0(E)},
\label{eq:T11LO}
\end{eqnarray}
\label{eq:T01,11LO}%
\end{subequations}
where $L_0(E)$, $W_1(E)$,  and $\mathcal{A}_C(E)$ are given in eqs.~\eqref{eq:L0-E}, eq.~\eqref{eq:W1-eta}, and \eqref{eq:ACoulomb}.

\subsection{Wino-pair annihilation}
\label{sec:Annihilation}

Wino-pair annihilation into specific final states with electroweak gauge bosons cannot be described within ZREFT, because the final states include particles with momenta of order $M$. However the inclusive annihilation rates can be calculated in ZREFT, because they can be expressed in terms of the entries of the $2\times 2$ submatrix of the complete T-matrix that corresponds to the wino-pair channels $w^0 w^0$ and $w^+ w^-$. By the optical theorem, the total cross sections for wino-wino scattering are proportional to the imaginary parts of the appropriate T-matrix elements:
\begin{equation}
\sigma_{i,\mathrm{tot}}(E) = \frac{1}{v_i(E)}  \mathrm{Im}\mathcal{T}_{ii}(E) .
\label{eq:sig01tot-calT}
\end{equation}
If there is wino-pair annihilation, the annihilation cross sections can be obtained by subtracting the cross sections into a neutral-wino pair and into a charged-wino pair in eqs.~\eqref{eq:sigE-calT}:
\begin{equation}
\sigma_{i,\mathrm{ann}}(E) =   \frac{1}{v_i(E)}
\left( \mathrm{Im}\mathcal{T}_{ii}(E) 
-  \frac{M^2}{8 \pi} \big| \mathcal{T}_{i0}(E) \big|^2v_0(E) -  \frac{M^2}{4 \pi} \big| \mathcal{T}_{i1}(E) \big|^2v_1(E)  \right).
\label{eq:sig01ann-calT}%
\end{equation}
In eq.~\eqref{eq:sig01ann-calT}, $\mathcal{T}_{10}(E)$ should be interpreted as 0 if $0 < E < 2 \delta$.

The expression for the annihilation cross section in eqs.~\eqref{eq:sig01ann-calT} involves large cancellations between the total cross section and the cross sections for scattering into wino pairs, because the annihilation cross sections are suppressed compared to wino-wino cross sections by a factor of $\alpha_2^2 m_W^2/M^2$. The annihilation cross sections can be expressed in a form that avoids such cancellations by using the unitarity condition in eq.~\eqref{eq:A-nonunitarity}:
\begin{equation}
\sigma_{i,\mathrm{ann}}(E) = \frac{1}{8 \pi v_i(E)}
\big[ \bm{W}(E)\, \bm{\mathcal{A}}_s(E)\, \bm{M}^{1/2}  \bm{\beta} \bm{M}^{1/2}\, \bm{\mathcal{A}}_s^*(E) \, \bm{W}^*(E) \big]_{ii},
\label{eq:sigmaiann-beta}
\end{equation}
where $\bm{W}(E)$ is the diagonal matrix in eq.~\eqref{eq:Wmatrix}, $\bm{\mathcal{A}}_s(E)$ is the short-distance amplitude matrix in eq.~\eqref{eq:AsmatrixLOC2}, and $\bm{\beta}$ is the real-valued matrix in eq.~\eqref{eq:betaphi}. Since there is a term in $\bm{\beta}$ proportional to $1/a_v$, there are LO contributions to $\sigma_{i,\mathrm{ann}}$ from the term in $\bm{\mathcal{A}}_s$ proportional to $a_v$ in eq.~\eqref{eq:AsmatrixLOC2}. After taking the limit $a_v \to 0$, the annihilation rate reduces to
\begin{eqnarray}
2v_i\sigma_{i,\mathrm{ann}}(E) &=&16 \pi 
\bigg[ \bm{W}\, \bm{M}^{-1/2} \bigg( 
\frac{1}{|L_u|^2} \mathrm{Im}[\gamma_u] \, \bm{\mathcal{P}}_u \, \mathrm{Re}[\bm{\mathcal{P}}_u]\, \bm{\mathcal{P}}_u^*
\nonumber\\
&&\hspace{0cm}
-  \frac{1}{L_u}\bm{\mathcal{P}}_u \, \mathrm{Im}[\bm{\mathcal{P}}_v] \, \bm{V}^*\bm{V}^\dagger
-  \frac{1}{{L_u}^*}  \bm{V}  \bm{V}^T \, \mathrm{Im}[\bm{\mathcal{P}}_v] \, \bm{\mathcal{P}}_u^*
\bigg)  \bm{M}^{-1/2} \, \bm{W}^* \bigg]_{ii}.~~~
\label{eq:sigmaiann-beta2}
\end{eqnarray}
There is no term proportional to $\mathrm{Re}[\gamma_u]\, \mathrm{Im}[\bm{\mathcal{P}}_u]$, because it vanishes when sandwiched between $\bm{\mathcal{P}}_u$ and $\bm{\mathcal{P}}_u^*$. Similarly, there is no term proportional to Im$[1/a_v]$. Using the properties of the matrices $\bm{\mathcal{P}}_u$ and $ \bm{\mathcal{P}}_v$ and the expression for the vector $\bm{V}$ in eq.~\eqref{eq:Vvector}, this expression can be reduced to
\begin{eqnarray}
2v_i\sigma_{i,\mathrm{ann}}(E) &=&\frac{16 \pi }{|L_u|^2}
\bigg[ \bm{W}\, \bm{M}^{-1/2} \bigg( 
\mathrm{Im}[\gamma_u] \,(\bm{u}^T \bm{u}^*) \, \bm{u} \bm{u}^\dagger
\nonumber\\
&&\hspace{-1.5cm}
+ \frac{1}{2i} (\bm{u}^T \bm{v}^*) \Big( \big[ -\gamma_u\mathds{1}+\bm{\tilde K} \big]\, \bm{v}  \bm{u}^\dagger
+ \bm{u} \bm{v}^\dagger \big[ -\gamma_u^*\mathds{1}+\bm{\tilde K}^* \big]
\Big)  \bigg)\bm{M}^{-1/2} \, \bm{W}^* \bigg]_{ii},~~~
\label{eq:sigmaiann-beta3}
\end{eqnarray}
where $\bm{\tilde K}$ is the matrix in eq.~\eqref{eq:Ktilde}. The inner products in eq.~\eqref{eq:sigmaiann-beta3} are $\bm{u}^T \bm{u}^* = \cos(\phi - \phi^*)$ and $\bm{u}^T \bm{v}^*  = \sin(\phi - \phi^*)$.

The predictions of ZREFT at LO for the inclusive neutral-wino-pair annihilation rate and the inclusive charged-wino-pair annihilation rate are
\begin{subequations}
\begin{eqnarray}
2v_0\sigma_{0,\mathrm{ann}}(E) &=& \frac{16\pi/M}{ |L_0(E)|^2} \mathrm{Im}
\Big[ \gamma_0  - \left(t_\phi^2-|t_\phi^2|\right)  \big[ K_1(E) - K_1(0) \big] \Big],
\label{eq:sigma0ann-beta}
\\
2v_1\sigma_{1,\mathrm{ann}}(E) &=& 
\frac{(8\pi/M) C^2(E)}{|L_0(E)|^2} \mathrm{Im}
\Big[   (t_\phi^2)^* \gamma_0   - \left( t_\phi^2-|t_\phi^2|\right)^*\kappa_0(E)  \Big],
\label{eq:sigma1ann-beta}
\end{eqnarray}
\label{eq:sigma01ann-beta}%
\end{subequations}
where $L_0(E)$, $K_1(E)$, $\kappa_0(E)$, and $C^2(E)$ are given in eqs.~\eqref{eq:L0-E}, \eqref{eq:K1-E}, \eqref{eq:kappa0}, and eq.~\eqref{eq:Sommerfeld}. These annihilation rates depend on the complex parameters  $\gamma_0$ and $t_\phi^2$ explicitly and through $L_0(E)$. The term $ t_\phi^2 -|t_\phi^2|$ in the numerator is approximately $i\, \mathrm{Im}[t_\phi^2]$. The parameter $\gamma_0$ is given as a function of $M$ by the Pad\'e approximant in eq.~\eqref{eq:gamma0Pade}. The parameter $t_\phi^2$ is determined below in eq.~\eqref{eq:r0match} as a function of $M$ by matching neutral-wino elastic scattering amplitudes in ZREFT at LO and NREFT.

A Sommerfeld enhancement factor is the ratio of an annihilation rate to the leading-order annihilation rate. We now present an analytic expression for the Sommerfeld enhancement factor $S(v)$ for the inclusive annihilation rate of a neutral-wino pair. At leading order in $\alpha_2$, the annihilation rate of a neutral wino pair is given in eq.~\eqref{eq:sig0ann-thresh}. In ZREFT at LO, the inclusive annihilation rate is given in eq.~\eqref{eq:sigma0ann-beta}. The Sommerfeld enhancement factor is conventionally expressed as a function of the relative velocity $v = 2 v_0(E)$ of the neutral winos. Keeping only the terms in the numerator through first order in the imaginary parts of $\gamma_0$ and $t_\phi^2$, the Sommerfeld enhancement factor reduces to
\begin{equation}
S(v) = \frac{8 M}{\alpha_2^2\, |L_0(E)|^2}
\Big[ \mathrm{Im} [\gamma_0]  
- \, \mathrm{Im}[t_\phi^2] \,
\big(  \mathrm{Re}[K_1(E)] -  K_1(0)\big) \Big],
\label{eq:SommerfeldZREFTinc}
\end{equation}
where $K_1(E)$ and $L_0(E)$ are the functions of $E = Mv^2/4$ in eqs.~\eqref{eq:K1-E} and \eqref{eq:L0-E}. The function $i\eta(E)$ in the expression for $K_1(E)$  is $-\alpha(8\delta/M - v^2)^{-1/2}$ if $0<v<\sqrt{8\delta/M}$ and $-i\alpha(v^2-8\delta/M)^{-1/2}$ if $v>\sqrt{8\delta/M}$.

The unitarization of wino-pair annihilation  is taken into account in the expression for the Sommerfeld enhancement factor in eq.~\eqref{eq:SommerfeldZREFTinc} through the imaginary parts of $\gamma_0$ and $t_\phi^2$ in the function $L_0(E)$ in the denominator. In most previous calculations of the Sommerfeld enhancement factor, the unitarization of wino-pair-annihilation  has not been taken into account. This approximation can be  made in eq.~\eqref{eq:SommerfeldZREFTinc} by replacing the complex parameters $\gamma_0$ and $t_\phi^2$ in $L_0(E)$ by their real parts.

In ref.~\cite{Blum:2016nrz}, Blum, Sato, and Slatyer analyzed the effects of the unitaritization of  wino-pair annihilation on the Sommerfeld enhancement factor  in a single-channel model. They concluded that unitaritization can be taken into account by dividing the conventional Sommerfeld factor $S(v)$ by a denominator that involves two functions of both $v$ and $M$ that can be calculated numerically using the Schr\"odinger equation without an imaginary delta function potential. The analytic prediction for the Sommerfeld enhancement factor from ZREFT at LO in eq.~\eqref{eq:SommerfeldZREFTinc} is much simpler, because it depends only on the two complex parameters $\gamma_0$ and $t_\phi$ that can be determined numerically as functions of $M$ only.

\subsection{Matching with NREFT}
\label{sec:MatchingLO}

The interaction parameters of ZREFT can be determined by matching T-matrix elements in ZREFT with low-energy T-matrix elements in NREFT. The dimensionless T-matrix elements $T_{ij}(E)$ for wino-wino scattering in NREFT can be calculated numerically by solving the coupled-channel Schr\"odinger equation in eq.~\eqref{eq:radialSchrEqann}, in which the potential includes an imaginary delta-function potential. The T-matrix elements ${\cal T}_{ij}(E)$ for wino-wino scattering in ZREFT at LO are given analytically in eqs.~\eqref{eq:T00LO} and \eqref{eq:T01,11LO}. For $E>2 \delta$, the relation between the T-matrix in NREFT and the T-matrix in ZREFT is
\begin{equation}
\frac{1}{2M} \, \bm{v}(E)^{-1/2} \, \bm{T}(E)  \, \bm{v}(E)^{-1/2}= 
\frac{1}{8\pi} \, \bm{M}^{1/2} \, \bm{\mathcal{T}}(E) \, \bm{M}^{1/2},
\label{eq:TNR-TZR}
\end{equation}
where $\bm{M}$ is the diagonal matrix of masses in eq.~\eqref{eq:Mmatrix} and $\bm{v}(E)$ is the diagonal matrix of the velocities defined in eq.~\eqref{eq:v0,1-E}:
\begin{equation}
\bm{v}(E) = 
\begin{pmatrix} v_0(E) & 0 \\ 0 & v_1(E) \end{pmatrix}.
\label{eq:vmatrix}
\end{equation}
For $0<E<2 \delta$, the relation between the T-matrix elements for neutral-wino scattering is
\begin{equation}
\frac{1}{2Mv_0(E)} T_{00}(E) = \frac{M}{8 \pi}\, {\cal T}_{00}(E).
\label{eq:TNR-TZR00}
\end{equation}

In the absence of wino-pair annihilation, the interaction parameters of ZREFT at LO are $\alpha=1/137$ and the real parameters $\gamma_0$ and $t_\phi$. In ref.~\cite{Braaten:2017kci}, $\gamma_0(M)$ and $t_\phi(M)$ were determined as functions of the wino mass $M$ in the region near the critical mass $M_*=2.39$~TeV for $\delta = 170$~MeV by matching results from solving the Schr\"odinger equation for neutral-wino elastic scattering in NREFT. The inverse scattering length $\gamma_0(M)$ was set equal to its real value calculated as a function of $M$ in NREFT, which can be accurately approximated for $M$ in the region near $M_*$ by the Pad\'e approximant in eq.~\eqref{eq:Regamma0Pade}. The parameter $t_\phi(M)$ was determined by matching the prediction in eq.~\eqref{eq:r0LO} for the effective range $r_0$ from ZREFT at LO with the effective range calculated in NREFT. The explicit expression for $t_\phi^2$ is
\begin{equation}
t^2_\phi(M) = -  \frac{\Delta/2}{z_0^2\, \psi'(z_0) - \frac12 - z_0}r_0(M),
\label{eq:r0match}
\end{equation}
where $\Delta = \sqrt{2 M \delta}$ and $z_0 = - \alpha M/(2 \Delta)$. The real effective range $r_0(M)$ in NREFT can be accurately approximated for $M$ in the region near $M_*$ by the Pad\'e approximant in eq.~\eqref{eq:Rer0Pade}. The parameter $t_\phi$ is determined by eq.~\eqref{eq:r0match} up to a sign. In ref.~\cite{Braaten:2017kci}, the sign was chosen so that  $t_\phi$ is positive. At the unitarity mass $M_* = 2.39$~TeV, the parameters of ZREFT at LO are $\gamma_0=0$ and $t_\phi =0.877$. Having determined $t_\phi(M)$ and $\gamma_0(M)$, ZREFT at LO gives good predictions for other low-energy wino-pair observables as functions of the wino mass $M$ and the wino-pair energy $E$, including wino-wino cross sections and the binding energy of a wino-pair bound state \cite{Braaten:2017kci}.

\begin{figure}[t]
\centering
\includegraphics[width=0.48\linewidth]{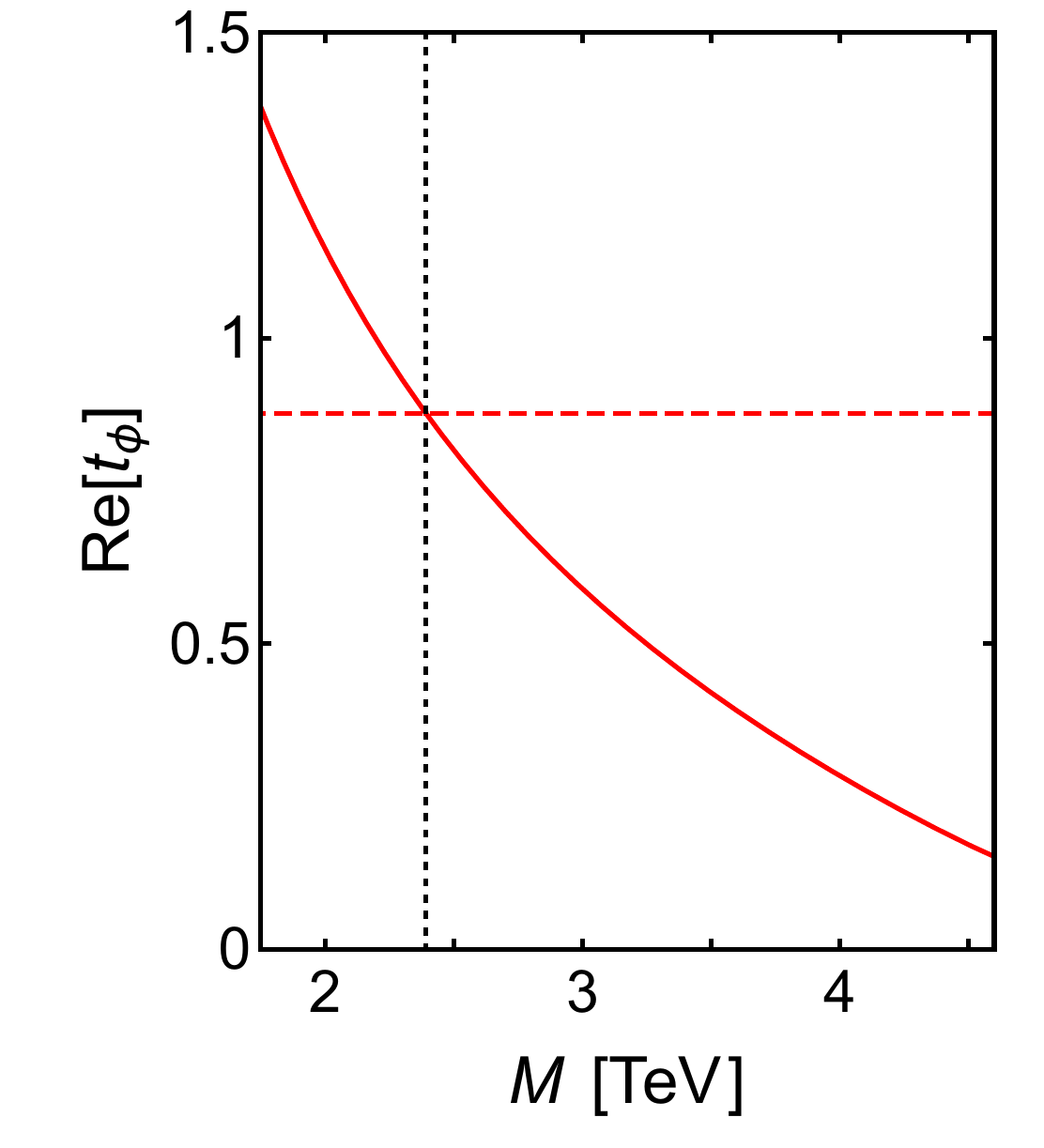}
~
\includegraphics[width=0.48\linewidth]{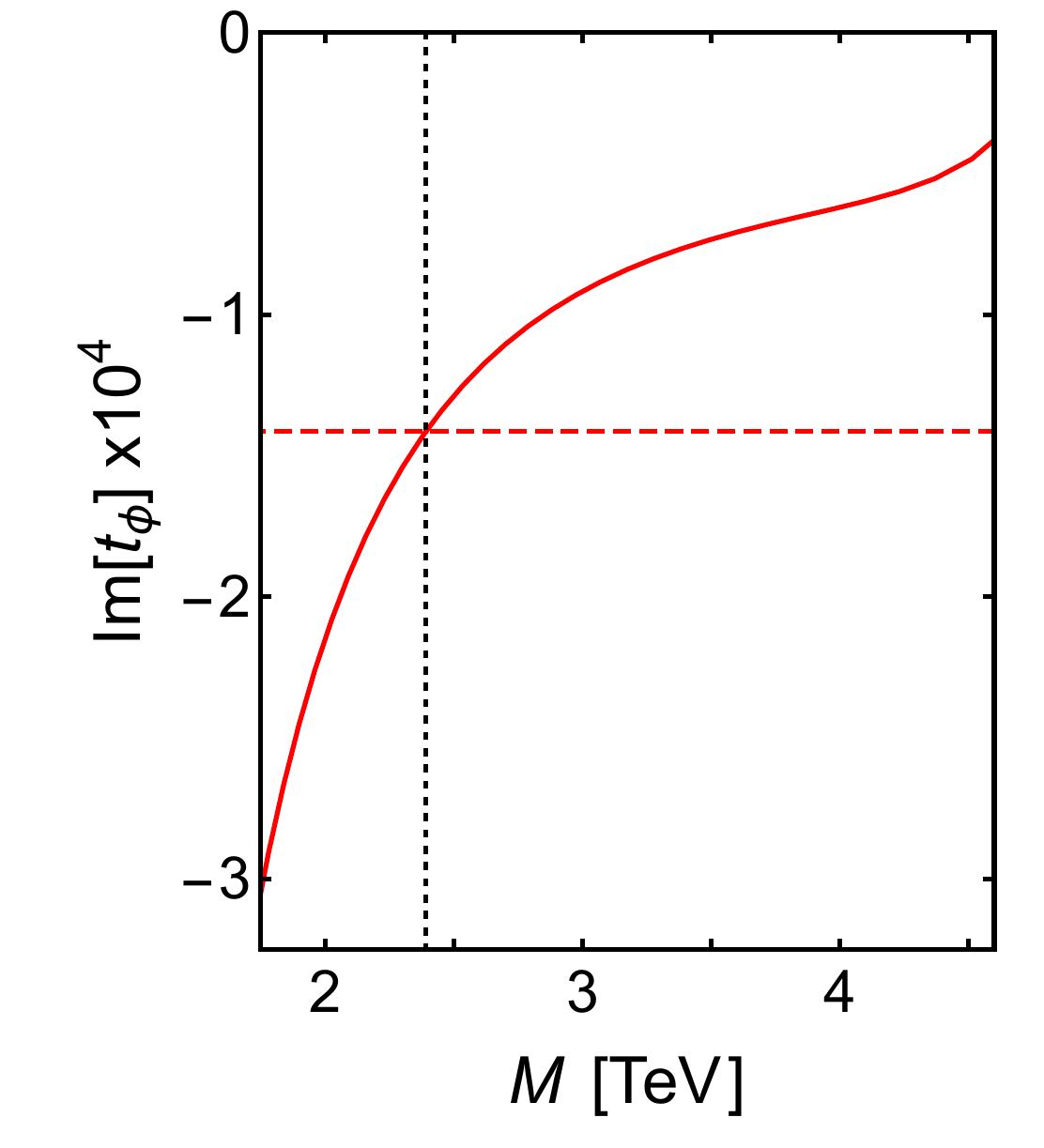}
\caption{Real and imaginary parts (left and right panels) of the interaction parameter $t_\phi = \tan\phi$ for ZREFT at LO as functions of the wino mass $M$. The complex parameter $t_\phi(M)$ (solid red curves) is determined from the complex effective range $r_0$ in NREFT by the matching condition in eq.~\eqref{eq:r0match}. The vertical dotted lines indicate the  unitarity mass $M_* = 2.39$~TeV.}
\label{fig:tanphivsM}
\end{figure}

If wino-pair annihilation is taken into account, the interaction parameters of ZREFT at LO are $\alpha=1/137$ and the complex parameters $t_\phi$ and $\gamma_0$, which have small imaginary parts. We set $\gamma_0(M)$ equal to the complex inverse scattering length calculated as a function of $M$ in NREFT. Its real and imaginary parts  can be accurately approximated by the Pad\'e approximants  in eqs.~\eqref{eq:gamma0Pade}. We determine  $t_\phi(M)$ from the complex effective range $r_0(M)$ calculated in NREFT as a function of $M$ by using the matching condition in eq.~\eqref{eq:r0match}. The real and imaginary parts of $r_0(M)$ in NREFT can be accurately approximated by the Pad\'e approximants  in eqs.~\eqref{eq:r0Pade}. The parameter $t_\phi$ is determined by eq.~\eqref{eq:r0match} up to a sign. We choose the sign so the real part of $t_\phi$ is positive. At the unitarity mass $M_*=2.39$~TeV, the complex parameters of ZREFT at LO are the pure imaginary inverse  scattering length $\gamma_0$ given in eq.~\eqref{eq:gamma0*} and the complex parameter $t_\phi = \tan(\phi)$:
\begin{equation}
t_\phi(M_*) =  0.877  - 1.41 \times 10^{-4}\, i .
\label{eq:tphi*}
\end{equation}
The imaginary part of $t_\phi$ is smaller than the real part by about 4 orders of magnitude, consistent with suppression by a factor of $\alpha_2 m_W/M$. Its real and imaginary parts are shown as functions of $M$ in figure~\ref{fig:tanphivsM}. They both vary significantly with $M$ within the range of validity of ZREFT. In the predictions of ZREFT at LO away from the unitarity mass $M_*$, it is therefore essential to use the $M$-dependent value of $t_\phi(M)$ from matching at the mass $M$ rather than the constant value in eq.~\eqref{eq:tphi*} from matching at unitarity.

\subsection{Predictions of ZREFT at LO}
\label{sec:PredictLO}

\begin{figure}[t]
\centering
\includegraphics[width=0.48\linewidth]{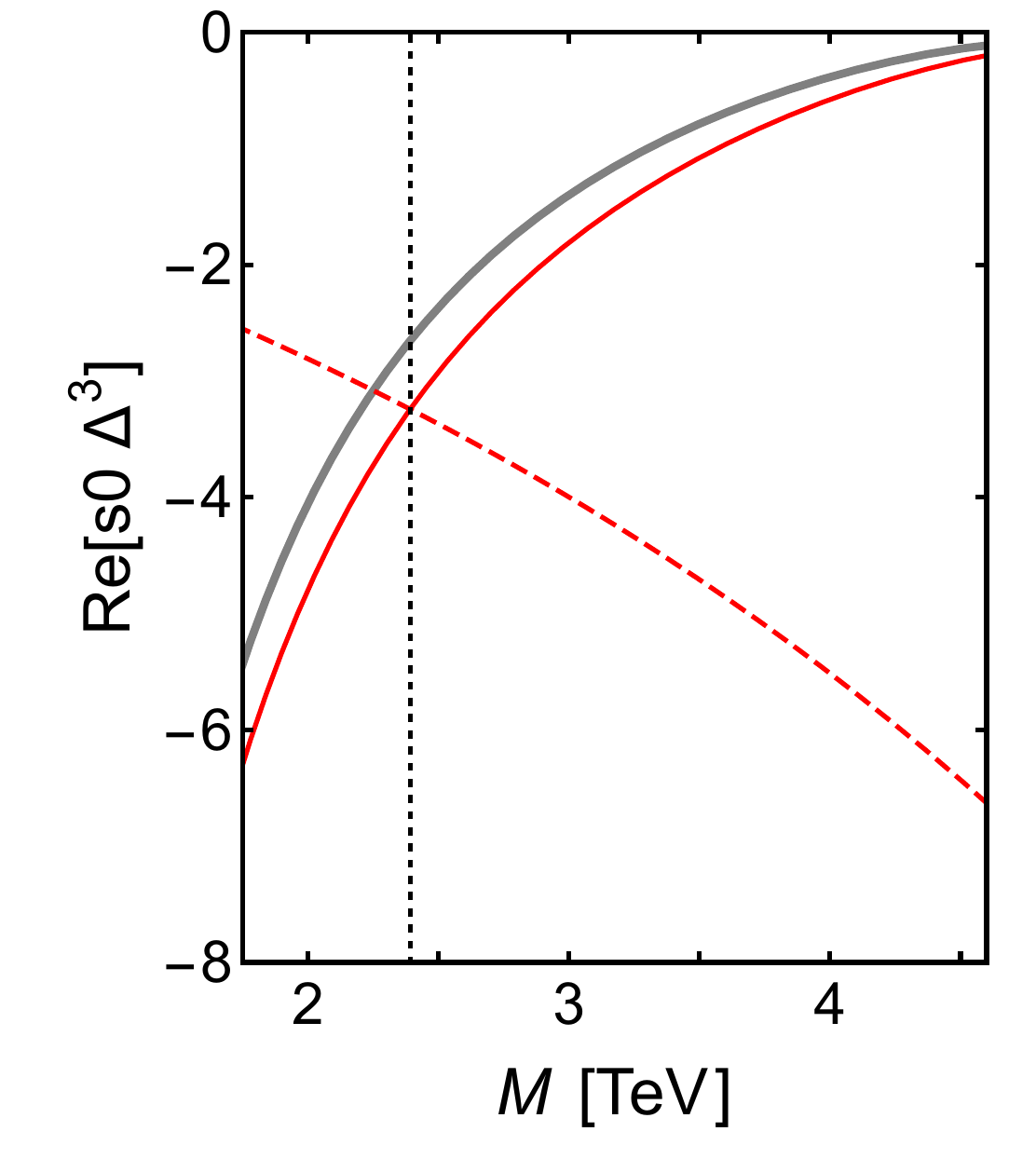}
\includegraphics[width=0.48\linewidth]{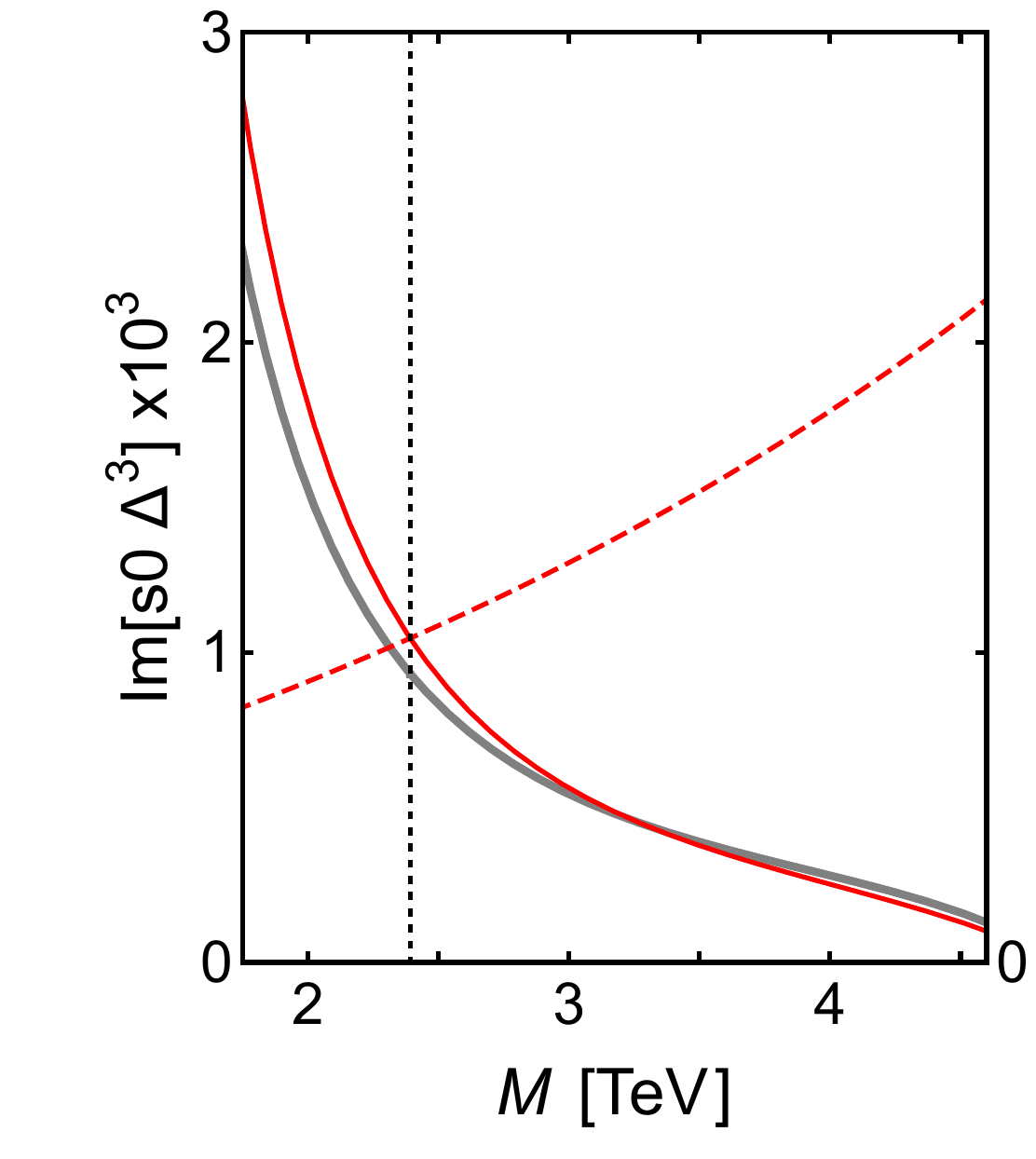}
\caption{Real and imaginary parts (left and right panels) of the neutral-wino shape parameter $s_0$ as functions of the wino mass $M$: NREFT (thicker grey curves), ZREFT at LO  with $M$-dependent parameter $\tan \phi(M)$ (solid red curves), and ZREFT at LO with  constant parameter $\tan \phi(M_*)$ (dashed red curves). The vertical dotted lines indicate the  unitarity mass $M_* = 2.39$~TeV.}
\label{fig:r0s0vsM}
\end{figure}

Having determined the complex inverse scattering length $\gamma_0(M)$ and the complex parameter $t_\phi(M)$ as functions of the wino mass $M$, we  can use ZREFT at LO to predict other observables. They include the complex shape parameter $s_0$ and the wino pair annihilation rates.

The analytic prediction for the shape parameter $s_0$ in ZREFT at LO is given in eq.~\eqref{eq:s0LO}. In figure~\ref{fig:r0s0vsM}, the predictions of ZREFT at LO for the real and imaginary parts of $s_0(M)$ as functions of $M$ are compared to the results from NREFT. At unitarity, the predictions for  Re$[s_0]$ and Im$[s_0]$ differ from the results from NREFT in eq.~\eqref{eq:s0*EM} by the multiplicative factors 0.82 and 0.89, respectively. The accuracy of the predictions remains comparable at other values of $M$ within the range of  validity of ZREFT. If $M$ is very close to the unitarity mass $M_* = 2.39$~TeV, we can use the constant  parameter $t_\phi(M_*)$ in eq.~\eqref{eq:tphi*}, which was determined by matching $r_0$ at unitarity. However, as shown in figure~\ref{fig:r0s0vsM}, the resulting predictions  for Re$[s_0]$ and Im$[s_0]$ as functions of $M$ have the wrong slopes. For predictions away from unitarity, it is  essential to use the $M$-dependent parameters $\gamma_0(M)$ and $t_\phi(M)$.

\begin{figure}[t]
\centering
\includegraphics[width=0.48\linewidth]{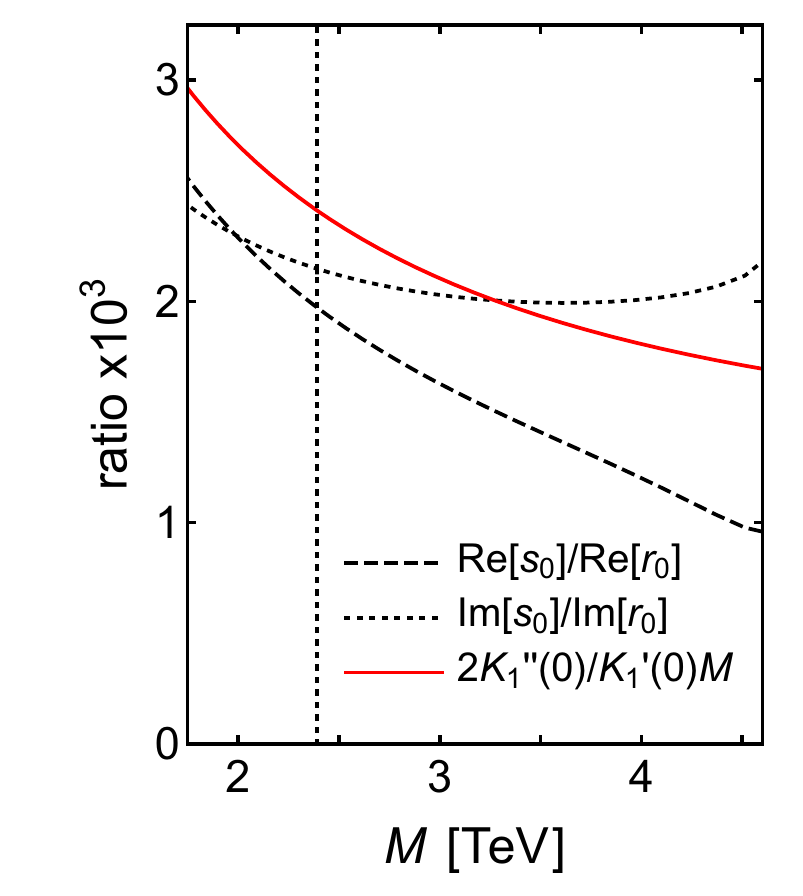}
\caption{The ratio of the real parts of $s_0$ and $r_0$ (dashed curve) and the ratio of the imaginary parts of $s_0$ and $r_0$ (dotted curve) as functions of the wino mass $M$. The solid line is the parameter-free prediction of ZREFT at LO in eq.~\eqref{eq:s0r0ratio}. The vertical dotted line indicates the unitarity mass $M_*=2.39$~TeV.}
\label{fig:s0r0ratios}
\end{figure}

ZREFT at LO gives the parameter-free predictions in eq.~\eqref{eq:s0r0ratio} for the ratio of Re$[s_0]$ to Re$[r_0]$ and the ratio of Im$[s_0]$ to  Im$[r_0]$ as functions of $M$. The ratios and the predictions are plotted as functions of $M$ in figure~\ref{fig:s0r0ratios}. At unitarity, the ratio of the real parts and the ratio of the imaginary parts differ from the prediction in eq.~\eqref{eq:s0r0ratio} by multiplicative factors of $0.82$ and $0.89$, respectively. The accuracy of the  parameter-free predictions remains comparable at other values of $M$ within the range of  validity of ZREFT.

The observables that depend most dramatically on the wino mass $M$ are the neutral-wino elastic cross section $\sigma_{0 \to 0}$ at zero energy, which is shown in figures~\ref{fig:sigma00vsM} and \ref{fig:sigma00vsMlog}, and the neutral-wino annihilation rate $2v_0 \sigma_{0,\mathrm{ann}}$ at zero energy which is shown in figures~\ref{fig:2v0sigma0annvsM} and \ref{fig:2v0sigma0annvsMlog}. Having chosen the complex inverse scattering length $\gamma_0(M)$ as one of the interaction parameters of ZREFT at LO, its predictions for $\sigma_{0 \to 0}$ at $E=0$ and $2v_0\sigma_{0,\mathrm{ann}}$ at $E=0$ are exact.

Near a unitarity mass, the neutral-wino  elastic cross section $\sigma_{0 \to 0}$ has dramatic dependence on the energy $E$ in the low-energy limit. The $E$-dependence of the NREFT cross section at $M_*=2.39$~TeV is shown as a log-log plot in figure~\ref{fig:sigma00-NREFT-lowE}. In the low-energy limit and in the scaling region, the prediction of ZREFT at LO is so accurate that it cannot be distinguished from the  NREFT result. The $E$-dependence of the NREFT cross section above the scaling region is shown in figure~\ref{fig:sigma00-NREFT}. In the region near the charged-wino-pair threshold, the effects of wino-pair annihilation are extremely small, because they are suppressed by $\alpha_2 m_W/M$. The predictions of ZREFT at LO with wino-pair annihilation taken into account cannot be distinguished from the predictions without wino-pair annihilation. The prediction of ZREFT at LO was compared to the NREFT result without wino-pair annihilation in ref.~\cite{Braaten:2017kci}. The plots look the same with wino-pair annihilation taken into account.

\begin{figure}[t]
\centering
\includegraphics[width=0.8\linewidth]{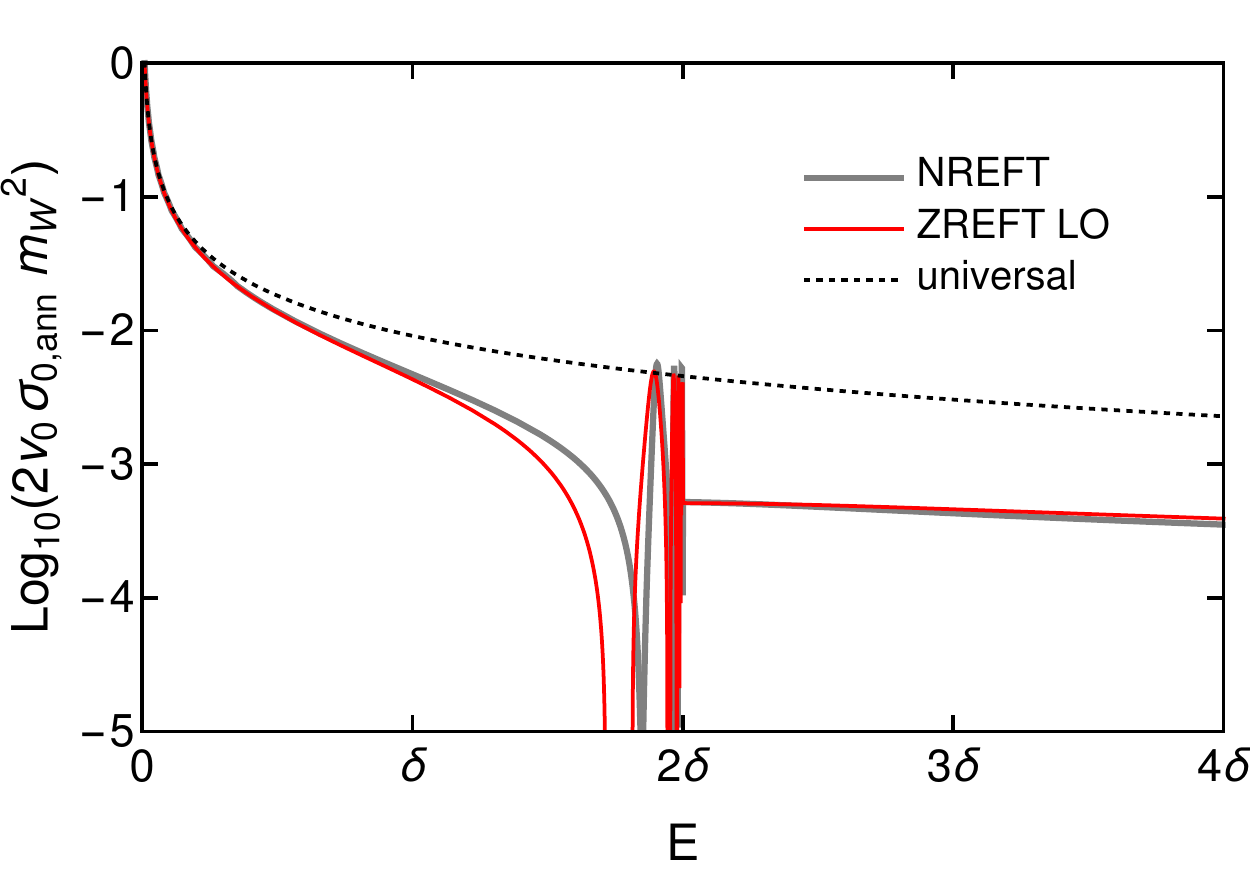}
\caption{Neutral-wino annihilation rate $2v_0\sigma_{0,\mathrm{ann}}$ (solid curves) as a function of the energy $E$. The annihilation rate at the unitarity mass $M_*=2.39$~TeV is shown for NREFT (thicker grey solid curve), for ZREFT at LO (red solid curve), and for the universal approximation in Eq.~\eqref{eq:sigma0ann-uni} (dotted curve).}
\label{fig:sigma0annvsE}
\end{figure}

Near a unitarity mass, the neutral-wino annihilation rate $2v_0\sigma_{0,\mathrm{ann}}$ also has dramatic dependence on the energy $E$  in the low-energy limit. The $E$-dependence of the NREFT annihilation rate at $M_*=2.39$~TeV is shown a log-log plot in figure~\ref{fig:2v0sigma0ann-Elog}. In the low-energy limit and in the scaling region, the prediction of ZREFT at LO is so accurate that it cannot be distinguished from the  NREFT result. The $E$-dependence of the NREFT annihilation rate above the scaling region is shown in figure~\ref{fig:2v0sigma0ann-E}. In figure~\ref{fig:sigma0annvsE}, the annihilation rate at $M_*=2.39$~TeV in NREFT is compared to the prediction of ZREFT at LO and to the universal approximation in Eq.~\eqref{eq:sigma0ann-uni}. The prediction of ZREFT at LO differs from the NREFT result by less than 5\% for $E/2\delta$ less than $0.42$. The universal approximation differs from the NREFT result by less than 5\% for $E/2\delta$ less than $0.06$.

\begin{figure}[t]
\centering
\includegraphics[width=0.48\linewidth]{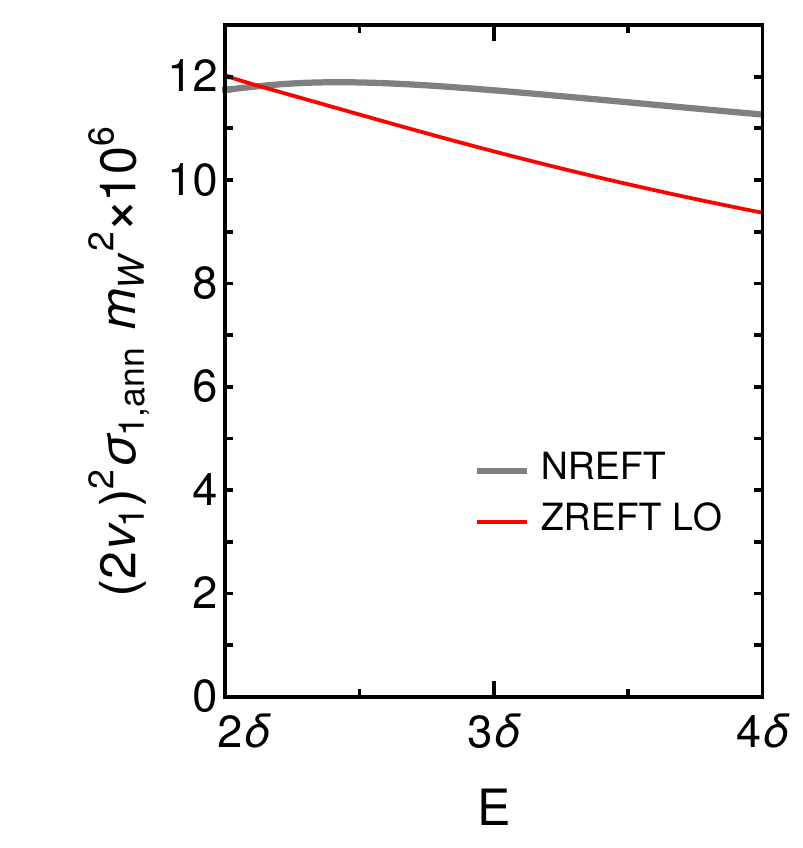}
\caption{Charged-wino annihilation cross section $\sigma_{1,{\rm ann}}$ multiplied by two powers of the relative velocity as a function of the energy $E$. The annihilation rate at the unitarity mass $M_*=2.39$~TeV is shown for NREFT (thicker grey curve) and for ZREFT at LO (red curve).}
\label{fig:sigma1annvsE}
\end{figure}

The charged-wino annihilation cross section $\sigma_{1,\mathrm{ann}}$ is  a smooth function of the energy that diverges as $1/(E-2 \delta)$ as $E$ approaches the threshold $2 \delta$ from above. If it is multiplied by two powers of the relative velocity $2v_1$, it has a finite limit as $E\to 0$. In figure~\ref{fig:sigma1annvsE}, the energy dependence of $(2v_1)^2 \sigma_{1,\mathrm{ann}}$ at $M_*=2.39$~TeV in NREFT is compared to the prediction of ZREFT at LO in eq.~\eqref{eq:sigma1ann-beta}. At the charged-wino pair threshold, the ratio of the prediction and the NREFT result is 1.03. The ratio  decreases slowly as $E$ increases.

\subsection{Wino-pair bound state}
\label{sec:BoundStateLO}

If the wino mass $M$ is larger than the unitarity mass where the real part of the neutral-wino inverse scattering length $\gamma_0(M)$ vanishes, the S-wave resonance is a bound state below the neutral-wino-pair threshold. The bound state is a superposition of a neutral-wino pair and a charged-wino pair, and we denote it by $(ww)$. In NREFT, the energy of the bound state is a discrete eigenvalue of the coupled-channel radial Schr\"odinger equation in eq.~\eqref{eq:radialSchrEq} or eq.~\eqref{eq:radialSchrEqann}. In the absence of wino-pair annihilation, the energy of the bound state is $- E_{(ww)}$, where $E_{(ww)}$ is the positive binding energy. If wino-pair annihilation is taken into account, the bound state has a complex energy $- E_{(ww)} -i \Gamma_{(ww)}/2$, where $ \Gamma_{(ww)}$ is the decay width of the bound state.

In ZREFT, the binding energy $E_{(ww)}$ and the decay width $ \Gamma_{(ww)}$ of the wino pair bound state can be obtained by solving an analytic equation numerically. If Re$[\gamma_0] > 0$, each of the transition amplitudes ${\cal A}_{ij}(E)$ given by the entries of the matrix in eq.~\eqref{eq:AmatrixLOC} has a pole at an energy  $-E_{(ww)} -i \Gamma_{(ww)}/2$ whose real part is below the neutral-wino-pair threshold. The pole in $E$ is at a zero of the function $L_0(E)$ in eq.~\eqref{eq:L0-E}. The complex energy $-E_{(ww)} -i \Gamma_{(ww)}/2$ can be expressed as $-\gamma^2/M$, where the complex binding momentum $\gamma$ is a solution to the equation
\begin{equation}
0=  \gamma - \gamma_0   + t_\phi^2 \, \big[ K_1(-\gamma^2/M) -K_1(0) \big].
\label{eq:gammaLO-eq}
\end{equation}
The correct root of this equation is the one that approaches $\gamma_0$ as $\gamma_0$ approaches 0 with a positive real part. The binding energy and the decay width can be expressed as
\begin{subequations}
\begin{eqnarray}
E_{(ww)}  &=& \mathrm{Re}[\gamma^2]/M,
\label{eq:E-gamma}
\\
\Gamma_{(ww)} &=& 2\, \mathrm{Im}[\gamma^2]/M.
\label{eq:Gamma-gamma}
\end{eqnarray}
\label{eq:EGamma-gamma}
\end{subequations}

\begin{figure}[t]
\centering
\includegraphics[width=0.48\linewidth]{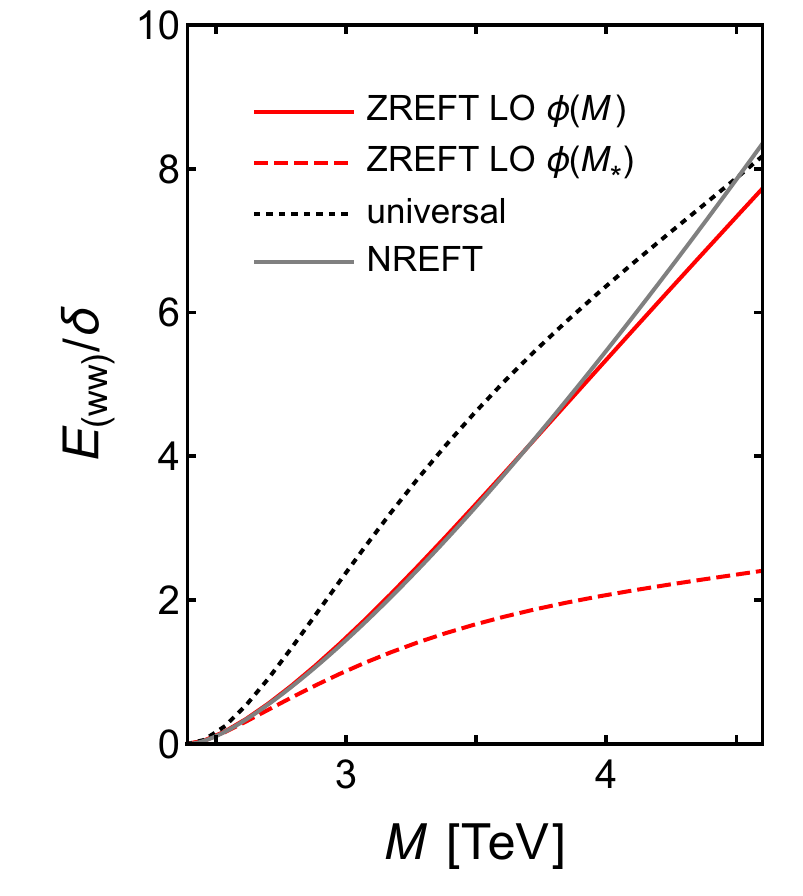}
\includegraphics[width=0.48\linewidth]{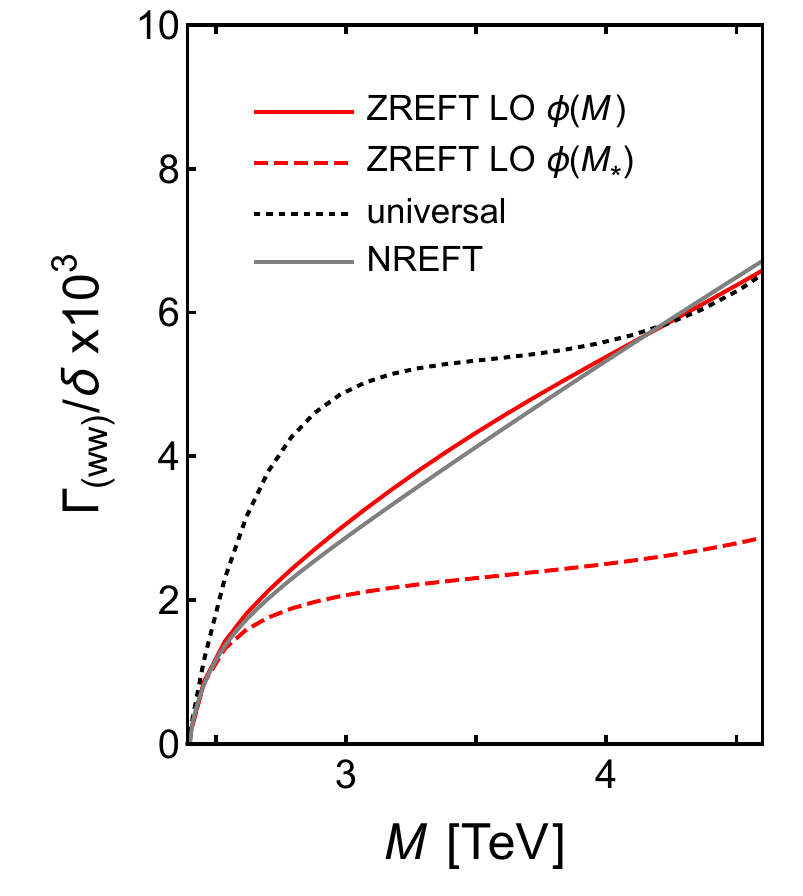}
\caption{Binding energy $E_{(ww)}$ (left panel) and decay width $\Gamma_{(ww)}$ (right panel) of the wino-pair bound state as functions of the wino mass $M$: NREFT  (thicker grey curves), ZREFT at LO with $M$-dependent parameter $\tan\phi(M)$ (solid red curves), and ZREFT at LO with constant parameter $\tan\phi(M_*)$ (dashed red curves), and the universal approximations in eqs.~\eqref{eq:EGamma-gamma} (dotted curves).}
\label{fig:bindingwidth}
\end{figure}

The binding energy $E_{(ww)}$ and the decay width $\Gamma_{(ww)}$ are shown as functions of $M$ in the left and right panels of figure~\ref{fig:bindingwidth}, respectively. The horizontal range of $M$ is from the unitarity mass $M_* = 2.39$~TeV to the upper end of the region of applicability of ZREFT, which is about 4.6~TeV. The NREFT results for the binding energy and decay width go to zero as $M$ approaches $M_*$ from above. The predictions of ZREFT at LO from eq.~\eqref{eq:EGamma-gamma} agree very well with the NREFT results. The errors in $E_{(ww)}$ and $\Gamma_{(ww)}$ remain below 8\% over the entire range of $M$ in figure~\ref{fig:bindingwidth}.

Particles with short-range interactions that produce an S-wave resonance sufficiently close to their scattering threshold have universal low-energy behavior that is completely determined by their S-wave scattering length $a_0$ \cite{Braaten:2004rn}. If the real part of $1/a_0$ is positive, the S-wave bound state closest to the threshold is universal. The universal approximation to the complex energy $-E_{(ww)} -i \Gamma_{(ww)}/2$ is $-\gamma_0^2/M$, where $\gamma_0$ is the complex inverse scattering length. The universal approximations to the binding energy and the decay width are 
\begin{subequations}
\begin{eqnarray}
E_{(ww)}  &=& \big( \mathrm{Re}[\gamma_0]^2 - \mathrm{Im}[\gamma_0]^2 \big)/M,
\label{eq:E-gamma0}
\\
\Gamma_{(ww)} &=& 4\, \mathrm{Re}[\gamma_0]\, \mathrm{Im}[\gamma_0]/M.
\label{eq:Gamma-gamma0}
\end{eqnarray}
\label{eq:EGamma-gamma0}%
\end{subequations}
For neutral winos with mass near  the unitarity mass $M_*=2.39$~TeV, the universal approximations in eqs.~\eqref{eq:EGamma-gamma0} are applicable for $M$ inside the region between $M_*$ and 2.9~TeV. The universal approximations in eqs.~\eqref{eq:EGamma-gamma0} imply that $\Gamma_{(ww)}$ decreases to 0 as $M$ decreases to the unitarity mass $M_*$ defined by $\mathrm{Re}[\gamma_0]=0$, but that $E_{ww}$ decreases to 0 and changes sign before $M$ reaches $M_*$. However the binding energy and decay width are not physically well-defined when $\mathrm{Re}[\gamma_0]$ is comparable to $\mathrm{Im}[\gamma_0]$. For NREFT, the real and imaginary parts of the inverse scattering length $\gamma_0(M)$ can be approximated by the Pad\'e approximants in eq.~\eqref{eq:gamma0Pade}. In the left and right panels of figure~\ref{fig:bindingwidth}, the universal approximations to the binding energy and decay width in eq.~\eqref{eq:EGamma-gamma0} are compared with the results from NREFT. The universal approximations for $E_{(ww)}$ and $\Gamma_{(ww)}$ remain reasonably accurate  as $M$ increases from $M_*$ to about 2.5~TeV, but this range of $M$ is much narrower than the expected range of validity, which is up to about 2.9~TeV.

\subsection{Partial annihilation rates}
\label{sec:PartialRates}

The effects of wino-pair annihilation can be taken into account in NREFT through the imaginary delta-function potential in the Schr\"odinger equation in eq.~\eqref{eq:radialSchrEqann}. The coefficient matrix $\bm{\Gamma}$  is given to order $\alpha_2^2$ in eq.~\eqref{eq:Gamma}. It is obtained by calculating the annihilation contribution to the discontinuity in wino-wino scattering amplitudes from the diagrams in figure~\ref{fig:AnnihilationVertex}. Perturbative corrections to $\bm{\Gamma}$ can be obtained by calculating the annihilation contribution to the discontinuity to higher orders. If $M \gg m_W$, the corrections involve the large logarithm $\log (M/m_W)$. The leading logarithms can be summed to all orders by replacing $\alpha_2$ in eq.~\eqref{eq:Gamma} by the running $SU(2)$ coupling constant $\bar\alpha_2$ at the scale of the wino mass $M$. In the fundamental theory we consider, the winos are either in a single $SU(2)$ multiplet added to the Standard Model or they are the lightest supersymmetric partner particles in the MSSM. In either case, there are no other additional particles below the scale $M$, so the running of gauge couplings above the electroweak scale is the same as in the Standard Model. The one-loop running of $\alpha_2$ is given by
\begin{equation}
\bar\alpha_2 = \frac{1}{1/\alpha_2 + (19/12\pi) \log(M/m_W)},
\label{eq:alpha2-running}
\end{equation}
where $\alpha_2 = 1/29.5$. At the first unitarity mass  $M=2.39$~TeV, the running $SU(2)$ coupling constant is $\bar\alpha_2 = 1/31.2$.

The leading-order $2\times2$ matrix $\bm{\Gamma}$ in eq.~\eqref{eq:Gamma} can be decomposed into its exclusive contributions from each of the pairs of electroweak gauge bosons, which are $\gamma\gamma$, $\gamma Z^0$, $Z^0 Z^0$, and $W^+ W^-$  \cite{Hisano:2004ds}:
\begin{equation}
\bm{\Gamma} = 
\bm{\Gamma}^{(\gamma\gamma)}+\bm{\Gamma}^{(\gamma Z)} +\bm{\Gamma}^{(Z Z)}+\bm{\Gamma}^{(WW)}.
\label{eq:Gammatotal}
\end{equation}
The decay modes that produce monochromatic photons are $\gamma\gamma$ and $\gamma Z^0$. The $2 \times 2$ matrices corresponding to these final states are \cite{Hisano:2004ds}
\begin{subequations}
\label{eq:Gamma-VV}
\begin{eqnarray}
\bm{\Gamma}^{(\gamma\gamma)} &=&
\frac{\pi  \alpha_2^2}{2M^2} \begin{pmatrix} ~0~  & 0 \\ 0 & 2s_w^4 \end{pmatrix},
\label{eq:Gamma-2gamma}
\\
\bm{\Gamma}^{(\gamma Z)} &=&
\frac{\pi \alpha_2^2}{2M^2} \begin{pmatrix} ~0~  & 0 \\ 0 & 4 s_w^2 c_w^2 \end{pmatrix},
\label{eq:Gamma-gammaZ}
\end{eqnarray}
\end{subequations}
where $s_w = \sin \theta_w$ and $c_w = \cos \theta_w$. The $2 \times 2$ matrix in the semi-inclusive rate for production of a monochromatic photon is
\begin{equation}
\bm{\Gamma}^{(\gamma X)} \equiv 2\bm{\Gamma}^{(\gamma\gamma)}+\bm{\Gamma}^{(\gamma Z)}
=\frac{2\pi \alpha_2^2 s_w^2}{M^2} \begin{pmatrix} ~0~  & ~0~ \\ 0 & 1 \end{pmatrix}.
\label{eq:GammagammaX}
\end{equation}

Radiative corrections to the annihilation rates into exclusive final states, such as $\gamma\gamma$ and $\gamma Z^0$, have coefficients with large logarithms of $M/m_W$. There are up to two such logarithms for each additional power of $\alpha_2$. Several groups have used soft collinear effective theory (SCET) to sum large logarithms  of $M/m_W$ to all orders in $\alpha_2$. Baumgart, Rothstein, and Vaidya calculated the semi-inclusive annihilation rate into final states with a monochromatic photon to leading-double-logarithm accuracy for wino dark matter \cite{Baumgart:2014vma,Baumgart:2014saa}. The calculation was subsequently extended to include single logarithms at fixed order \cite{Baumgart:2015bpa}. Ovanesyan, Slatyer, and Stewart calculated the exclusive annihilation rates into $\gamma \gamma$ and $\gamma Z^0$ to next-to-leading-logarithm accuracy for wino dark matter \cite{Ovanesyan:2014fwa}. Bauer, Cohen, Hill and Solon calculated the  exclusive annihilation rates into $\gamma \gamma$ and $\gamma Z^0$ to next-to-leading-logarithm accuracy for scalar dark matter \cite{Bauer:2014ula}. To leading-double-logarithm accuracy, the $2\times2$ matrix associated with the semi-inclusive final state $\gamma + X$  is \cite{Baumgart:2014vma}
\begin{equation}
\bm{\Gamma}^{(\gamma X)} \approx
\frac{\pi  \alpha_2^2s_w^2}{3M^2} 
\begin{pmatrix} 2 f_-  &  \sqrt{2} f_- \\  \sqrt{2} f_- & 3f_+ \end{pmatrix},
\label{eq:Gamma-gamma+X}
\end{equation}
where
\begin{equation}
f_\pm = 1 \pm  \exp\big(-(3\alpha_2/\pi) \log^2(M/m_W)\big).
\label{eq:fpm}
\end{equation}
The scale of $\alpha_2$ in $f_\pm$ can be determined only by also summing the leading single logarithms to obtain next-to-leading-logarithm accuracy as in ref.~\cite{Ovanesyan:2014fwa}. The matrix  in eq.~\eqref{eq:Gamma-gamma+X} reproduces the structure of the annihilation rate in eq.~(12) of ref.~\cite{Baumgart:2014vma}. In the limit $\alpha_2 \to 0$, the matrix reduces to the leading-order matrix in eq.~\eqref{eq:GammagammaX}.

The predictions of ZREFT at LO for the inclusive wino-pair annihilation rates are given in eq.~\eqref{eq:sigma01ann-beta}. To leading order in $\alpha_2 m_W/M$, each rate is the sum of a term proportional to  Im$[\gamma_0]$ and a term proportional to Im$[t_\phi^2]$, which by the matching condition in eq.~\eqref{eq:r0match} is proportional to Im$[r_0]$. The small imaginary parts of $\gamma_0$ and $r_0$ are linear in the entries of the matrix $\bm{\Gamma}$ in the Schr\"odinger equation for NREFT in eq.~\eqref{eq:radialSchrEqann}. It is convenient to define a dimensionless matrix $\hat{\bm{\Gamma}}$ by dividing $\bm{\Gamma}$ by $\pi \alpha_2^2/2M^2$:
\begin{equation}
\bm{\Gamma} = \frac{\pi \alpha_2^2}{2M^2} \, \hat{ \bm{\Gamma}}, \quad
\hat{ \bm{\Gamma}}=\begin{pmatrix} 2  & \sqrt{2} \\ \sqrt2 & 3 \end{pmatrix}.
\label{eq:Gamma-hat}
\end{equation}
We have given the explicit form of $\hat{\bm{\Gamma}}$ obtained from the leading order expression for $\bm{\Gamma}$ in eq.~\eqref{eq:Gamma}. If the entries of $\hat{ \bm{\Gamma}}$ are allowed to vary, the imaginary parts of $\gamma_0$ and $r_0$ are linear in $\hat \Gamma_{00}$, $\hat \Gamma_{01}= \hat \Gamma_{10}$, and $\hat \Gamma_{11}$. They can be expressed as
\begin{subequations}
\begin{eqnarray}
\mathrm{Im}[\gamma_0(M,\hat\Gamma)] &=&
\left[ c_0(M)\, \hat \Gamma_{00} +c_1(M)\, \hat \Gamma_{01} +c_2(M)\, \hat \Gamma_{11} \right]
\mathrm{Im}[\gamma_0(M_*)],
\label{eq:Imgamma0}
\\
\mathrm{Im}[r_0(M,\hat\Gamma)] &=&
\left[d_0(M)\, \hat \Gamma_{00} \,+d_1(M)\, \hat \Gamma_{01} \,+ d_2(M)\, \hat \Gamma_{11} \right]
\mathrm{Im}[r_0(M_*)] ,
\label{eq:Imtphi2}
\end{eqnarray}
\label{eq:Imgamma0tphi2}%
\end{subequations}
where $\mathrm{Im}[\gamma_0(M_*)]=3.40\times 10^{-4} \,m_W$ and $\mathrm{Im}[r_0(M_*)]=1.50\times 10^{-3}\, m_W^{-1}$ are given in Eqs.~\eqref{eq:gamma0*} and \eqref{eq:r0*EM}. The dimensionless coefficients $c_n(M)$ and $d_n(M)$ in eqs.~\eqref{eq:Imgamma0tphi2} are smooth functions of $M$ within the region of validity of ZREFT, which is roughly from 1.8~TeV to 4.6~TeV. The prefactors of the imaginary parts of $\gamma_0(M)$ and $t_\phi^2(M)$ on the right sides of eqs.~\eqref{eq:Imgamma0tphi2} reduce to 1 at $M_*=2.39$~TeV if we set $\hat \Gamma_{00}=2$, $\hat \Gamma_{01}=\sqrt{2}$, and $\hat \Gamma_{11}=3$. Within the region of validity of ZREFT, the functions $c_n(M)$ and $d_n(M)$ can be approximated by quartic polynomials in $(M-M_*)/M_*$:
\begin{subequations}
\begin{eqnarray}
c_n(M) &=& \sum_{j=0}^4 c_{nj}  \big[(M-M_*)/M_*\big]^j,
\\
d_n(M) &=& \sum_{j=0}^4 d_{nj}  \big[(M-M_*)/M_*\big]^j.
\label{eq:dnpoly}
\end{eqnarray}
\end{subequations}
The fitted coefficients of the polynomials are given in table~\ref{tab:Coefficients}.

\begin{table}[t]
\begin{center}
\begin{tabular}{l|ccccc|ccccc|}
       &  \multicolumn{5}{c}{$c_{nj}$} & \multicolumn{4}{c}{$d_{nj}$} \\
\hline
$n$~\textbackslash~$j$
    &   0  &   1   &  2   &   3   &  4   &   0  &   1   &  2   &   3   & 4 \\
\hline
0   & 0.090 & -0.302 & 0.822 & -1.111 & 0.565 & 0.153 & -0.775 & 2.304 & -3.311 & 1.697 \\
1   & 0.240 & -0.620 & 1.460 & -1.807 & 0.894 & 0.279 & -1.154 & 2.954 & -3.950 & 1.946 \\
2   & 0.160 & -0.298 & 0.587 & -0.638 & 0.308 & 0.100 & -0.323 & 0.685 & -0.837 & 0.388 \\
\end{tabular}
\end{center}
\caption{
Fitted coefficients $c_{nj}$ and $d_{nj}$ of  $[(M-M_*)/M_*]^j$ in the polynomials $c_n(M)$ and $d_n(M)$ that appear in the decompositions of the imaginary parts of $\gamma_0$ and $t_\phi^2$ in eqs.~\eqref{eq:Imgamma0tphi2}.}
\label{tab:Coefficients}
\end{table}

The partial annihilation rates into monochromatic photons in ZREFT can be obtained from the inclusive annihilation rates in eqs.~\eqref{eq:sigma01ann-beta} by replacing the imaginary parts of $\gamma_0$ and $t_\phi^2$ in the numerator by the contributions from the monochromatic photon channels. Using the matching condition in eq.~\eqref{eq:r0match}, we can write
\begin{equation}
\mathrm{Im}[t_\phi^2]^{(\gamma X)} = -  \frac{\Delta/2}{z_0^2\, \psi'(z_0) - \frac12 - z_0}\mathrm{Im}[r_0]^{(\gamma X)},
\label{eq:r0match-gammaX}
\end{equation}
where $\Delta = \sqrt{2 M \delta}$ and $z_0 = - \alpha M/(2 \Delta)$. The contribution to Im$[\gamma_0]$ and Im$[r_0]$ are
\begin{subequations}
\begin{eqnarray}
\mathrm{Im}[\gamma_0(M)]^{(\gamma X)} &=&
\left[c_0(M)\, \hat \Gamma_{00}^{(\gamma X)} + c_1(M)\, \hat \Gamma_{01}^{(\gamma X)} 
+c_2(M)\, \hat \Gamma_{11}^{(\gamma X)} \right] \mathrm{Im}[\gamma_0(M_*)],
\label{eq:Imgamma0gammaX}
\\
\mathrm{Im}[r_0(M)]^{(\gamma X)} &=&
\left[ d_0(M)\, \hat \Gamma_{00}^{(\gamma X)} \,+d_1(M)\, \hat \Gamma_{01}^{(\gamma X)} \,
+ d_2(M)\, \hat \Gamma_{11}^{(\gamma X)} \right] \mathrm{Im}[r_0(M_*)],~~~~~~
\label{eq:Imr0gammaX}
\end{eqnarray}
\label{eq:Imgamma0r0gammaX}%
\end{subequations}
where the coefficient functions $c_n(M)$ and $d_n(M)$ are the same as in eqs.~\eqref{eq:Imgamma0tphi2} and where $\hat \Gamma_{ij}^{(\gamma X)}$ are the entries of the dimensionless matrix obtained by dividing $\bm{\Gamma}^{(\gamma X)}$ by $\pi \alpha_2^2/2M^2$:
\begin{equation}
  \bm{\Gamma}^{(\gamma X)}\ = \frac{\pi \alpha_2^2}{2M^2} \hat{ \bm{\Gamma}}^{(\gamma X)}, \quad
 \hat{ \bm{\Gamma}}^{(\gamma X)} \approx \frac{2s_w^2}{3} 
\begin{pmatrix} 2 f_-  &  \sqrt{2} f_- \\  \sqrt{2} f_- & 3f_+ \end{pmatrix}.
\label{eq:Gamma-hatgammaX}
\end{equation}
We have given the explicit form of $\hat{\bm{\Gamma}}^{(\gamma X)}$ in the leading-double-logarithm approximation.

\begin{figure}[t]
\centering
\includegraphics[width=0.8\linewidth]{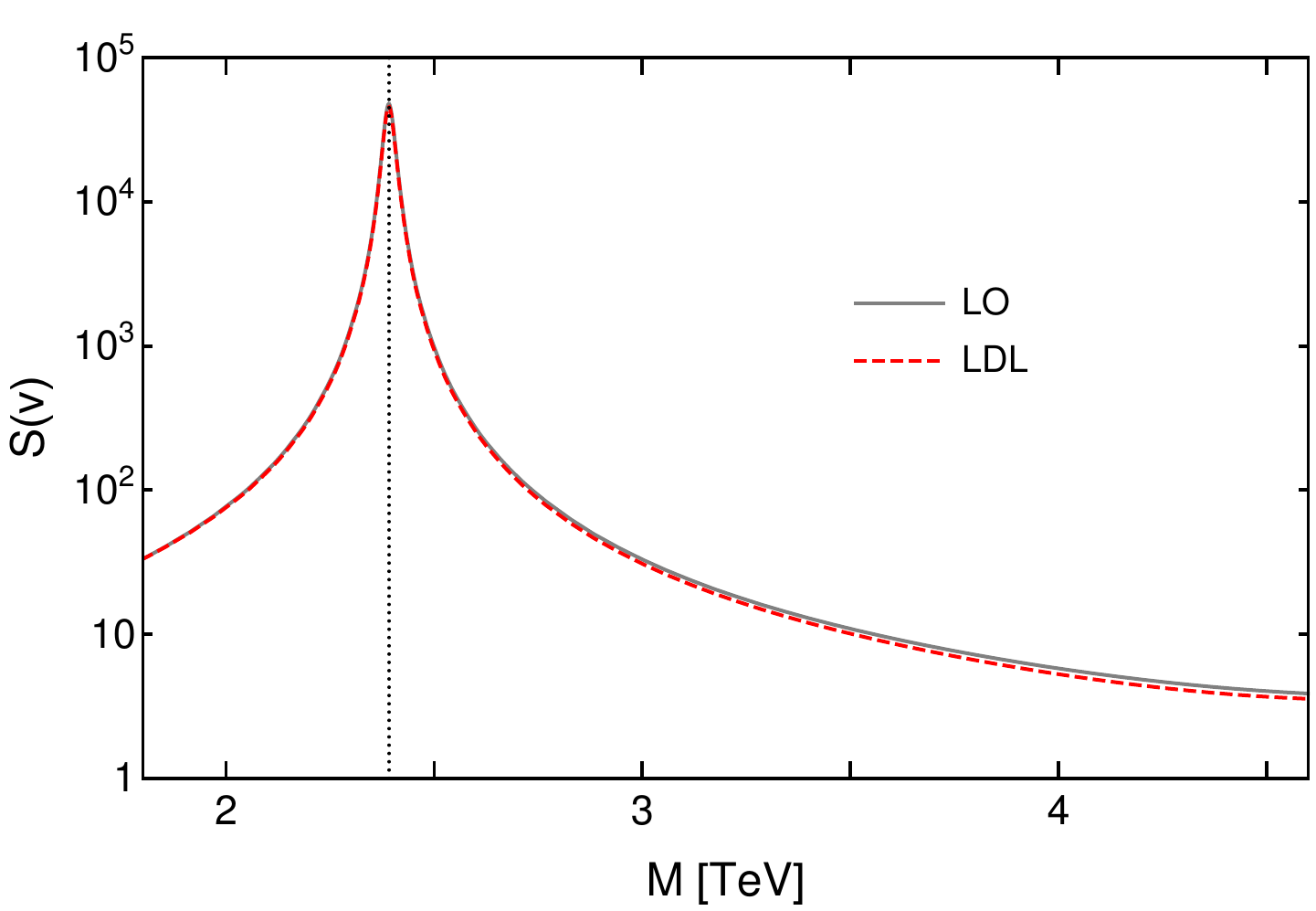}
\caption{Sommerfeld enhancement factor $S(v)$ as a function of the wino mass $M$ for the relative velocity $v=2 v_0=10^{-3}$:
 leading-order approximation  (solid curve) and leading-double-logarithm approximation (dashed curve).}
\label{fig:SommerfeldRelV}
\end{figure}

A Sommerfeld enhancement factor is the ratio of an annihilation rate to the leading-order annihilation rate. We now present an analytic expression for the Sommerfeld enhancement factor $S(v)$ for the semi-inclusive annihilation rate of a neutral-wino pair into a monochromatic photon. At leading order in $\alpha_2$, the annihilation rate of a neutral wino pair into monochromatic photons is the sum of twice eq.~\eqref{eq:sig0-gammagamma} and eq.~\eqref{eq:sig0-gammaZ}:
\begin{equation}
2v_0\sigma[w^0 w^0 \to \gamma +X]  =   \frac{8\pi  \alpha_2^4 s_w^2}{m_W^2}.
\label{eq:sig1-gammaX}
\end{equation}
In ZREFT at LO, the semi-inclusive annihilation rate into a monochromatic photon is obtained from the inclusive annihilation rate in eq.~\eqref{eq:sigma0ann-beta} by replacing Im$[\gamma_0]$ and Im[$t_\phi^2$] in the numerator by the contributions from the monochromatic photon channels in eqs.~\eqref{eq:r0match-gammaX} and \eqref{eq:Imgamma0r0gammaX}. The Sommerfeld enhancement factor is conventionally expressed as a function of the relative velocity $v = 2 v_0(E)$ of the neutral winos. Keeping only the terms in the numerator through first order in the imaginary parts of $\gamma_0$ and $t_\phi^2$, the Sommerfeld enhancement factor reduces to
\begin{equation}
S(v) = \frac{2 m_W^2}{\alpha_2^4 s_w^2 M\, |L_0(E)|^2}
\Big[ \mathrm{Im} [\gamma_0]^{(\gamma X)}  
- \, \mathrm{Im}[t_\phi^2]^{(\gamma X)} \,
\big(  \mathrm{Re}[K_1(E)] -  K_1(0)\big) \Big],
\label{eq:SommerfeldZREFT}
\end{equation}
where $K_1(E)$ and $L_0(E)$ are the functions of $E = Mv^2/4$ in eqs.~\eqref{eq:K1-E} and \eqref{eq:L0-E}. Specifically, the values of $\mathrm{Im} [\gamma_0]$ and $\mathrm{Im} [t_\phi^2]$ in $L_0(E)$ are those for inclusive annihilation, rather than $\mathrm{Im} [\gamma_0]^{(\gamma X)}$ and $\mathrm{Im}[t_\phi^2]^{(\gamma X)}$. The coefficient $\mathrm{Im} [\gamma_0]^{(\gamma X)}$ is given as a function of $M$ in eq.~\eqref{eq:Imgamma0gammaX}. The coefficient $\mathrm{Im}[t_\phi^2]^{(\gamma X)}$ is given as a function of $M$ by eqs.~\eqref{eq:r0match-gammaX} and \eqref{eq:Imr0gammaX}. In figure~\ref{fig:SommerfeldRelV}, the Sommerfeld enhancement factor is shown as a function of $M$ for the relative velocity $v=2 v_0=10^{-3}$. The Sommerfeld enhancement factor using the leading-double-logarithm approximation in Eq.~\eqref{eq:Gamma-hatgammaX} is compared to the leading-order approximation, which is obtained by setting $f_+=2$ and $f_-=0$. The ratio of the leading-double-log and leading-order approximations to $S(v)$ decreases from about 100\% at $M = 1.8$~TeV to about 96\% at $M_*$ and to about 91\% at $M = 4.3$~TeV, and then it increases slowly.

\section{Summary}
\label{sec:Conclusion}

One of the options for a wimp is the neutral wino $w^0$, which belongs to an $SU(2)$ multiplet that also includes the charged winos $w^+$ and $w^-$. The splitting $\delta$ between a charged wino and a neutral wino is small compared to the mass $M$ of the wino. A fundamental description of winos is provided by a relativistic quantum field theory. The most important dimensionless coupling constants are the $SU(2)$ coupling constant $\alpha_2$ and the electromagnetic coupling constant $\alpha$. The physics of nonrelativistic winos involves many momentum scales, including
\begin{itemize}
\item 
the wino mass $M$, which is also the momentum scale of the electroweak gauge bosons 
produced by wino-pair annihilation,
\item 
the weak gauge boson mass scale $m_W$,
\item 
the scale  $\alpha_2 M$  of nonperturbative effects from exchange of weak gauge bosons,
\item 
the Bohr momentum  $\alpha M$, which is the scale of nonperturbative effects from the Coulomb interaction,
\item 
the scale $\sqrt{2M\delta}$ associated with the transition between a neutral-wino pair and a charged-wino pair,
\item 
the real and imaginary parts of the complex inverse scattering length 
$\gamma_0$ of the neutral wino. 
\end{itemize}
Nonrelativistic effective field theories provide simpler descriptions for low-energy winos  in which some of the larger momentum scales are not described explicitly.

If the winos are nonrelativistic, the momentum scale $M$ does not need to be treated explicitly. If $M$ is large enough that $\alpha_2 M$ is comparable to  $m_W$, interactions between nonrelativistic winos from the exchange of the $W^\pm$ and $Z^0$ are nonperturbative. The effects of Coulomb interactions  between charged winos are also nonperturbative. The winos can be described by a nonrelativistic effective field theory  called  NREFT, in which winos interact instantaneously at a distance through a potential generated by the exchange of  weak gauge bosons and in which charged winos have local couplings  to the electromagnetic field. Calculations in NREFT require the numerical solution of a coupled-channel Schr\"odinger equation. The effects of wino-pair annihilation can be taken into account in NREFT by solving the Schr\"odinger equation in eq.~\eqref{eq:radialSchrEqann}, which includes  a delta-function potential with imaginary coefficients given by the matrix $\bm{\Gamma}$ in eq.~\eqref{eq:Gamma}. The exact solutions of this  Schr\"odinger equation take into account the unitarization of wino-pair annihilation effects. In most previous calculations of Sommerfeld enhancement factors $S(v)$ for wino-pair annihilation rates, the delta-function potential has been treated as a first-order perturbation. In this approximation, $S(v)$ has an unphysical divergence as the relative velocity $v$ approaches 0 if the wino mass is tuned to a unitarity mass where the real part of $\gamma_0$ vanishes.

If $M$ is near a unitarity mass, the real part of the inverse scattering length $\gamma_0$ is much smaller than the momentum scales $m_W$ and $\alpha_2 M$. If the relative momentum of winos is smaller than $m_W$ and $\alpha_2 M$, those momentum scales do not need to be described explicitly. In ref.~\cite{Braaten:2017gpq}, we developed a zero-range effective field theory called ZREFT to describe low-energy winos with mass $M$ near a unitarity mass. The effects of the exchange of weak gauge bosons between winos are reproduced by zero-range interactions between the winos, and charged winos have local couplings to the electromagnetic field. In the absence of electromagnetism, ZREFT is a systematically improvable effective field theory. The improvability is guaranteed by identifying a point in the parameter space in which the S-wave interactions of winos are scale invariant in the low-energy limit, so the winos can be described by an effective field theory that is a renormalization-group (RG) fixed point. At the RG fixed point, the mass splitting $\delta$ between the charged wino and the neutral wino is 0, and the electromagnetic coupling constant $\alpha$ is 0. If the wino mass splitting is $\delta = 170$~MeV and if $\alpha=0$, the first unitarity mass where $\gamma_0$ vanishes is $M = 2.88$~TeV. In ref.~\cite{Braaten:2017gpq}, we verified explicitly that, in the absence of electromagnetic interactions, ZREFT at NLO provides systematic improvements in the predictions of ZREFT at LO at $\delta = 170$~MeV and $M = 2.88$~TeV. 

In ref.~\cite{Braaten:2017kci}, we carried out the Coulomb resummation that is necessary to calculate the quantitative predictions  of ZREFT. The parameters of ZREFT at LO are the kinematic parameters $M$ and $\delta$ and the interaction parameters $\alpha = 1/137$, $\phi$, and $\gamma_0 = 1/a_0$. If the wino mass splitting is $\delta = 170$~MeV, the first unitarity mass is $M_* = 2.39$~TeV, and the window for the applicability of ZREFT is $M$ from about 1.8~TeV to about 4.6~TeV. The interaction parameters $\gamma_0(M)$ and $\phi(M)$ of ZREFT were determined as functions  of $M$ by matching scattering amplitudes for neutral winos that were calculated by solving the Schr\"odinger equation for NREFT. In ref.~\cite{Braaten:2017kci}, we verified that ZREFT at LO with Coulomb resummation gives reasonably accurate predictions for low-energy wino-wino cross sections and for the binding energy of a wino-pair bound state at $\delta = 170$~MeV.

In this companion paper to refs.~\cite{Braaten:2017gpq} and \cite{Braaten:2017kci}, we have taken into account the effects of wino-pair annihilation in ZREFT at LO by analytically continuing the interaction parameters $\phi$ and $\gamma_0$ to complex values. The complex  inverse scattering length $\gamma_0(M)$ was determined as a function of $M$ by the numerical solution of the Schr\"odinger equation for NREFT. An accurate Pad\'e approximant for $\gamma_0(M)$ is given in eq.~\eqref{eq:gamma0Pade}. The complex mixing angle $\phi(M)$ was determined as a function of $M$ by using the matching condition in eq.~\eqref{eq:r0match} and the complex effective range $r_0(M)$  calculated by the numerical solution of the Schr\"odinger equation for NREFT. An accurate Pad\'e approximant for $r_0(M)$ is given in eq.~\eqref{eq:r0Pade}. The real and imaginary parts of $\tan \phi(M)$ are shown as functions of $M$ in figure~\ref{fig:tanphivsM}. Sommerfeld enhancement factors  for inclusive annihilation rates can be calculated explicitly in ZREFT in terms of the complex parameters $\gamma_0$ and $\tan^2\phi$. An analytic expression for the Sommerfeld enhancement factor for the inclusive neutral-wino pair annihilation rate that is remarkably simple is given in Eq.~\eqref{eq:SommerfeldZREFTinc}.

Having taken the complex  inverse scattering length  $\gamma_0(M)$ as one of the interaction parameters, the predictions of ZREFT at LO for the neutral-wino elastic cross section at $E=0$ and the neutral-wino-pair annihilation rate at $E=0$ as functions of the wino mass $M$ are exact. The accuracy of the predictions of ZREFT at LO as functions of $M$ was illustrated by the real and imaginary parts of the neutral-wino shape parameter $s_0$ and by the binding energy $E_{(ww)}$ and width $\Gamma_{(ww)}$ of the wino-pair bound state. As shown in figures~\ref{fig:r0s0vsM} and \ref{fig:bindingwidth}, the predictions for  Re$[s_0]$, Im$[s_0]$, $E_{(ww)}$, and $\Gamma_{(ww)}$ are reasonably accurate for $M$ in the region of validity of ZREFT. As shown in figure~\ref{fig:s0r0ratios}, the parameter-free predictions  in eq.~\eqref{eq:s0r0ratio} for the ratio of Re$[s_0]$ and Re$[r_0]$ and the ratio of Im$[s_0]$ and Im$[r_0]$ are reasonably well satisfied. More accurate predictions for the dependence on $M$ could presumably be obtained by using ZREFT at NLO, which has two additional complex parameters that are functions of $M$.

The accuracy of the predictions of ZREFT at LO as functions of the energy $E$ was illustrated by using the wino-pair annihilation cross sections at the unitarity mass $M_*=2.39$~TeV. In the low-energy limit and in the scaling region of $E$, the predictions of ZREFT at LO for the neutral-wino-pair annihilation rate are determined by the imaginary constant $\gamma_0(M_*)$ and therefore cannot be distinguished from the results of ZREFT shown in the log-log plot in figure~\ref{fig:2v0sigma0ann-Elog}. As shown in figure~\ref{fig:sigma0annvsE}, the predictions in the region of $E$ near the charged-wino-pair threshold agree well with the results of NREFT. As shown in figure~\ref{fig:sigma1annvsE}, the predictions of ZREFT at LO for the charged-wino-pair annihilation cross section as a function of $E$ also agree reasonably well with the results of NREFT. More accurate predictions for the dependence on $E$ could presumably be obtained by using ZREFT at NLO, which has two additional complex parameters.

Partial annihilation rates of wino pairs into monochromatic photons is of particular interest, because monochromatic photons provide a signature for the indirect detection of dark matter. In NREFT, a partial annihilation rate can be calculated at leading order in the inclusive annihilation rate by solving a Schr\"odinger equation. The contribution to the matrix $\bm{\Gamma}$ in the imaginary part of the potential from final states with a monochromatic photon defines a matrix  $\bm{\Gamma}^{(\gamma X)}$. The partial annihilation rate into monochromatic photons can be calculated to leading order in the inclusive annihilation rate by replacing $\bm{\Gamma}$ in the potential by $\bm{\Gamma}^{(\gamma X)}$ and then solving the Schr\"odinger equation to leading order in $\bm{\Gamma}^{(\gamma X)}$. This prescription does not take into account the unitarization of wino-pair annihilation. It therefore gives an unphysical divergence in the annihilation rate at zero energy if the wino mass is tuned to a unitarity mass. It is not clear how partial annihilation rates could be calculated beyond leading order in the inclusive annihilation rate using NREFT. 

In ZREFT, it is straightforward to calculate a partial annihilation rate to all orders in the inclusive annihilation rate, provided the contributions from that partial annihilation channel to the matrix $\bm{\Gamma}$ in the imaginary part of the potential for NREFT are known. It is necessary to calculate the imaginary parts of the interaction parameters of ZREFT to first order in the entries $\Gamma_{ij}$ of the matrix $\bm{\Gamma}$. This can be accomplished by fitting the ZREFT parameters using results from solving the Schr\"odinger equation of NREFT for extremely small and variable  values of the entries $\Gamma_{ij}$. For ZREFT at LO, the imaginary parts of $\gamma_0$ and $t_\phi^2$  are expressed  as linear functions of the entries $\Gamma_{ij}$ in eqs.~\eqref{eq:Imgamma0tphi2}. The partial annihilation rate can be obtained from the inclusive annihilation rates in eqs.~\eqref{eq:sigma01ann-beta} by replacing Im$[\gamma_0]$ and  Im$[t_\phi^2]$ in the numerator by the contributions to these parameters from the partial annihilation channel while keeping the full imaginary parts of $\gamma_0$ and $t_\phi^2$ in the denominator. For ZREFT at LO, the contributions to Im$[\gamma_0]$ and  Im$[t_\phi^2]$ from final states with monochromatic photons are given by eqs.~\eqref{eq:r0match-gammaX} and \eqref{eq:Imgamma0r0gammaX}. The Sommerfeld enhancement factor $S(v)$ for the neutral-wino-pair annihilation rate into monochromatic photons is given by the analytic expression  in eq.~\eqref{eq:SommerfeldZREFT}. The unitarization of wino-pair annihilation is taken into account through the imaginary parts of the parameters $\gamma_0$ and $t_\phi^2$ in the denominator.

We have developed a ZREFT for winos with resonant S-wave interactions. The analytic results of ZREFT for the two-body problem, including wimp-wimp cross sections and wimp-pair annihilation rates, are convenient for exploring the effects of resonant interactions on dark matter. ZREFT can also simplify the numerical calculation of more complicated few-body reactions. One such reaction that could have dramatic resonant enhancement at low energy is three-body recombination, in which the collision of three neutral winos produces a wino-pair bound state and a recoiling wino. A ZREFT can also be developed for other wimp models in regions of parameter space with an S-wave resonance near the neutral-wimp-pair threshold. It would be worthwhile to develop a ZREFT for Higgsino wimps with resonant S-wave interactions. It would be a little more complicated than the ZREFT for winos, because there are three coupled spin-singlet S-wave channels instead of only two, but it should still be possible to obtain analytic results for the two-body problem. Finally ZREFT at LO with a wimp mass of about 15~GeV can also be used as a very predictive model for self-interacting dark matter that solves the small-scale structure problems of the universe \cite{Tulin:2017ara}.

\begin{acknowledgments}
This work was supported in part by the Department of Energy under grant DE-SC0011726. We thank T.~Slatyer for valuable discussions of the effects of Coulomb resummation. We thank M.~Baumgart for valuable comments on wino-pair annihilation. Some of our calculations were carried out on the T30 cluster at the Physics Department of Technische Universit\"at M\"unchen.
\end{acknowledgments}

\appendix

\section{Derivation of the Unitarity Condition}
\label{app:unitarity}

If wino-pair annihilation is taken into account, the $2 \times 2$ matrix  $\bm{\mathcal{A}}(E)$ of wino-pair transition amplitudes in a zero-range effective field theory does not satisfy the unitarity condition in eq.~\eqref{eq:A-unitarity}. If wino-pair annihilation is taken into account by analytically continuing the real parameters $\gamma_{ij}$ in the inverse T-matrix to complex values $\gamma_{ij} + i \beta_{ij}$, the unitarity condition is given in eq.~\eqref{eq:A-nonunitarity}. In this appendix, we derive that unitarity condition.

The matrix $\bm{\mathcal{A}}(E)$ of transition amplitudes can be expressed in the form in eq.~\eqref{eq:Amatrixann}, where $\bm{\mathcal{A}}_C(E)$ is the amplitude matrix for Coulomb scattering only  in eq.~\eqref{eq:ACmatrix}, $\bm{W}(E)$ is the diagonal matrix in eq.~\eqref{eq:Wmatrix}, and $\bm{\mathcal{A}}_s(E)$ is the matrix of short-distance amplitudes, whose inverse is given in eq.~\eqref{eq:Asinverse}. The expression  in eq.~\eqref{eq:Amatrixann} remains valid when wino-pair annihilation is taken into account. The unitarity condition can be derived using expressions for the discontinuities in the functions $\mathcal{A}_C(E)$, $W_1(E)$, and $K_1(E)$ at real energies $E$. The unitarity condition for the Coulomb transition amplitude $\mathcal{A}_C$ for $w^+w^-$ in eq.~\eqref{eq:ACoulomb} is
\begin{equation}
\label{eq:discAC}
\mathcal{A}_C(E) - \mathcal{A}_C(E)^* = 
- \frac{M}{4\pi} \mathcal{A}_C(E) \big[ \kappa_1(E) - \kappa_1(E)^* \big] \mathcal{A}_C(E)^*.
\end{equation}
The discontinuity in the amplitude $W_1(E)$ in eq.~\eqref{eq:W1-eta} for $w^+ w^-$ created at a point to have energy $E$ can be expressed as
\begin{equation}
W_1(E) - W_1(E)^* =
- \frac{M}{4\pi} \mathcal{A}_C(E) \big[ \kappa_1(E)  - \kappa_1(E)^* \big] W_1(E)^*.
\label{eq:discW1}
\end{equation}
The discontinuity in the  function $K_1(E)$ in eq.~\eqref{eq:K1-E}, which comes from Coulomb resummation in a $w^+w^-$ bubble diagram, is
\begin{equation}
K_1(E) - K_1(E)^*  = W_1(E) \big[ \kappa_1(E)  - \kappa_1(E)^* \big] W_1(E)^*.
\label{eq:discK1}
\end{equation}

To derive the unitarity condition with wino-pair annihilation taken into account, we begin with the expression that vanishes upon using the unitarity condition in the absence of wino-pair annihilation in eq.~\eqref{eq:A-unitarity}:
\begin{equation}
\bm{\mathcal{A}}(E)- \bm{\mathcal{A}}(E)^*
+ \frac{1}{8 \pi} \bm{\mathcal{A}}(E) \bm{M}^{1/2} \big[ \bm{\kappa}(E)  - \bm{\kappa}(E)^* \big] 
\bm{M}^{1/2} \bm{\mathcal{A}}(E)^*.
\label{eq:unitarity-1}
\end{equation}
Upon replacing $\bm{\mathcal{A}}$ by $\bm{\mathcal{A}}_C+\bm{W}\bm{\mathcal{A}}_s\bm{W}$, the expression in eq.~\eqref{eq:unitarity-1} can be written as
\begin{eqnarray}
\big(\bm{\mathcal{A}}_C- \bm{\mathcal{A}}_C^* \big)
+\bm{W}\bm{\mathcal{A}}_s(\bm{W} - \bm{W}^*) + \bm{W} (\bm{\mathcal{A}}_s - \bm{\mathcal{A}}_s^*) \bm{W}^* 
+ (\bm{W} - \bm{W}^*)\bm{\mathcal{A}}_s^* \bm{W}^*
\nonumber\\
 + \frac{1}{8 \pi} \big(\bm{\mathcal{A}}_C+\bm{W}\bm{\mathcal{A}}_s\bm{W}\big) \bm{M}^{1/2} 
\big[ \bm{\kappa}  - \bm{\kappa}^* \big] 
\bm{M}^{1/2} \big(\bm{\mathcal{A}}_C^*+\bm{W}^*\bm{\mathcal{A}}_s^*\bm{W}^*\big).
\label{eq:unitarity-2}
\end{eqnarray}
The term $\bm{\mathcal{A}}_C- \bm{\mathcal{A}}_C^*$ is cancelled by the term with factors of both $\bm{\mathcal{A}}_C$ and $\bm{\mathcal{A}}_C^*$ by the unitarity condition for $\mathcal{A}_C(E)$ in eq.~\eqref{eq:discAC}. The remaining terms can be rearranged as
\begin{eqnarray}
&& \left(\bm{W} - \bm{W}^* 
+ \frac{1}{8 \pi} \bm{\mathcal{A}}_C\bm{M}^{1/2} \big[ \bm{\kappa}  - \bm{\kappa}^* \big] \bm{M}^{1/2} \bm{W}^* \right)
\bm{\mathcal{A}}_s^* \bm{W}^*  
\nonumber\\
&& +  \bm{W} \bm{\mathcal{A}}_s
 \left( \bm{W} - \bm{W}^* 
+ \frac{1}{8 \pi} \bm{W} \bm{M}^{1/2} \big[ \bm{\kappa}  - \bm{\kappa}^* \big] \bm{M}^{1/2} \bm{\mathcal{A}}_C^*  \right)
\nonumber\\
&&- \bm{W} \bm{\mathcal{A}}_s  
\left( \bm{\mathcal{A}}_s^{-1} - {\bm{\mathcal{A}}_s^*}^{-1}  
- \frac{1}{8 \pi} \bm{W} \bm{M}^{1/2} \big[ \bm{\kappa}  - \bm{\kappa}^* \big] 
\bm{M}^{1/2} \bm{W}^* \right)  \bm{\mathcal{A}}_s^* \bm{W}^*.
\label{eq:unitarity-3}
\end{eqnarray}
The first and second lines vanish upon using the expression for the discontinuity in $W_1(E)$ in eq.~\eqref{eq:discW1}. Upon inserting the expression for $\bm{\mathcal{A}}_s^{-1}$ in eq.~\eqref{eq:Asinverse}, the expression in eq.~\eqref{eq:unitarity-3} reduces to
\begin{equation}
- \frac{1}{8 \pi} \bm{W} \bm{\mathcal{A}}_s  \bm{M}^{1/2}
 \big( [\bm{K}  - \bm{K}^*]  - 2 i \bm{\beta} - \bm{W} [\bm{\kappa}  - \bm{\kappa}^*] \bm{W}^* \big)
\bm{M}^{1/2}  \bm{\mathcal{A}}_s^* \bm{W}^*.
\label{eq:unitarity-4}
\end{equation}
The term with the factor $\bm{K}  - \bm{K}^*$ cancels the term with the factor $\bm{\kappa}  - \bm{\kappa}^*$ upon using the expression for the discontinuity in $K_1(E)$ in eq.~\eqref{eq:discK1}. The expression in eq.~\eqref{eq:unitarity-4} then reduces to
\begin{equation}
\frac{i}{4 \pi} \bm{W}(E)\,  \bm{\mathcal{A}}_s(E)  \, \bm{M}^{1/2}\bm{\beta} 
\bm{M}^{1/2}  \, \bm{\mathcal{A}}_s(E)^* \, \bm{W}(E)^*.
\label{eq:unitarity-5}
\end{equation}
The equality between this expression and the original expression in eq.~\eqref{eq:unitarity-1} is the unitarity condition in eq.~\eqref{eq:A-nonunitarity}.


\end{document}